\documentclass[aps,pra,preprint,showpacs,preprintnumbers,amsmath,amssymb,footinbib]{revtex4-1}
\usepackage{graphicx}
\usepackage[colorlinks=true, pdfstartview=FitV,linkcolor=red,citecolor=blue,
urlcolor=blue]{hyperref}
\usepackage{ulem}

\newcommand{\fg}{{\cal G}}
\newcommand{\Nc}{N}
\newcommand{\mf}{m_\text{qk}}
\newcommand{\minf}{m}

\newcommand{\mD}{M}

\newcommand{\Nf}{N_f}
\newcommand{\Acal}{{\cal A}}

\newcommand{\slashed}{ /\!\!\!}
\newcommand{\beq}{\begin{equation}}
\newcommand{\eeq}{\end{equation}}
\newcommand{\ba}{\begin{eqnarray}}
\newcommand{\ea}{\end{eqnarray}}
\newcommand{\no}{\nonumber \\}
\newcommand{\gsim}{\mathrel{\hbox{\rlap{\lower.55ex \hbox {$\sim$}}
                   \kern-.3em \raise.4ex \hbox{$>$}}}}
\newcommand{\lsim}{\mathrel{\hbox{\rlap{\lower.55ex \hbox {$\sim$}}
                   \kern-.3em \raise.4ex \hbox{$<$}}}}
\def\be{\begin{eqnarray}}
\def\ee{\end{eqnarray}}

\def\vk{{\vec k}}

\def\vp{{\vec p}}

\def\roughly#1{\mathrel{\raise.3ex\hbox{$#1$\kern-.75em%
\lower1ex\hbox{$\sim$}}}}
\def\lsim{\roughly<}
\def\gsim{\roughly>}

\def\vx{{\vec x}}

\def\({\left(}
\def\){\right)}
\def\[{\left[}
\def\]{\right]}

\def\lsim{\mathrel{\rlap{\lower3pt\hbox{\hskip1pt$\sim$}}
     \raise1pt\hbox{$<$}}} %less than or approx. symbol
\def\gsim{\mathrel{\rlap{\lower3pt\hbox{\hskip1pt$\sim$}}
     \raise1pt\hbox{$>$}}} %greater than or approx. symbol

\newcommand{\vgamma}{{\vec \gamma}}
\newcommand{\Tr}{\mathrm{tr}}
\newcommand{\cp}{g}
\newcommand{\nf}{\widetilde{n}}
\newcommand{\nb}{n}
\newcommand{\Slash}[1]{\ooalign{\hfil/\hfil\crcr$#1$}}
\newcommand{\lra}{\leftrightarrow}
\def\af{\alpha}
\def\bt{\beta}
\def\gm{\gamma}
\def\Gm{\Gamma}
\def\vGm{\varGamma}
\def\sig{\sigma}
\def\Sig{\Sigma}

\def\lam{\lambda}
\def\dlt{\delta}
\def\Dlt{\Delta}
\def\eps{\epsilon}

\def\para{\parallel}

\def\bp{{\bf k}}
\def\bq{{\bf q}}
\def\pl{k_\parallel}
\def\pt{k_\perp}
\def\sknew{p}
\def\lknew{P}
\def\spnew{k}
\def\lpnew{K}
\def\lqnew{Q}

\begin{document}
\preprint{RIKEN-QHP-155, RBRC-1121}

\title{Dilepton and photon production in the presence of a nontrivial
Polyakov loop}
\date{\today}
\author{Yoshimasa Hidaka}
\email{hidaka@riken.jp}
\affiliation{Theoretical Research Division,
Nishina Center, RIKEN, Wako 351-0198, Japan}
\author{Shu Lin}
\email{slin@quark.phy.bnl.gov}
\affiliation{RIKEN/BNL Research Center, Brookhaven National Laboratory,
Upton, NY 11973}
\author{Robert D. Pisarski}
\email{pisarski@bnl.gov}
\affiliation{
Department of Physics, Brookhaven National Laboratory,
Upton, NY 11973}
\affiliation{RIKEN/BNL Research Center, Brookhaven National Laboratory,
Upton, NY 11973}
\author{Daisuke Satow}
\email{daisuke.sato@riken.jp}
\affiliation{
European Centre for Theoretical Studies in Nuclear Physics
and Related Areas (ECT*) and Fondazione Bruno Kessler,\\
Villa Tambosi, Strada delle Tabarelle 286, I-38123 Villazzano (TN) Italy\\}

\begin{abstract}
We calculate the production of dileptons and photons 
in the presence of a nontrivial Polyakov loop in QCD.
This is applicable to the semi-Quark Gluon Plasma (QGP), 
at temperatures above but near
the critical temperature for deconfinement.  The Polyakov loop is
small in the semi-QGP, and near unity in the perturbative QGP.
Working to leading order in the coupling constant of QCD, 
we find that there is a mild enhancement, $\sim 20\%$, for dilepton
production in the semi-QGP over that in the perturbative QGP.
In contrast, we find that 
photon production is {\it strongly} suppressed in the semi-QGP,
by about an order of magnitude, relative to the perturbative QGP.
In the perturbative QGP photon production contains contributions 
from $2 \rightarrow 2$ scattering and collinear emission with the
Landau-Pomeranchuk-Migdal (LPM) effect. In the semi-QGP we show 
that the two contributions are modified differently.
The rate for $2\rightarrow 2$ scattering is suppressed by a 
factor which depends upon the Polyakov loop.
In contrast, in an $SU(\Nc)$ gauge theory
the collinear rate is suppressed by $1/\Nc$, 
so that the LPM effect vanishes at $\Nc = \infty$.
To leading order in the semi-QGP at large $\Nc$, we 
compute the rate from $2\rightarrow 2$ scattering
to the leading logarithmic order and 
the collinear rate to leading order.
\end{abstract}

\maketitle

\section{Introduction}

In many ways, the collisions of heavy ions at high energies 
appear to be well 
described by thermal properties of a Quark-Gluon Plasma (QGP).  Certainly the
bulk properties of hadrons are accurately modeled by a nearly ideal plasma,
using hydrodynamics
\cite{Heinz:2009xj, *Heinz:2013th, *Gale:2013da, Schenke:2010nt, Schenke:2010rr}.

It is also important to consider electromagnetic probes of a QGP
such as dilepton 
\cite{Baier:1991em, *Kapusta:1991qp, Aurenche:2000gf, Arnold:2002ja, *Arnold:2001ba,*Arnold:2001ms, *Arnold:2002ja, Lee:1998nz, Dusling:2006yv, *Dusling:2007su, *Dusling:2009ej, *Manninen:2010yf, *Staig:2010by, *Linnyk:2011vx, *Linnyk:2012pu, *Hohler:2013eba, *Lee:2014pwa, Rapp:2013nxa, Vujanovic:2013jpa}
and photon 
\cite{Chatterjee:2005de, *Bratkovskaya:2008iq, *vanHees:2011vb, *Basar:2012bp, *Bzdak:2012fr, *Fukushima:2012fg, *Liu:2012ax, *Shen:2013cca, *Shen:2013vja, *Linnyk:2013hta, *Linnyk:2013wma, *Muller:2013ila, Mamo:2013efa, *Basar:2014swa,*vanHees:2014ida, *McLerran:2014hza, *Monnai:2014kqa, Dion:2011pp} production.  
Theoretically, these
can be computed in the (resummed) perturbative QGP at high temperature
\cite{Haque:2014rua}, 
by using hadronic models at low temperature
\cite{Huovinen:2009yb, *Andronic:2012ut},
and using the AdS/CFT correspondence 
\cite{CaronHuot:2006te,Parnachev:2006ev,Atmaja:2008mt}.  
Neither applies directly to
Quantum ChromoDynamics (QCD) at temperatures near the 
the phase transition, at a temperature $T_c$.

The experiments demonstrated several phenomena which are difficult to
explain using these methods.  For dileptons, there is an enhancement at
invariant masses below that for the $\rho$-meson.  This is observed from
energies at the Super Proton Synchotron (SPS) at CERN, 
to the Relativistic Heavy Ion Collider (RHIC) at Brookhaven 
National Laboratory, and onto the 
Large Hadron Collider (LHC) at CERN \cite{Rapp:2013nxa}.

Another puzzle appears in the photon spectrum:  
there is
an unexpectedly large elliptic flow for photons at moderate momenta,
comparable to the elliptical flow observed for hadrons
\cite{Adare:2011zr, Lohner:2012ct}.  This large
elliptic flow for photons is very difficult to understand from either
a perturbative analysis or from AdS/CFT.

In this paper we consider electromagnetic signals in 
a matrix model of the semi-QGP,
which is constructed to describe QCD at temperatures near and
above $T_c$
\cite{Pisarski:2006hz, Hidaka:2008dr, Hidaka:2009hs, Hidaka:2009xh,
Hidaka:2009ma, Dumitru:2010mj, *Dumitru:2012fw, *Kashiwa:2012wa,
*Pisarski:2012bj, *Kashiwa:2013rm, *Lin:2013qu, Lin:2013efa}.
The relevant parameter is the
expectation value of the Polyakov loop: properly normalized, the
expectation value of the loop is near unity in the perturbative QGP
~\cite{Gava:1981qd, *Burnier:2009bk, *Brambilla:2010xn}.
Numerical simulations on the lattice
\cite{Bazavov:2009zn, *DeTar:2009ef, *Fodor:2009ax, *Petreczky:2012rq,
*Borsanyi:2013bia, *Bhattacharya:2014ara, *Sharma:2013hsa}
find that for QCD, there is no true phase transition, only a rapid increase
in the number of degrees of freedom.  For our purposes,
whether or not there is a true phase transition is irrelevant:
all that matters is that the (renormalized) Polyakov loop,
which from the lattice is $\langle \ell \rangle \sim 0.1$ at $T_c$, 
is small 
\cite{Bazavov:2009zn, *DeTar:2009ef, *Fodor:2009ax, *Petreczky:2012rq,
*Borsanyi:2013bia, *Bhattacharya:2014ara, *Sharma:2013hsa}.

A brief summary of the results of this analysis has appeared previously
\cite{previous}.  In this paper we
describe the computations in full.  These are straightforward,
simply a matter of computing in the presence of
a background field for the time-like
component of the gluon vector potential, $A_0$.  
We then compute to leading order in the QCD coupling $g$.  
These formalisms will be explained in Sec.~\ref{sec:model}.
For photons, we only compute to leading logarithmic order, 
which means that we regard the logarithm of 
some large number as much larger than unity.

In the semi-QGP, the production of colored particles is suppressed
by powers of the Polyakov loop as $T \rightarrow T_c$.  This is natural, 
as in the pure gauge theory, there are no colored particles in the
confined phase. 
Thus one might expect that dilepton production is suppressed in the semi-QGP,
relative to that in the perturbative phase.   
We make this comparison
at the same temperature, and the same value of the QCD coupling, so that
the ratio is only a function of the value of the 
Polyakov loop in Sec.~\ref{sec:dilepton}.
In contrast to the naive expectation above, we find a mild 
enhancement of dilepton
production in the semi-QGP, even into the confined phase.  This is because
for an off-shell photon, it can proceed directly through a color singlet
channel, a quark anti-quark pair.  While a single quark or anti-quark
is suppressed by a power of the Polyakov loop, a quark anti-quark
pair is not.  We also show that to leading order,
a Polyakov Nambu-Jona-Lasino
model \cite{Fukushima:2003fw,Hansen:2006ee, Ghosh:2014zra} 
gives essentially the same
result for dilepton production \cite{Islam:2014sea} as our matrix model.
As we discuss, this equality is not true beyond leading order.

The production of real photons, which will be analyzed 
in Sec.~\ref{sec:photon-2to2}, is very different.
Kinematically, a photon on its mass shell cannot decay directly 
into a quark anti-quark pair.  Therefore, 
the leading contribution is from a 2 to 2 scattering, which includes 
the Compton scattering of a quark with a gluon and the 
pair annihilation of a quark anti-quark pair.
These particles also can form a color singlet like the case 
of the dilepton production, but for an $SU(N)$ gauge theory, 
the ratio of the color singlet state to the number of all the 
states is suppressed by $1/N^2$ at large $N$.
Consequently, we find a strong suppression of real 
photon production in the semi-QGP.  
The contribution from the collinear emission of the photon, 
which also can contribute at the leading order to the 
photon production, is discussed in Sec.~\ref{sec:photon-collinear}.

%%%%%%%%%%%%%%%%%%%%%%%%%%%%%%%%%%%%%%%%%%%
\section{Semi-quark gluon plasma}
\label{sec:model}

\subsection{Double line notation}
It is useful to compute the color factors using the double line basis
\cite{Hidaka:2009hs}. In this basis, as usual
fundamental quarks carry a single index in the fundamental
representation, $a=1,\cdots,\Nc$.  Gluons, however,
carry a pair of fundamental indices, $(ab)$.
For an $SU(\Nc)$ group there are $\Nc^2$ such pairs, and so this basis
is overcomplete by one generator.  
This is compensated by introducing the operator
\begin{align}
{\cal P}^{ab}_{cd}=\dlt^a_c\dlt^b_d-\frac{1}{\Nc}\dlt^{ab}\dlt_{cd} \; .
\end{align}
This is a projection operator,
\begin{align}
{\cal P}^{ab}_{ef}{\cal P}^{ef}_{cd}={\cal P}^{ab}_{cd} \; .
\end{align}
In the double line basis, the vertex between a quark anti-quark pair and
a gluon is proportional to this projection operator,
\begin{align}
\label{eq:quark-gluon-vertex}
\(T^{ab}\)_{cd}=\frac{1}{\sqrt{2}}{\cal P}^{ab}_{cd} \; .
\end{align}
The other vertices are not relevant for the present discussion.

\subsection{The Polyakov loop in Euclidean spacetime}

To introduce the effect of nontrivial Polyakov loop in perturbative 
calculation, we work in an effective model introduced in 
Ref.~\cite{Hidaka:2009hs}.
The Lagrangian of that model is the same as that in QCD 
with $\Nc$ colors, but in a mean field type approximation,
we take the temporal component of the gluon field to
be a constant, diagonal matrix,
\begin{align} 
A_0^{a b} &=\frac{1}{\cp}\; \delta^{a b} \; Q^a \; ,
\label{mean_field}
\end{align}
where $\cp$ is the coupling constant.
There is no background field for the spatial components of the gluon, $A_i$.
As the gauge group is $SU(\Nc)$, $A_0$ is traceless, and
the sum of the $Q$'s vanishes, $\sum_a Q^a=0$.

The Wilson line 
in the temporal direction is
%
%\begin{align}
%{\bf L}(\vx)&\equiv \left\langle{\cal{P}}
%\exp\left(i \, g \, \int^{1/T}_0 d\tau \; A_0(\tau,\vx)\right) \right\rangle
%\end{align}
\begin{align}
{\bf L}(\vx)&\equiv {\cal{P}}
\exp\left(i \, g \, \int^{1/T}_0 d\tau \; A_0(\tau,\vx)\right) \; ,
\end{align}
where $\cal P$ denotes path ordering and $\tau$ is the imaginary time,
$\tau: 0 \rightarrow 1/T$.

To leading order in the coupling constant, a mean field approximation
implies that we can neglect fluctuations in $A_0$.  
The variable $Q$ is naturally proportional to the temperature, so
it is useful to introduce a dimensionless variable $q$, where
\begin{equation}
Q^a = 2 \pi T \; q^a \; .
\label{define_q_Q}
\end{equation}
In %expressions 
this paper we shall use both the $Q^a$'s and the $q^a$'s.  For
intermediate expressions the $Q^a$'s are more convenient, but
final expressions are simpler in terms of the $q^a$'s.

Traces of powers of the the Wilson line are Polyakov loops,
\begin{align}
\begin{split}
\ell_n(Q) &\equiv \frac{1}{\Nc}\; \langle\Tr \; {\bf L}^n\rangle
= \frac{1}{\Nc}\; \sum_{a=1}^{\Nc} e^{i \, n \, Q^a/T} \; ,
\end{split}
\label{define_polyakov_loop}
\end{align}
and are gauge invariant.
Since it arises frequently we write a loop without the subscript as the
first Polyakov loop, $\ell= \ell_1$.

In general there are $\Nc - 1$ independent $Q^a$'s.  
For the problems of interest in this paper, though, we can perform
a global color rotation to enforce that the expectation value of
the loop $\ell$ is real.  
This implies that the eigenvalues pair up as
\begin{equation}
\label{eq:Q-distribution}
Q^a = (- Q^j , \, -Q^{j-1}  \ldots -Q^{1} , \, 0 , \, Q^{1} 
\ldots Q^{j-1} , \, Q^j ) \; ,
\end{equation}
where we assume that $\Nc$ is odd, and
$j = (\Nc - 1)/2$.  When $\Nc$ is even the zero eigenvalue is dropped,
and there are $j = \Nc/2$ pairs.  
Thus in general there are $j$ independent eigenvalues.
For an arbitrary value of the loop, there is no simple relation between
these eigenvalues.

Nevertheless, there are two exceptions.  One is the perturbative QGP, where all $Q^a$
vanish.  The other is the confined phase of a pure gauge theory, 
\begin{equation}
Q^a_{{\rm conf}}
 %= 2 \pi T \; \frac{k}{\Nc} 
%\;\; , \;\; k = -j, -j+1 \ldots - 1, 0, 1 \ldots j-1, j \; .
= \pi T \; \frac{\Nc+1-2k}{\Nc} 
\;\; , \;\; k = 1, \dots, \Nc \; .
\label{confined_vac_odd}
\end{equation}
That is, in the confined phase the eigenvalues are evenly distributed
on the unit circle.  
The loops in the confined phase are
\begin{align}
\label{eq:ln-confined}
\ell_n(Q_{{\rm conf}})=\left\{ \begin{array}{ll}
(-1)^{j(\Nc+1)} & \;\; , \;\; n=j\Nc \; ; \\
0 & \;\; , \;\; n \neq j \Nc \; ,
\end{array} \right.
\end{align}
for general $\Nc$.
This behavior is easy to understand.  Loops 
which carry $Z(\Nc)$ charge
vanish in the confined phase of the pure gauge theory, 
while those which are $Z(\Nc)$ neutral
do not.  

For three colors, 
\begin{equation}
Q^a = (-Q , \, 0 , \, Q) = 2 \pi T \; (-q, 0, q) \; .
\label{three_colors_eigenvalues}
\end{equation}
The first Polyakov loop is then
\begin{equation}
\ell = \frac{1}{3} \left( 1 + 
2\, \cos\left(2 \pi q\right) \right) \; .
\label{eq:l-Q}
\end{equation}
In the confined phase of the pure gauge theory $q_{{\rm conf}} = 1/3$.
Similarly, 
\begin{equation}
\label{eq:ln-Q}
l_n=\frac{1}{3}(1+2\cos(2\pi \, n\, q)) \; .
\end{equation}

In the presence of dynamical quarks there is no rigorous definition of a
confined phase.  Dynamical quarks act as a background $Z(\Nc)$ field,
so that any Polyakov loop is nonzero at nonzero temperature.  Nevertheless,
numerical simulations on the lattice find that $\ell$ is small,
$\langle \ell \rangle \sim 0.1$, at
the phase transition, at least for three colors and three light flavors.
Thus we shall find it very convenient to compare results in the perturbative
QGP to those in the confined phase of the pure gauge theory, as a
limiting case of how large the effects can possibly be.

\subsection{Analytic continuation to Minkowski space-time}
\label{sec:analytic}

Expanding %about
around the background field in Eq.~(\ref{mean_field}), in Euclidean spacetime the
energy of a quark becomes 
\begin{equation}
p_0 \rightarrow p_0 + Q^a \; , 
\end{equation}
while that of a gluon becomes
\begin{equation}
p_0 \rightarrow p_0 +  Q^{ab} \; ;\;  \;\;\; Q^{a b} \equiv Q^a - Q^b \; ,
\end{equation}
where $a$ and $b$ are color indices of the quark and the gluon in the double line basis~\cite{Hidaka:2009hs}.
Because of the usual
boundary conditions in imaginary time, the energy $p_0$ for
a fermion is an odd multiple of $\pi T$, while that for a boson is
an even multiple of $\pi T$.

Although the momenta for fermions and bosons are rather different in
Euclidean spacetime, it was argued previously
that the proper procedure for
analytic continuation to Minkowski spacetime is to continue the entire
Euclidean energy to $- i E$, where $E$ is a continuous energy variable
\cite{Hidaka:2009hs}. 

This has a simple but profound implication.  
In kinetic theory
a given process is given by
an integral over phase space of the square of a matrix elements
times products of statistical distribution functions.
Since the energies in Minkowski spacetime are as usual, then,
for processes in which all the momenta are hard, the only change is
in the $Q$-dependence of the statistical distribution functions.  
For processes involving soft momenta, 
it is also necessary to include the $Q$-dependence of the hard thermal
loops as well \cite{Hidaka:2009hs}.
We shall illustrate these general expectations by our explicit computations.
It also suggests that it may be useful to treat the semi-QGP in
kinetic theory, as for the perturbative QGP \cite{Blaizot:2001nr,Baym:2006qf}.

The background gluon field acts as an imaginary 
chemical potential for colored particle, so that 
the statistical distribution functions for the quark, anti-quark, and gluon
are, respectively, 
\begin{align}
\nf_a(E)&= \frac{1}{e^{(E-iQ^a)/T}+1} \;\; , \;\; 
\nf_{\overline{a}} (E) = \frac{1}{e^{(E+iQ^a)/T}+1} \; ,\no
\nb_{ab}(E)&=  \frac{1}{e^{(E-i(Q^a - Q^b))/T}-1} \; .
\label{q_distribution_functions}
\end{align}
Notice that the sign of $Q$ for the anti-quark, $+ i Q^a$
in $\nf_{\overline{a}} (E)$, is opposite to that for the quark,
$- i Q^a$ in $\nf_{a} (E)$.
This is just like the change in sign for a quark chemical potential
which is real.  When the $Q^a = 0$, of course these reduce to the usual
Fermi-Dirac and Bose-Einstein distribution functions.

For future reference, it is useful to compute the statistical
distribution functions, summed over all colors, in the confined
phase of a pure gauge theory,
Eqs.~(\ref{confined_vac_odd}) and (\ref{eq:ln-confined}).
For the quark distribution function, this is
\begin{align}
\frac{1}{\Nc} \sum_{a = 1}^{\Nc} \; 
\frac{1}{e^{(E-iQ^a_{{\rm conf}})/T}+1} = \frac{1}{e^{\Nc E/T}+1} \; ,
\label{sum_qk_dist}
\end{align}
while that of the gluon distribution function is
\begin{align}
\frac{1}{\Nc^2} \sum_{a,b = 1}^{\Nc} \; 
\frac{1}{e^{(E-iQ^a_{{\rm conf}} +i Q^b_{{\rm conf}})/T}-1} \; 
= \frac{1}{e^{\Nc E/T}-1} \; .
\label{sum_gluon_dist}
\end{align}
In the confined phase of the pure gauge theory, the only loops which
contribute are those which wrap around a multiple of $\Nc$ times.
These can be considered as a type of ``baryon'',
albeit in the pure gauge theory.  Consequently, the energy which enters
in the right hand side of Eqs.~(\ref{sum_qk_dist}) and (\ref{sum_gluon_dist})
is not $E$, but $\Nc$ times $E$.  
This rescaling of the energy
will be seen to help explain the suppression of photon production at
large $\Nc$, Eq.~(\ref{ratio_pert_semi_photon}).

\subsection{Relation to lattice results}

Strictly speaking, $A^0$ and thus $Q^a$ should be determined 
dynamically from %a complete 
our model itself.  
Instead, in this paper, we determine these quantities from numerical simulations
on the lattice, following Ref.~\cite{Lin:2013efa}.
First, in order to extract the Polyakov loop determined by nonperturbative dynamics, it is necessary to remove perturbative corrections from the
expectation value of the loop ~\cite{Gava:1981qd},
\begin{align}
\label{eq:correction-Polyakov}
\ell(Q=0)&= 1+\delta \ell(Q=0) \; ,\\
\nonumber
\delta \ell(Q=0)&= \frac{\cp^2 C_f m_D}{8\pi T}
+\frac{\cp^4 C_f}{(4\pi)^2}
\Bigl[-\frac{\Nf}{2}\ln 2
+ \Nc\left(\ln \frac{m_D}{T}+\frac{1}{4}\right)\Bigr]
+{\cal O}(\cp^5) \; ,
\end{align}
where $C_f\equiv(\Nc^2-1)/(2\Nc)$ is the Casimir for the
fundamental representation, 
$m_D$ is the Debye mass of the gluon, and 
$\Nf$ is number of quark flavors.
We use the running coupling constant calculated in the 
modified minimal subtraction scheme at two-loop order, 
and the expression of the Debye mass at one-loop order 
\cite{Burnier:2009bk}:
\begin{align} 
\nonumber
g^2&= 24\pi^2\biggl[(11\Nc-2\Nf)\left\{\ln
\left(\frac{4\pi T}{\varLambda_{\overline {\text{MS}}}}\right)
-\gamma_E\right\} 
+\Nf(4\ln 2-1)-\frac{11\Nc}{2}\biggr]^{-1} \; ,\\
\nonumber
m^2_D&= (2\Nc+\Nf)4\pi^2T^2\biggl[(11\Nc-2\Nf)
\left\{\ln\left(\frac{4\pi T}{\varLambda_{\overline {\text{MS}}}}\right)
-\gamma_E\right\}\\
&+4\Nf \ln 2-\frac{5\Nc^2+\Nf^2+9\Nf/(2\Nc)}{2\Nc+\Nf}\biggr]^{-1} \; .
\end{align}
Here $\varLambda_{\overline {\text{MS}}}$ is the 
renormalization mass scale, in the modified minimal subtraction scheme, 
and $\gamma_E\simeq 0.57721$ is Euler's constant.

\begin{figure*}[t] 
\begin{center}
\includegraphics[width=1.0\textwidth]{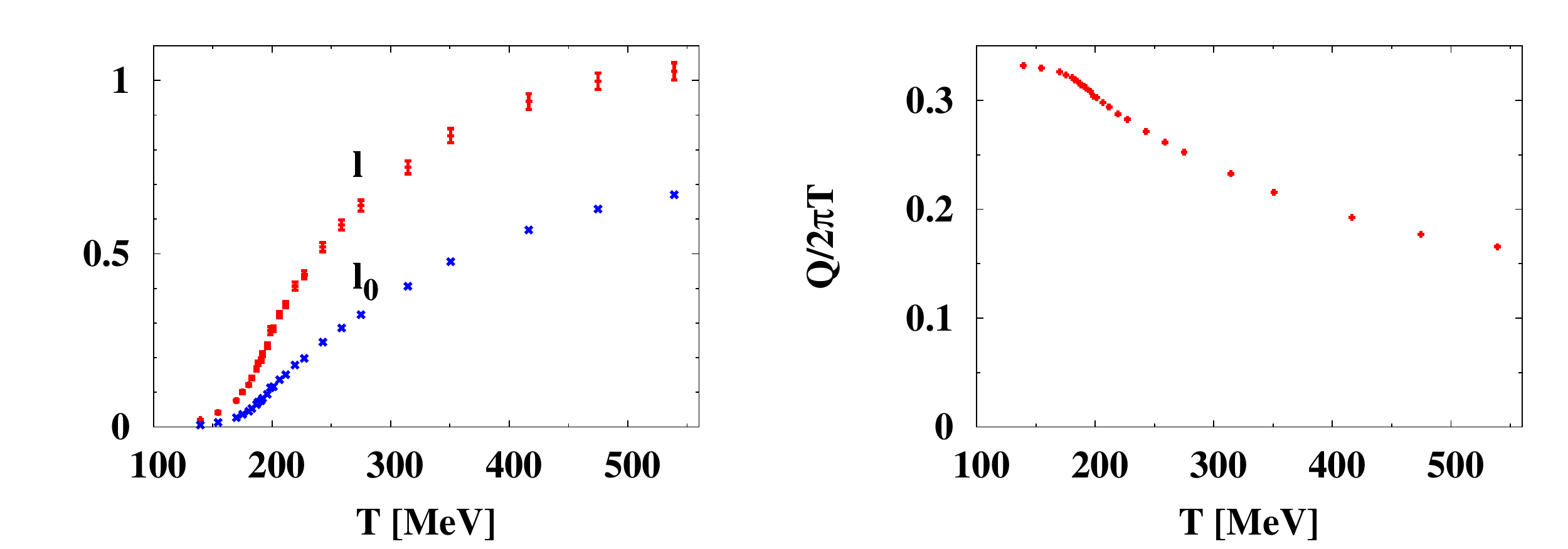}
\caption{Left panel: The Polyakov loop ($l$) determined 
from the lattice calculation~\cite{Bazavov:2009zn}, the 
Polyakov loop in which the perturbative correction is removed ($l_0$) as 
a function of $T$.
Right panel: $Q$ as a function of $T$.
We set $\varLambda_{\overline {\text{MS}}}=T_c/1.35$, where $T_c=170$ MeV.
}
\label{fig:Polyakov-Q}
\end{center}
\end{figure*}

Equation~(\ref{eq:correction-Polyakov}) shows that a finite renormalization
gives $\ell(Q=0) > 1$.  
We assume that perturbative corrections exponentiate,
\begin{align}
\label{eq:correction-Polyakov-Q}
\ell(Q)&= e^{\delta \ell(Q=0)} \ell_{0}(Q) \; .
\end{align}
We take $\ell$ from numerical simulations of 
lattice QCD~\cite{Bazavov:2009zn}, 
and calculate $\ell_0$ from Eq.~(\ref{eq:correction-Polyakov-Q}), 
to obtain $Q$ from Eq.~(\ref{eq:l-Q}).
These quantities are plotted in Fig.~(\ref{fig:Polyakov-Q}), 
by setting $\varLambda_{\overline {\text{MS}}}=T_c/1.35$.
We see that $\ell_0$ is different from unity even around 
$\sim 3T_c$, where $T_c \sim  170$~MeV is the pseudo-critical temperature
of the phase transition ~\cite{Bazavov:2009zn}.

%%%%%%%%%%%%
%%%%%%%%%%%%

%%%%%%%%%%%%%%%%%%%%%%%%%%%%%%%%%%%%%%%%%%%
\section{Dilepton production rate}
\label{sec:dilepton}

\subsection{Computation to leading order}

We calculate the production rate of dileptons when
$Q_a\neq 0$ in this subsection.
To leading order in $\alpha_\text{em}$, the production rate is %proportional to the retarded photon self-energy
%,
\begin{align} 
\frac{d\varGamma}{d^4P}
%&=-\; \frac{\alpha_\text{em}}{96\pi^5 P^2}
&=-\frac{\alpha_\text{em}}{24\pi^4 P^2}
\; W^\mu_\mu(P) \; ,
%&=\frac{\alpha_\text{em}}{12\pi^4 P^2}
%\; \frac{1}{e^{E/T}-1}\; {\text{Im}}\; \varPi^{\mu}_{\mu}(P) \; ,
\end{align}
where $W_{\mu \nu}(P)$ is the Wightman correlator for two 
electromagnetic currents,
\begin{align}\label{W_def}
W_{\mu\nu}(P)
=\int d^4x \; e^{i \, P\cdot x}\; \langle \; j_\nu(0)
\; j_\mu(x) \; \rangle \; ,
\end{align}
where $j^\mu\equiv e\sum_f \overline{\psi}_f\gamma^\mu\psi_f$, 
with $\psi$ is the quark operator with flavor index $f$.
In thermal equilibrium, $W_{\mu\nu}$ is 
related to the imaginary part of the retarded %correlator for the
photon self-energy as
\begin{align}
\label{eq:W-ImPiR}
W_{\mu\nu}(P) =- \; 2 \; \nb(E) \; {\rm Im} \; 
\varPi^R_{\mu\nu}(P) \; ,
\end{align}
with
\begin{align}
\varPi^R_{\mu\nu}=
-i\int d^4x \; e^{i P \cdot x}
\; \theta(x^0) \; \langle \; \left[ j_\mu(x), j_\nu(0)
\right] \; \rangle \; .
\end{align}
Here $P\equiv P_1+P_2$ with $P_1$ and $P_2$ 
being the momenta of the two leptons. 

\begin{figure}[t]
\begin{center}
\includegraphics[width=0.3\textwidth]{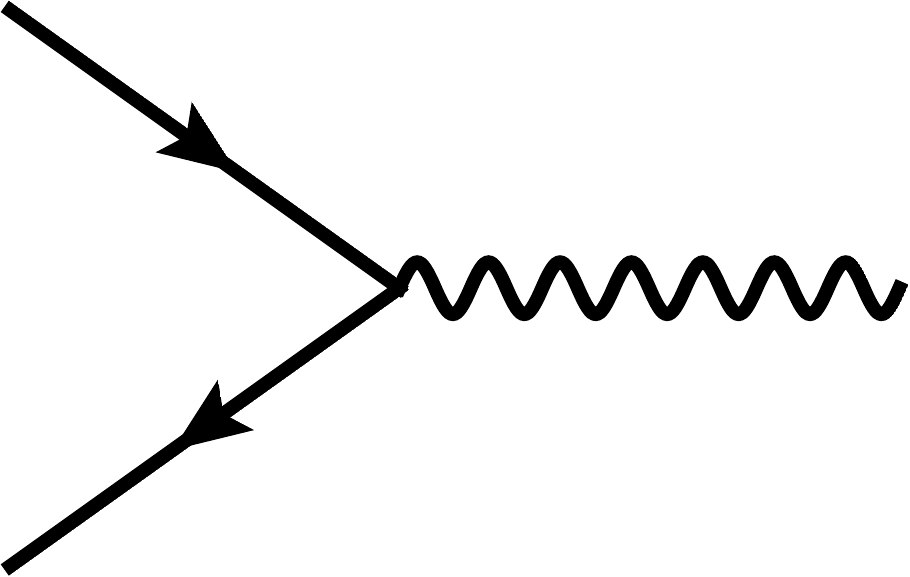}
\caption{The 1 to 2 process which results the production of dilepton.
The solid line denotes the quark while the wavy line denotes the (virtual) photon.}
\label{fig:2to1}
\end{center}
\end{figure}

At the leading order in the QCD coupling constant $\cp$, 
the contribution is obtained by $1 \rightarrow 2$ processes,
illustrated in Fig.~(\ref{fig:2to1}). 
In this process, a
quark anti-quark pair becomes a virtual
photon, which then decays to a dilepton
pair.  This gives
\begin{align}
\begin{split}
\frac{d\varGamma}{d^4P}&=\frac{\alpha_\text{em}}{24\pi^4P^2}
 \sum_{f, {\text{spin}}}
\; \int\frac{d^3k_1}{(2\pi)^3}\frac{1}{2E_1}
\; \int\frac{d^3k_2}{(2\pi)^3}\frac{1}{2E_2} 
(2\pi)^4\delta^{(4)}(P-K_1-K_2) \\
&~~~\times |{\cal M}|^2 \; 
\sum_{a=1}^{\Nc} \; \nf_a(E_1) \; \nf_{\overline{a}}(E_2) \; ,
\label{dilepton_exp}
\end{split}
\end{align}
where $f$ is a subscript for flavor running from $1$ to $\Nf$.
We use the spacetime signature $(+\, -\, -\, -)$ in this paper;
four-momenta are denoted by capital letters,
$P^\mu=(E,\vec{p})$, $K_1^\mu=(E_1,\vec{k}_1)$, 
$K_2^\mu=(E_2,\vec{k}_2)$.
The quark anti-quark pair is produced on it mass shell,
$K_1^2 = K_2^2 = 0$, so 
$E_1 = |\vec{k}_1|\equiv k_1$, $E_2 = |\vec{k}_2|\equiv k_2$,
and $P$ is time-like, $P^2 > 0$.  
Without loss of generality we can assume that the (virtual) photon energy is
positive, $E > 0$.
Here $\nf_a(E_1)$ and $\nf_{\overline{a}}(E_2)$ are the
statistical distribution functions for the quark and anti-quark in
Eq.~\eqref{q_distribution_functions}.  
The square of matrix element is 
\begin{align}
\sum_{{\text{spin}}} |{\cal M}|^2
&=8\, e^2 q^2_f \; K_1\cdot K_2 = 4\, e^2 q^2_f \; P^2 \; ,
\label{matrix_dilepton}
\end{align}
where we have used $K^2_1=K^2_2=0$, and 
$q_f$ is the electromagnetic charge of the quark with flavor 
$f$ in the unit of $e$.

The result when $Q=0$ is well known \cite{Vujanovic:2013jpa}:
\begin{align}
\label{eq:dilepton-result-Q=0}
\begin{split}
\left.  \frac{d\varGamma}{d^4P}\right |_{Q=0}
&= \frac{\alpha_\text{em}^{2}}{12\pi^4}
\; \sum_{ f} q^2_f \; \Nc\; \nb(E)
\; h(E,p) \; .
\end{split}
\end{align} 
For three flavors of quarks, $\sum_{ f} q^2_f = 2/3$.
Here
\begin{equation}
h(E,p) \equiv
1 - \frac{2T}{p}\ln\left(\frac{1 + e^{-p_-/T}}{1 + e^{-p_+/T}}\right) \; ,
\end{equation}
and
\begin{align}
p_\pm = \frac{1}{2 } ( E \pm p)\; 
\end{align}
is the range of the quark momenta.
Especially, when the dilepton pair is produced at rest, $\vp = 0$,
the quark anti-quark pair are produced back to back, with
$\vk_1=-\vk_2$.  Their energies are equal,
$E_1=E_2=E/2$, and there is no integral over the quark momentum.
The expression then reduces to
\begin{align} 
\label{eq:f-dilepton-p=0}
\left.  \frac{d\varGamma}{d^4P}\right |_{Q=0}
&= \frac{\alpha_\text{em}^{2}}{12\pi^4}\; \sum_{ f} q^2_f 
\;\; \Nc \; \nf^2(E/2) \; .
\end{align}
This is natural, as the product of a Fermi-Dirac distribution
function for the quark and anti-quark appears.  

Equation~(\ref{dilepton_exp}) illustrates our comment in Sec. \ref{sec:analytic},
that for hard momenta the only change when $Q^a \neq 0$ is in the
change in the statistical distribution functions.  To handle the
$Q$-dependence of the $\widetilde{n}$'s it is useful to note that
\begin{align}
\nf_a(E_1) \; \nf_{\overline{a}}(E_2) 
= \nb(E) \left(
1 - \nf_a(E_1) - \nf_{\overline{a}}(E_2) \right) \; ,
\label{product_fermi}
\end{align}
remembering that $E = E_1 + E_2$.  
This identity is familiar from when $Q^a = 0$.

Using this, we can derive
\begin{equation}
{\text {Im}} \; \Pi^{R \mu}_{\mu}
= \alpha_{{\text em}} \; \sum_f q_f^2 \;
\left(\frac{E^2 - p^2}{p}\right) \;
\sum_{a = 1}^{\Nc}
\int^{p_{+}}_{p_{-}} dk \left(
1 - \widetilde{n}_a(k) - \widetilde{n}_{\overline{a}}(E - k) \right)
\; .
\label{photon_self_energy}
\end{equation}
Here we wrote the photon retarded self-energy instead of the dilepton production rate for future convenience.
We use energy-momentum conservation to write
$E_1 = k$ and $E_2 = E-k$.  This expression
is useful when we compare to the results of Ref.~\cite{Islam:2014sea} 
at the end of this section, see Eq.~(\ref{PNJL}).

To leading order, we can write the rate for dilepton production when
$Q^a \neq 0$ as a momentum dependent factor times that for $Q^a = 0$,
\begin{align}
\label{define_fll}
\left. \frac{d\varGamma}{d^4P} \right|_{Q \neq 0}
&= \left. f_{l\overline{l}}(Q)  \; \frac{d\varGamma}{d^4P}\right |_{Q=0} \; ,
\end{align}
where
\begin{align}
\label{eq:dilepton-f-result}
h(E, p) \; f_{l\overline{l}}(Q)&= 
\frac{1}{\Nc} \; \sum^{\Nc}_{a=1}
\left( 1 - \frac{2T}{p}
\ln\left(\frac{1+e^{-(p_- - \, iQ^a)/T}}{1+e^{-(p_+ -\, iQ^a)/T}}\right) 
\right) \; .
\end{align}
This result can be evaluated by expanding
in powers of $\exp(-(p_\mp - i Q^a)/T)$, and performing the sum over
$a$ to obtain a series of Polyakov loops.  For general $\Nc$ all 
independent Polyakov loops, which run from 
$\ell_1$ to $\ell_{N-1}$, enter.
The resulting expression is not very illuminating.

There are two cases in which one can obtain simple results.  One 
is the confined phase of the pure gauge theory, Eqs.
(\ref{confined_vac_odd}) and (\ref{eq:ln-confined}).
Then only loops which are a multiple of $\Nc$ contribute, so
that
\begin{align}
\label{eq:dilepton-f-result-arbN}
h(E,p) \; f_{l\overline{l}}(Q_{{\rm conf}})
=  1 -\frac{2\, T}{\Nc \, p}
\ln\left(\frac{1+ e^{- \Nc p_-/T}}{1+e^{-\Nc p_+/T}}\right)
\; .
\end{align}

Another special case is three colors.  Then one can rewrite 
$f_{l\overline{l}}$ so that only 
the first Polyakov loop appears,
\begin{align}
\label{eq:dilepton-f-result-N=3}
h(E,p) \; f_{l\overline{l}}(Q)=  1 -\frac{2\, T}{3\, p}
\ln\left(\frac{1+3 \, \ell \, e^{-p_-/T}+3 \, \ell \, e^{-2 p_-/T}+e^{-3 p_-/T}}{1+3 \, \ell\, e^{- p_+/T}+3 \, \ell \,e^{-2 p_+/T}+e^{-3 p_+/T}}
\right)
\; .
\end{align}
Of course $f_{l\overline{l}}(0)=1$ in the perturbative QGP,
when $\ell = 1$.  In the confined phase where $\ell = 0$, this agrees
with the result in Eq.~(\ref{eq:dilepton-f-result-arbN}).

%%%%%%%%%%%%

\begin{figure}[t]
\begin{center}
\includegraphics[width=0.6\textwidth]{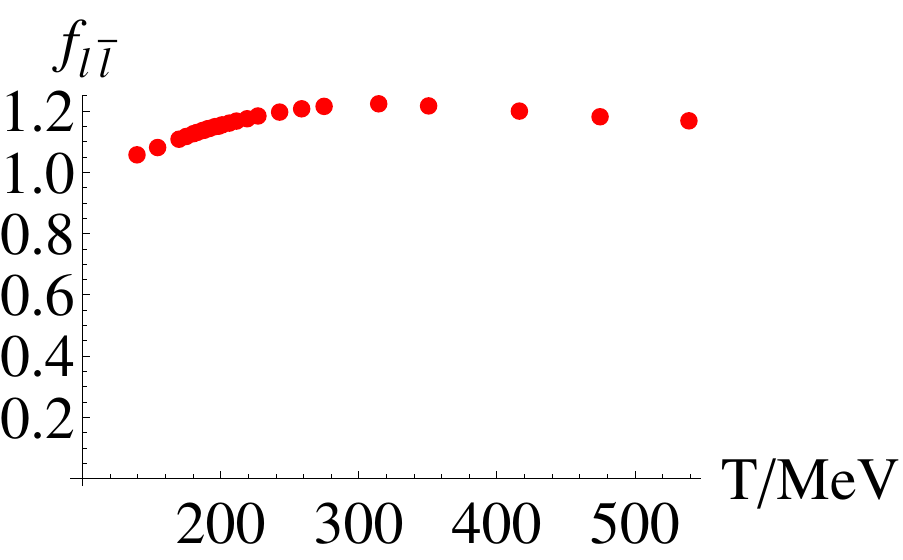}
\caption{The ratio of dilepton production in the semi-QGP
versus the perturbative QGP, 
$f_{l\overline{l}}$ in
Eq.~(\ref{eq:dilepton-f-result-N=3}), as a function of temperature.
The dileptons are back to back, $p=0$, with a total energy $E = 1$~GeV.}
\label{fig:dilepton}
\end{center}
\end{figure}

In Fig.~(\ref{fig:dilepton}) 
we plot $f_{l\overline{l}}(Q)$ as a function of temperature for
three colors.  
We do this for back to back dileptons, $p = 0$, with $E = 1$~GeV.
We see that the production of dileptons is not suppressed by
the effect of the Polyakov loop, but moderately {\it enhanced}, by $\sim 20\%$,
at {\it low} temperatures $T \sim 300$~MeV in the semi-QGP.

This enhancement is rather unexpected.  
While the probability to produce either
a {\it single} quark or anti-quark is small when the loop is small,
that to produce a quark anti-quark pair is greater in the semi-QGP than
the perturbative QGP.

%%%%%%%%%%%%%%

%%%%%%%%%%%%

%%%%%%%%%%%%%%%%%%%%%
\subsection{Enhancment of dilepton production in the confined phase
versus the perturbative Quark-Gluon Plasma}
\label{more_dilepton_sec}

To better understand the enhancement of dilepton production in the
semi-QGP, relative to that in the perturbative QGP,
we consider dilepton production for infinite
$\Nc$, comparing the confined phase to the perturbative QGP.

To simplify the analysis we consider dileptons which are produced
back to back.  This is most useful, because if
the total spatial momentum of the pair vanishes,
$p = 0$, then each dilepton carries the
same energy, $E/2$, and we can ignore the integral over phase space as a
common factor, independent of the $Q^a$.  
The effects of confinement, represented by 
$Q^a \neq 0$, can then be included just by computing the sum 
over the statistical distribution functions in
Eq.~(\ref{dilepton_exp}),
%{\bf{\color{green}{SL: I removed the $1/\Nc$ in (33), (34) and (35) to match
%    the text sum instead of average.}}}
%
\begin{equation}
%\frac{1}{\Nc}
\sum_{a=1}^{\Nc} \; \nf_a(E/2) \; \nf_{\overline{a}}(E/2) 
= \; %\frac{1}{\Nc}
\sum_{a=1}^{\Nc} 
\frac{1}{e^{(E/2-iQ^a)/T}+1} \; 
\frac{1}{e^{(E/2+iQ^a)/T}+1} \; .
\label{sum_dilepton}
\end{equation}
We note that, since the background field acts like a chemical potential for color, albeit
imaginary, the sign for $Q^a$ is opposite 
between the quark and the anti-quark.

In the perturbative phase, $Q^a = 0$, and Eq.~(\ref{sum_dilepton}) is
just $= \Nc \, \nf(E/2)^2$, as appears in Eq.~(\ref{eq:f-dilepton-p=0}).

In the semi-QGP, Eq.~(\ref{sum_dilepton}) is computed by expanding each
statistical distribution function in powers of $\exp((E/2 \mp i Q^a)/T)$,
\begin{equation}
\label{sum_qs}
%\frac{1}{\Nc}
\sum_{a=1}^{\Nc} \sum_{m=1}^\infty \sum_{m'=1}^\infty
(-)^{m + m'} \; \exp\left(- \left( (m + m')E/2 + i (m-m')Q^a\right)/T\right) \; .
\end{equation}
This sum is especially easy
to compute in the confined phase at infinite $\Nc$.  
In that case, if $m \neq m'$ the sum over $a$ gives 
$\sum \exp(i (m-m')Q^a/T)$; this is the Polyakov loop
$\ell_{|m - m'|}$, whose contribution vanishes at large $\Nc$.  
The only nonzero contributions are from terms where $m = m'$.
For the terms in Eq.~(\ref{sum_qs}) where $m = m'$, though, the
dependence on $Q^a$ drops out, cancelling identically between the quark
and the anti-quark.  The sums over $a$ and $m$ are then independent,
and easy to do,
\begin{equation}
\label{sum_qs_B}
%\frac{1}{\Nc}
\sum_{a=1}^{\Nc} \sum_{m=1}^\infty 
\; e^{- m E/T} = \frac{\Nc }{e^{E/T} - 1} 
=  \Nc\; n(E) \; ,
\end{equation}
which does not vanish at large $\Nc$.
That is, while we start with
only Fermi-Dirac distribution functions with $Q^a \neq 0$, in the confined
phase at infinite $\Nc$ we end up with a {\it Bose-Einstein} distribution
function, which corresponds to the mesonic distribution function instead of the quark and the anti-quark.  
We also note that, previously we showed that the cancellation of the phases of the quark and the anti-quark are essential for the non-suppression of the dilepton rate at large $\Nc$ by using the Boltzmann approximation~\cite{previous}.
From the discussion above, we see that the cancellation ($m=m'$) is important also in the case that we do not use the Boltzmann approximation.

This is a type of statistical confinement.  Our simple
model does not have true bound states, but there is a remnant
of a bound state from the statistical sum over the $Q^a$'s.
It is this sum in Eq.~(\ref{sum_qs}) 
which generates the Bose-Einstein distribution function in Eq.
(\ref{sum_qs_B}).

Thus in the confined phase at infinite $\Nc$,
\begin{equation}
f_{l\overline{l}}(Q_{{\rm conf}})_{\Nc = \infty} =
\frac{\nb(E)}{\nf^2(E/2) } \; .
\label{dilepton_ratio_confined}
\end{equation}
We note that this result also can be obtained by taking $p\rightarrow 0$ limit in Eq.~(\ref{eq:dilepton-f-result}).
This demonstrates a few interesting features.
First, $f_{l\overline{l}}(Q_{{\rm conf}})_{\Nc = \infty}$ is always larger than unity.
Second, at low energy, the Bose-Einstein distribution function is enhanced,
$n(E) \sim T/E$, while the Fermi-Dirac distribution function is constant, so
\begin{equation}
\label{eq:dilepton-result-BE-enhance}
f_{l\overline{l}}(Q_{{\rm conf}})_{\Nc = \infty} = 
\frac{4 \, T}{E} \; , \;\;\; E \ll T \; .
\end{equation}
Thus under the given assumptions, at small energies
dilepton production in the confined phase {\it dominates} that 
from the perturbative Quark-Gluon Plasma.  This occurs because
statistical confinement generates confined ``bosons'' from quark
anti-quark pairs, and these confined bosons become over-occupied when their energies are much smaller than temperature.
This occurs even though the probability to produce a single quark, or anti-quark, is strictly zero in the confined phase at infinite $\Nc$.  
Nevertheless, we note that, when $E\leq gT$, we need to calculate with the HTL resummation~\cite{Braaten:1990wp} instead of our calculation, so our result Eq.~(\ref{eq:dilepton-result-BE-enhance}) can be altered in that energy region.

More generally, that the ratio of dilepton production in the
confined phase to that in the perturbative QGP, $f_{l\overline{l}}(Q)$,
is of order one, indicates that at all temperature dilepton production is
of order $\Nc$.  This is one example 
of quark-hadron duality \cite{Shifman:2000jv}.

A similar enhancement of dilepton production in the confined phase was
found previously by
Lee, Wirstram, Zahed, and Hansson \cite{Lee:1998nz}.
They considered a condensate for $A_0^2$.  We can take our
result in Eq.~(\ref{eq:dilepton-f-result}), 
and expand up to quadratic order in the $Q^a$'s, to obtain
\begin{align}
h(E, p) \; f_{l\overline{l}}(Q) &\approx  
h(E,p)
+\frac{1}{\Nc pT} \sum_{a=1}^{\Nc} Q_a^2
\; \left(\nf(p_-)(1-\nf(p_-)) - \nf(p_+)(1-\nf(p_+))\right)
\; ,
\end{align}
in agreement with Eq.~(7) of Ref.~\cite{Lee:1998nz}.
These authors suggested that the enhancement
of dilepton production in the confined phase may be related to the
excess seen in heavy ion collisions for dilepton masses below that for
the $\rho$-meson \cite{Rapp:2013nxa}.

We can also make contact with results from Polyakov Nambu--Jona-Lasino 
(PNJL) models
\cite{Fukushima:2003fw,Hansen:2006ee, Ghosh:2014zra},
especially with the computation of dilepton production by Islam, Majumder,
Haque, and Mustafa \cite{Islam:2014sea}.  
To do so we need a simple identity.
For three colors the sum of the Fermi-Dirac distribution function, with
the $Q^a$ and $\ell$ as in Eqs.~(\ref{three_colors_eigenvalues})
and (\ref{eq:l-Q}), obeys
\begin{equation}
\frac{1}{3} \; \sum_{a=1}^3 \; \widetilde{n}_a(E)
= \frac{1}{3} \; \sum_{a=1}^3 \; \widetilde{n}_{\overline{a}}(E)
= \frac{ 
\ell \, {\rm e}^{- E/T} + 2 \, \ell \, {\rm e}^{- 2E/T} + {\rm e}^{- 3E/T}
}
{
1 + 3 \, \ell\, {\rm e}^{- E/T} 
+ 3 \, \ell\, {\rm e}^{- 2 E/T}+ {\rm e}^{- 3E/T}
} \; .
\label{PNJL}
\end{equation}

In the PNJL models of 
Refs.~\cite{Fukushima:2003fw,Hansen:2006ee, Ghosh:2014zra, Islam:2014sea}, 
when $\ell \neq 1$ the effective statistical distribution function 
is {\it defined} as the right hand side
of Eq.~(\ref{PNJL}); {\it e.g.}, Eqs.~(67) and (68)
of Ref.~\cite{Hansen:2006ee}.
In Refs.~\cite{Hansen:2006ee, Ghosh:2014zra, Islam:2014sea} 
this effective distribution function was obtained
by taking the derivative of the
free energy, when $\ell \neq 1$, with respect to a given energy $E$.
Since the free energy involves a sum over all colors, it is clear
that defining the effective statistical distribution function in this way 
automatically gives a sum over all 
$\widetilde{n}_a(E)$ (or $\widetilde{n}_{\overline{a}}(E)$), 
which appears on
the left hand side of Eq.~(\ref{PNJL}).
(The identity of Eq.~\eqref{PNJL}
holds for the case of zero quark chemical potential.  Then we
can define $\ell$ to be real, and 
$\sum_a \widetilde{n}_{a}(E)$ and 
$\sum_a \widetilde{n}_{\overline{a}}(E)$ are equal.
At nonzero quark chemical potential the loops in the fundamental
and anti-fundamental representations are not equal,
$\ell_{\mathbf 3} \neq \ell_{\overline{\mathbf 3}}$ 
\cite{Dumitru:2005ng}.
In this instance, identities similar to Eq.~\eqref{PNJL} hold
for $\sum_a \widetilde{n}_{a}(E)$ and 
$\sum_a \widetilde{n}_{\overline{a}}(E)$ separately,
and are again equal to those in the PNJL model,
\cite{Hansen:2006ee, Ghosh:2014zra, Islam:2014sea}.)

In our matrix model the sum over the statistical distribution functions
with all $Q^a$, 
$\sum_a \widetilde{n}_a(E)$ and
$\sum_a \widetilde{n}_{\overline{a}}(E)$, 
enters naturally when we sum over all quark colors, 
Eqs.~(\ref{product_fermi}) and (\ref{eq:dilepton-f-result}).
In the PNJL model calculation done in Ref.~\cite{Islam:2014sea}, 
${\text {Im}} \, \varPi^R_{00}$ is given in
Eq.~(4.36),  and ${\text {Im}} \, \varPi^R_{ii}$ by Eq.~(4.46).
Taking the quarks to be massless, and using the
fact that $\int^{p_{+}}_{p_{-}} dp \; \widetilde{n}_{\overline{a}}(E-p)
= \int^{p_{+}}_{p_{-}} dp \; \widetilde{n}_{\overline{a}} (E)$,
it can be shown that their result 
for ${\text {Im}}\, \Pi^{R \mu}_{\mu}$ coincides identically with our
Eq.~(\ref{photon_self_energy}).  

We emphasize that the equality between our results and the PNJL model
\cite{Islam:2014sea} is valid {\it only} to leading order.  In both cases,
at leading order
dilepton production is only a function of the Polyakov loop and the
temperature.  
(As well as the quark mass and chemical potential, if one chooses to add 
them.)  The results will certainly 
differ beyond leading order, and depend strongly upon
the details of each effective model.  

We note that, results for dilepton production at nonzero quark masses were
computed in Ref.~\cite{Islam:2014sea} and by Satow and Weise
\cite{Satow-Weise}.  There is a relatively mild dependence on
the quark masses, apart from obvious kinematical constraints, such as
the energy of the photon has to be greater than twice the quark mass.

\section{Photon production}

\subsection{Overview}\label{sec:photon-2to2}

To leading order in $\alpha_\text{em}$,
the photon rate in the QGP is
\begin{align}
\label{eq:photon-W}
p\frac{d\vGm_\gm}{d^3p}=
{-\frac{1}{2(2\pi)^3}\; g^{\mu\nu}} \; 
 W_{\mu\nu}(P) \; .
\end{align}
Since a photon on its mass shell cannot decay directly to a quark
anti-quark pair, this quantity vanishes at one-loop order.

\begin{figure}
\includegraphics[width=0.4\textwidth]{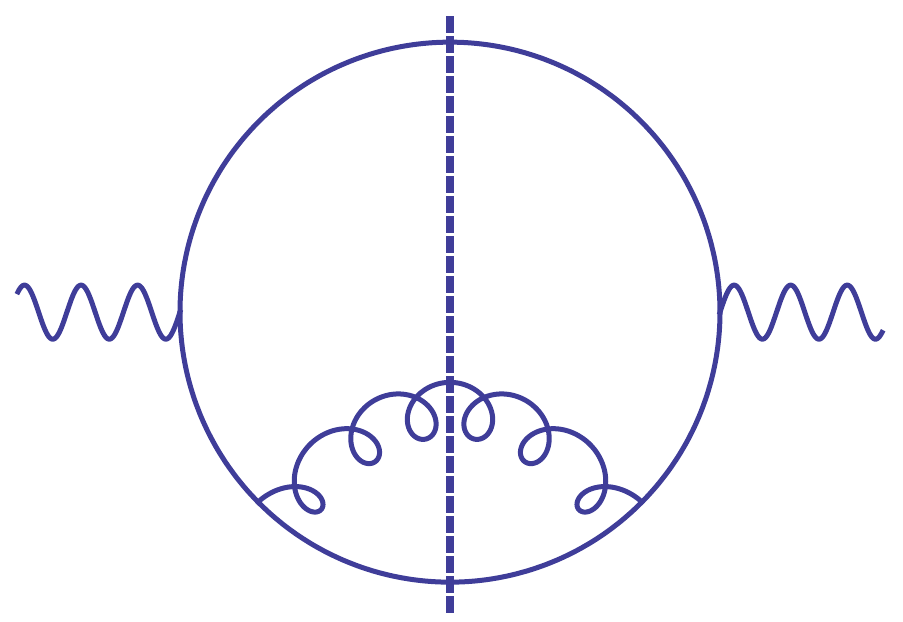}
\includegraphics[width=0.4\textwidth]{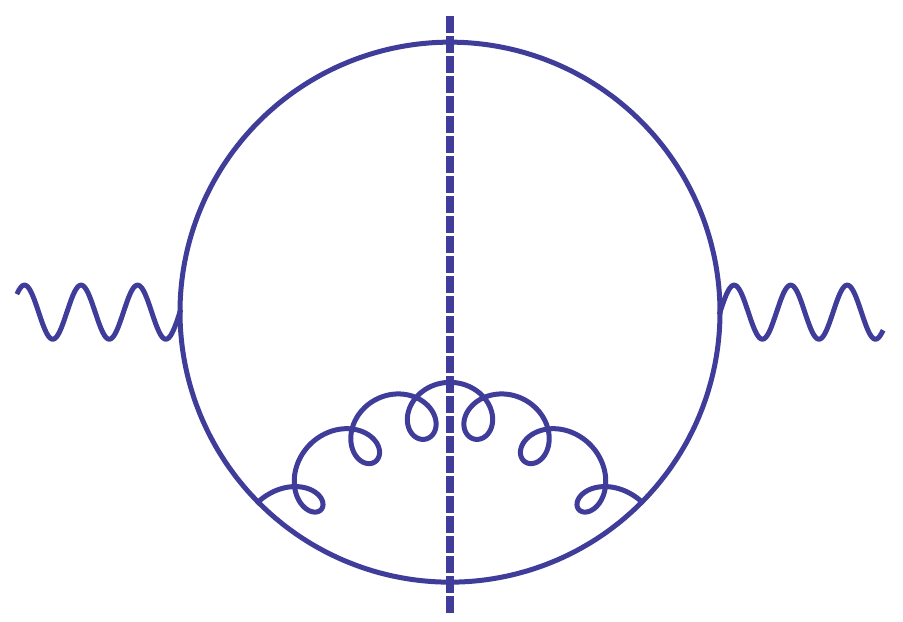}
\caption{Two loops diagrams contributing to photon self-energy.}
\label{fig_2to2}
\end{figure}

In our model the first nonzero contribution occurs at two-loop order,
from the diagrams shown in Fig.~(\ref{fig_2to2}). Cutting the diagrams
we obtain $2 \rightarrow 2$ processes, which are Compton scattering
and pair annihilation, both of order $e^2 g^2$.
We note that, consequently, and unlike the case of dilepton production to leading
order, the results which we find have no direct correspondence
in a PNJL model.  One could compute photon production in a PNJL model,
but since these models do not have dynamical gluons, the results will
be very different from our matrix model.
Each of these two processes has an infrared divergence when the momentum exchanged becomes
soft~\cite{Baier:1991em}. The divergence is removed by using a 
resummed quark propagator for soft momenta, corresponding to the uncutted
lower quark line in the left diagram of Fig.~(\ref{fig_2to2}), for example.

It was later realized that there exists another kinematic regime which
contributes at the same order 
\cite{Arnold:2002ja, *Arnold:2001ba,*Arnold:2001ms, *Arnold:2002ja}. 
This corresponds to the case when the
photon becomes collinear with quarks in the loop in the two diagrams
of Fig.~(\ref{fig_2to2}),  more precisely, when the longitudinal momenta
(defined with respect to photon momentum) of quarks remain
hard,  $\sim T$, and the transverse momenta are soft,  
$\sim gT$. 
Despite the reduced phase space, 
due to collinear enhancement, this regime was found to 
contribute equally as the $2 \rightarrow 2$ processes, in the analysis by 
Arnold, Moore and Yaffe (AMY) \cite{Arnold:2001ba}:
%We can view the exchanged soft gluon ladder as dressing of the quark-photon vertex on the left. 
%The dressed vertex, as has been shown in the analysis by Arnold, Moore and Yaffe (AMY) \cite{Arnold:2001ba}, is of order $eg$, just as the undressed photon-quark vertex on the right. 
The collinear regime in this diagram also gives an
overall $e^2g^2$ contribution to the photon emission rate.
To clarify terminology, we will refer $2\lra2$ rate as the contribution from 
Fig.~(\ref{fig_2to2}), excluding the rate in the collinear regime. 
We refer to the rest of the contributions $\sim e^2 g^2$ 
as the collinear rate.

It turns out that the collinear rate goes beyond two-loop order:
%{\it i.e.} further dressing of the quark-photon vertex by additional soft gluons ladders in the collinear regime still gives a dressed quark-photon vertex of order $eg$. 
{\it i.e.} further additional soft gluons ladders in the collinear regime still contributes at the same order. 
%Therefore a subset of diagrams with arbitrary gluon ladders contribute to order $e^2g^2$.
Thus, the story is further complicated by interference among different diagrams.
Physically, it is because that, the formation time of a photon,
$t_F\sim 1/(g^2T)$, is comparable to the mean free path for quarks,
$\lam\sim 1/(g^2T)$. 
Since these two scales are similar, interference effects
between scattering with multiple gluons must be included, which
is the Landau-Pomeranchuk-Migdal (LPM) 
effect~\cite{Arnold:2001ba,*Arnold:2001ms}.
Different diagrams add destructively, so that
the LPM effect leads to additional suppression of 
collinear photon rate by $p^{-1/2}$ at large photon momentum $p\gg T$.

In this section we compute the production of real photons with large momentum
in the presence of a nontrivial Polyakov loop.  We begin by reviewing
the computation of photon production to leading order in perturbation
theory for $2 \rightarrow 2$ processes.  We then generalize this
to $Q^a \neq 0$.  In contrast to dilepton production, we find that
photon production is {\it strongly} suppressed in the confined phase.
We give a simple explanation for this in terms of the initial state of
the scattering.  

We then give a detailed computation of the leading contributions to the
collinear rate when $Q^a \neq 0$.   
In the presence of a nontrivial loop, the thermal mass of the quark
is suppressed by a loop dependent factor, but it remains $\sim g \sqrt{\Nc}T$ (here we explicitly wrote the $\Nc$ dependence in the large $\Nc$ limit).
In contrast, the damping rate is suppressed by a factor of $1/\Nc$.
Consequently, the mean free path of a quark or gluon is much larger,
$\lam \sim 1/(g^2 T)$, not $1/(g^2\Nc T)$ as in the $Q_a=0$ case. This implies that the LPM effect can be 
neglected at large $\Nc$.

We compute the collinear processes 
when $Q^a \neq 0$ at large $\Nc$.  Doing so, we find that for three colors,
the result is not that small, at least for physically reasonable values
of the QCD coupling constant.  Nevertheless, we find the result illuminating,
to show how results can change in the semi-QGP.

\subsection{Hard momentum exchange with trivial Polyakov loop}
\label{ssc:photon-2to2-hard}

\begin{figure}
\includegraphics[width=0.8\textwidth]{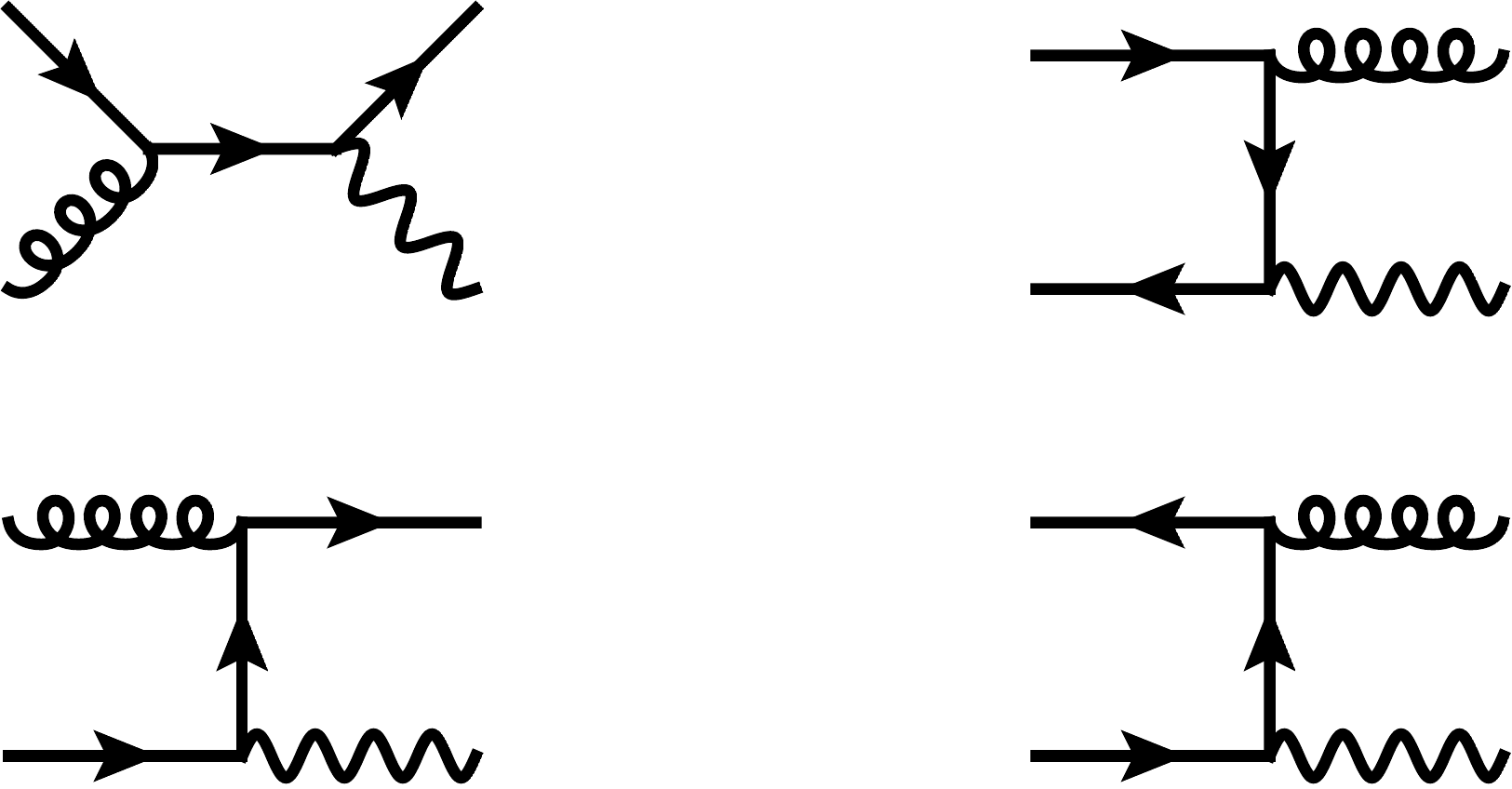}
\caption{The diagrams for the Compton scattering (left) and the pair 
annihilation (right). 
The solid line corresponds to a quark, 
the wavy line to the photon, and the curly line to the gluon, respectively.
The Compton scattering includes $s$ and $t$ channel processes 
while pair annihilation includes the $t$ and $u$ channel processes.
}\label{fig:2to2-nocolor}
\end{figure}

To establish notation on kinematics, we first review the
computation of the differential photon rate for 
$2\lra2$ processes at hard momentum exchange, in the case of 
$Q^a=0$~\cite{Baier:1991em}.  In kinetic theory, this is given by
\begin{align}
p\frac{d\vGm}{d^3p}=\sum_{i=1,2}
\int\frac{d^3k_1\, d^3k_2d^4P'}{(2\pi)^8\, 8\, E_1\, E_2} &
\dlt^{(4)}(K_1+K_2-P-P')\; \dlt((P')^2)\theta(E')\no
& \times n(E_1)\; n(E_2)\; (1\pm n(E')) \; |{\cal M}|_i^2 \; .
\label{hard_matrix}
\end{align}
The summation $i$ represents the contribution of
Compton scattering and pair annihilation, whose diagrams are shown in 
Fig.~(\ref{fig:2to2-nocolor}). 
The statistical factors $n(E_1)$, $n(E_2)$, and $n(E')$
can refer to either Fermi-Dirac or Bose-Einstein factors,
depending upon the particular process.
For Compton scattering, the statistical factor above is
$1 - \nf(E')$, which corresponds to Pauli blocking;
for pair annihilation, the corresponding factor
$1 + \nb(E')$, which represents Bose enhancement.

The incoming momenta are $K_1 = (E_1,\vec{k}_1)$ and
$K_2 = (E_2,\vec{k}_2)$ the outgoing mometum
$P'=(E',\vec{p}\,')$, and $P=(E,\vec{p}\,)$ is the photon
momentum.  We assume all particles are massless, so 
$E_1 = |\vec{k}_1|$, {\it etc.}  Whether the incoming or outgoing momenta
are quarks or gluons depends upon the process considered.
In this paper, we consider the case that the photon energy is 
much larger than temperature, $E\gg T$.

It is convenient to introduce the Mandelstam variables,
\begin{align}
\begin{array}{l}
s=(K_1+K_2)^2 \; ,\\
t=(K_1-P)^2 \; ,\\
u=(K_2-P)^2\;  . \\ 
\end{array}
\end{align}
With our kinematics,
\begin{align}
s\ge 0 \;\; ; \;\; t\; , u \; \le \; 0 \; .
\end{align}
We decompose the incoming momenta $\vec{k}_1$ and $\vec{k}_2$ into components 
parallel and perpendicular to the photon momentum $\vec{p}$, with
\begin{align}
k_1^{\para}=\frac{t}{2p}+E_1\;,\;\; 
(k_1^{\perp})^2=-\frac{t^2}{4p^2}-\frac{tE_1}{p}\; , \no
k_2^{\para}=\frac{u}{2p}+E_2\;,\;\; 
(k_2^{\perp})^2=-\frac{u^2}{4p^2}-\frac{uE_2}{p} \; , 
\end{align}
and where $\vec{k_1}^{\perp} \cdot \vec{k_2}^{\perp}
= k_1^{\perp} k_2^{\perp} \cos(\phi_1 - \phi_2)$.
We can then convert the variables of integration as
\begin{align}
d^3k_1\, d^3k_2=
\frac{1}{4}\, d\phi_1\, d\phi_2\, dk_1^{\para}\, dk_2^{\para}\, 
d(k_1^{\perp})^{2}\, d(k_2^{\perp})^{2}=\, \frac{1}{4}
\, d\phi_1 \, d\phi_2 \, \frac{E_1E_2}{p^2}\, dt\, du\, dE_1 \, dE_2 \; .
\end{align}
The integrand only depends on 
$\phi\equiv\phi_1-\phi_2$ through $\dlt((P')^2)$:
\begin{align}
\dlt(P'{}^2)=\dlt(2E_1E_2-2k_1^\para k_2^\para-2k_1^\perp k_2^\perp\cos\phi-s)
 \; .
\end{align}
The angular integrals are easily done to give the following result:
\begin{align}
\int\frac{d^3k_1d^3k_2}{8E_1E_2}\dlt((P')^2)=
%\int\frac{dt\, du\, dx\, dy(2\pi)\, p}{32}\frac{1}{\sqrt{ay^2+by+c}},
\int\frac{dt\, du\, dx\, dy(2\pi)}{32p}\frac{1}{\sqrt{ay^2+by+c}} \; ,
\end{align}
where we define 
\begin{align}
x=E_1+E_2 \;\;\; ; \;\;\; y=E_1-E_2 \; ,
\end{align}
and
\begin{align}
a&=-\frac{s^2}{4}\; ,\no
b&=(\frac{x}{2}-p)(t^2-u^2)\; , \no
c&=-\frac{1}{4}(t-u)^2 \, x^2+ \, p \, s^2 \, x- \, p^2 \, s^2-\, u\, t\, s \; .
\end{align}

We start with the integral over $y$.  
Let $y_\pm$ be the solutions of the quadratic form in $y$, 
$a y_\pm^2 + b y_\pm + c = 0$.  The integral over $y$ runs from
$y_-$ to $y_+$, where $ay^2+by+c \geq 0$.

In considering the quadratic form in $y$, we assumed that 
$b^2 - 4 a c \geq 0$.  A bit of algebra shows that this determines the range
for $x$ to be $x \ge p+s/(4p)$. 
As the energy $E' = E_1 + E_2 - p = x - p$, we automatically
satisfy the condition that this particle has positive energy,
$E' > 0$, and can set $\theta(E') = 1$ in Eq.~(\ref{hard_matrix}).

Since we assume that the incoming momenta are hard, the
distribution functions, $n(E_1)$ and $n(E_2)$, can be replaced
by their Boltzmann forms, $\exp(- E_1/T)$ and $\exp(- E_2/T)$.  
Consequently, the product of 
statistical distribution functions in Eq.~(\ref{hard_matrix}) reduces to 
\begin{align}\label{boltzmann}
n(E_1)n(E_2)(1\pm n(E'))\sim e^{-(E_1+E_2)/T}
\(1\pm\frac{1}{e^{E'/T}\mp1}\)=e^{-x/T}\(1\pm\frac{1}{e^{(x-p)/T}\mp1}\)
\; .
\end{align}
This vastly simplifies the integral over phase space.  In general, the
product in Eq.~(\ref{boltzmann}) is a function of both 
sum and difference of the energies, $x$ and $y$.
For hard momenta, though, this reduces just to a function of the sum,
of $x$.  In appendix A, we show that corrections to 
Eq.~(\ref{boltzmann}) are in fact exponentially suppressed, as one would
expect.

This allows us to immediately perform the integral over $y$.  Although
the coefficients $b$ and $c$, and $y_\pm$, are all functions of $x$,
in the end we obtain simply 
\begin{align}
\int^{y_+}_{y_-} \frac{dy}{\sqrt{ay^2+by+c}}=
\frac{1}{\sqrt{-a}} \sin^{-1} \left. 
\left(\frac{2 a y + b}{\sqrt{b^2 - 4 a c}}\right)
\right|^{y_+}_{y_-} = \frac{\pi}{\sqrt{-a}} \; .
\end{align}
We can then readily evaluate the integral over $x$,
\begin{align}
\int_{p+\frac{s}{4p}}^\infty dx
\; e^{-x/T}\(1\pm\frac{1}{e^{(x-p)/T}\mp1}\)
=\mp \; T \; e^{- p/T} \ln\(1\mp e^{-s/(4 p T)}\) \; .
\end{align}
Therefore, the phase space integrals give
\begin{align}\label{ps_int}
&\int\frac{d^3k_1\, d^3k_2}{8E_1E_2}
\; \dlt((P')^2)\; n(E_1)\; n(E_2)\; (1\pm n(E')) \no
=&\int dt\, du\; \frac{\pi^2}{8\, p\, s}
\(\mp \, T \, e^{-p/T}\)\ln\(1\mp e^{-s/(4p T)}\) \; .
\end{align}

To proceed, we consider Compton scattering and pair 
annihilation separately, since it involves a calculation of the 
matrix element squared. 
For Compton scattering off of quarks and antiquarks,
the squared amplitude is given by
\begin{align}\label{msq_comp}
|{\cal M}|^2
=2\; \sum_f q^2_f\; (4\pi)^2 \af_\text{em}\; \af_s \;
\frac{\Nc^2-1}{2}\; (-8)\; \(\frac{s}{t}+\frac{t}{s}\) \; .
\end{align}
The first term in Eq.~\eqref{msq_comp}, $\sim s/t$, is logarithmically
divergent when integrated over $t$.  The second term, $\sim t/s$, does
not produce a logarithmic divergence. As we show below, 
it is the logarithmic divergence that gives rise to 
leading logarithmic results in photon production, and
we can ignore the second term.

Remembering that $t$ is negative, the logarithmic divergence 
happens for small $-t$, and invalidates the kinetic theory description.
The standard treatment is to introduce an IR cutoff $\mu$ for the 
spatial component of the exchanged momentum, $|{\vec k}_1-{\vec p}\, |>\mu$. 
We assume that this cutoff lies between the hard and soft scales in the
problem, $\mu \gg gT $ and $\mu \ll T$. 
Near zero, the integral over $t$ is modified as follows:
\begin{align}\label{y_excise}
|{\vec k}_1-{\vec p}\,|\; > \mu \; \Rightarrow\; t+\mu^2\le(E_1-p)^2 \; .
\end{align}
On the other hand, the integration range of $y$ is given by 
$ay^2+by+c\ge0$, which as $t\to0$ takes the following form
\begin{align}\label{y_domain}
2|E_1-p|=|x+y-2p|\le 2 \; \sqrt{\frac{t}{s}(4p^2+s-4px)} \; .
\end{align}
Comparing Eqs.~\eqref{y_excise} and \eqref{y_domain} and noting $x>p$, 
the lower cutoff on $-t$ is
\begin{align}
- t\ge\frac{s}{4p(x - p)}\,  \mu^2 \; .
\label{bound_t}
\end{align}
Since we compute only to leading logarithmic accuracy, 
in the integral over $-t$ we can simply
take the lower limit to be $\mu^2$, to obtain
\begin{align}
\int^{s}_{\mu^2}d(-t)\frac{s}{t}
=- s\ln\left(\frac{s}{\mu^2}\right) \; .
\end{align}
This leaves an integral over $u$.  However, since $s = -t -u$, we can trade
this for an integral over $s$.  
The final $s$-integral becomes
\begin{align}
\label{estimate}
&\int_{\mu^2}^\infty\frac{ds}{s}\ln\(1+e^{-s/(4pT)}\)
(-s)\ln\(\frac{s}{\mu^2}\) \no
& \sim -\ln\left(\frac{p T}{\mu^2}\right) \int_{0}^\infty
ds\; \ln\(1+e^{-s/(4pT)}\)
= - \frac{\pi^2}{12}\; (4\, p \, T )\; 
\ln\left(\frac{p\, T}{\mu^2}\right) \; ,
\end{align}
where we have replaced $\ln(s/\mu^2)$ by $\ln(pT/\mu^2)$ 
and extend the lower bound of the integration to zero.
This is justified as to leading logarithmic order 
the region of integration is $s\sim pT\gg\mu^2$.

Collecting everything together, we obtain
\begin{align}
p\frac{d\vGm}{d^3p}
\simeq \sum_f q^2_f \; \frac{\, \af_\text{em}\, \af_s}{48\pi^2}(\Nc^2-1)
\; T^2\, e^{-p/T}\ln\left(\frac{p\, T}{\mu^2}\right) \; .
\label{zeroQ_compton}
\end{align}

The case of annihilation proceeds similarly. The squared amplitude is given by
\begin{align}
|{\cal M}|^2= \sum_f q^2_f\;
(4\pi)^2\af_\text{em}\af_s
\frac{\Nc^2-1}{2}\; 8\(\frac{u}{t}+\frac{t}{u}\) \; .
\end{align}
Since the integrand is symmetric in $t$ and $u$, 
both $t$ and $u$-channels 
contribute the same to leading logarithmic order. 
The integral in the $t$-channel becomes
\begin{align}
\int_{s}^{\mu^2}d(-t)\; \frac{u}{t}=\mu^2-s+s\ln\frac{s}{\mu^2} \; .
\end{align}
We again keep only the logarithm and use the same trick as in 
Eq.~\eqref{estimate} to obtain the leading logarithmic result. 
Note that there is Bose-Einstein enhancement for the annihilation process:
\begin{align}
&\int_{\mu^2}^\infty\frac{ds}{s}\ln\(1-e^{-s/(4pT)}\)\, 
s\, \ln(\frac{s}{\mu^2}) \no
& \sim  \ln  \left(\frac{ p \, T}{\mu^2} \right)\int_{0}^\infty
ds\ln\(1-e^{-s/(4pT)}\)
=  - \frac{\pi^2}{6}\; (4 \, p \, T )\; 
\ln\left(\frac{p\, T}{\mu^2}\right) \; .
\end{align}
The $u$-channel gives an identical contribution. 

Collecting everything together, the combination of Compton scattering in
the $t$ channel, and pair annhilation in the $t$ and $u$ channels, is
\begin{align}
p\frac{d\vGm}{d^3p}
\simeq \sum_f q^2_f \; \frac{\, \af_\text{em}\, \af_s}{16\pi^2}(\Nc^2-1)\;
T^2\, e^{- p/T}\ln\left(\frac{p\,T}{\mu^2}\right) \; .
\label{zeroQ_hard_total}
\end{align}

\subsection{Hard momentum exchange with nontrivial Polyakov loop}
\label{hard_photon_prod}
\begin{figure}
\includegraphics[width=0.4\textwidth]{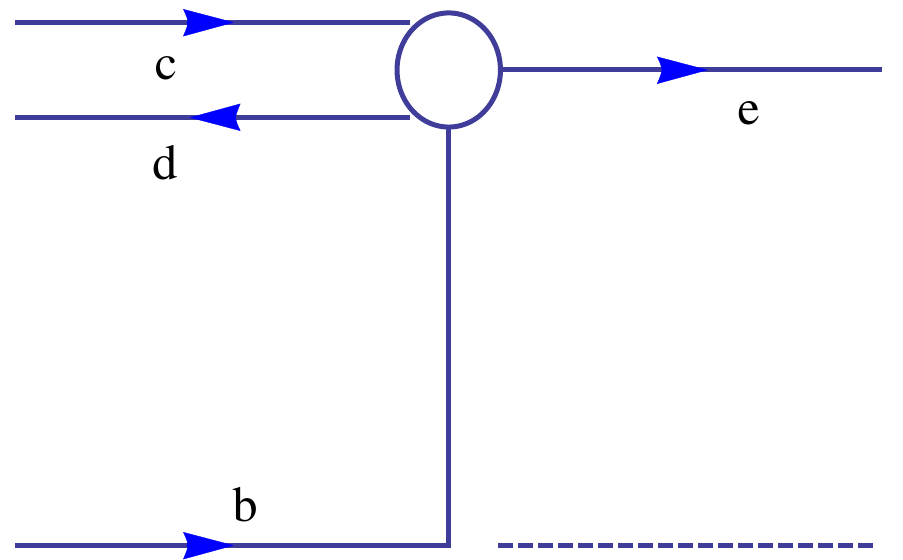}
\includegraphics[width=0.4\textwidth]{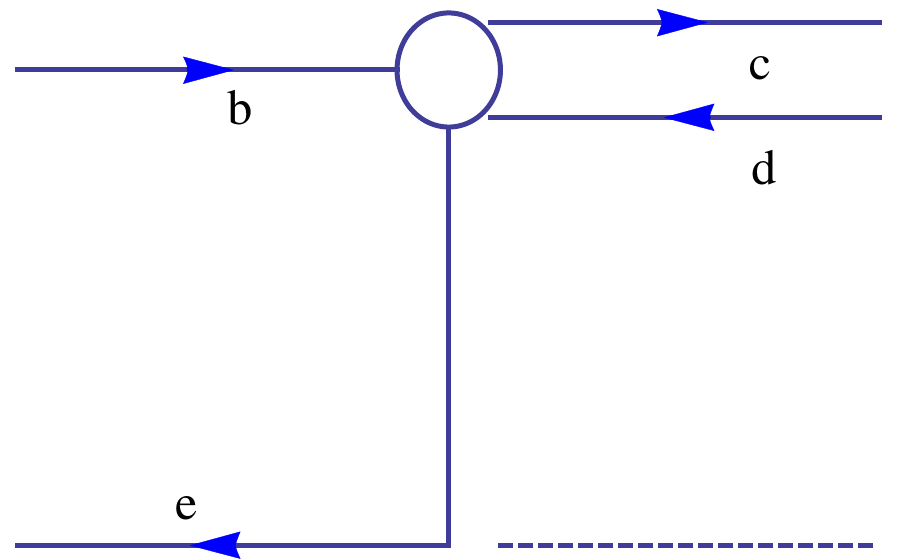}
\caption{The color labeling of Compton scattering (left) and pair 
annihilation (right). The double line corresponds to a gluon,
a single line to a quark or antiquark. The color flow does not 
involve color neutral photon, 
which we still indicate with a dashed line. 
The quark-gluon vertices are drawn as an empty circle. 
They have their own graphic representation \cite{Hidaka:2009hs}, 
but this is not needed here. The Feynman diagrams obtained 
by crossing symmetry are identical in the flow of colors.}\label{color_flow}
\end{figure}

In the previous section, we computed the matrix elements for the diagrams
which contribute to photon production at leading logarithmic order.  Once we work
in terms of Minkowski variables, there is no change in computing in
the presence of a background field for the Polyakov loop.

The {\it only} change in a background field arises from the modification
of the statistical distribution functions.   We start with the case
of Compton scattering, as illustrated in the figure in the left hand side
of Fig.~(\ref{color_flow}).  In this case, the incoming momenta are those of
a gluon, with momentum $K_1$, and a quark, with momentum $K_2$.  
Consequently, in the statistical distribution functions we replace
the gluon energy as $E_1 \rightarrow E_1 + i(Q^c - Q^d)$, while the
quark energy $E_2 \rightarrow E_2 + i Q^b$.  Similarly, the energy
of the outgoing quark becomes $E' \rightarrow E' + i Q^e$.

With the color labeling in Fig.~(\ref{color_flow}), the
thermal distribution functions when $Q^a \neq 0$ are
\begin{align}
\int_{p+s/(4p)}^\infty \; dx \; & e^{- (x+i(Q_b+Q_c-Q_d))/T}
\(1-\frac{1}{e^{(x-p+iQ_e)/T}+1}\) \no
= \; & e^{- i (Q_b+Q_c-Q_d)/T}
\int_{p + s/4p}^{\infty} 
dx \; \sum_{n=0}^\infty e^{-x/T}(-1)^ne^{-n(x-p+iQ_e)/T} \no
= \; & e^{- i (Q_b+Q_c-Q_d)/T}\; \sum_{n=1}^\infty
\frac{(-1)^{n+1}}{n}
\; T \, e^{-p/T} \; e^{-n\, s/(4\,p\, T)}
e^{-i \, (n-1) \, Q_e/T}\; .
\label{product_dist_fncs}
\end{align}
To obtain the leading logarithmic result, we 
recall Eq.~\eqref{estimate}: Eq.~\eqref{product_dist_fncs} should 
be integrated over $s$. The $s$-dependent factor 
gives rise to an additional factor of $1/n$:
\begin{align}
\int_0^\infty ds \; \exp\left(- \; \frac{n}{4\,p\,T} \; s \right)
=\frac{4pT}{n} \; .
\end{align}
It is sufficient to calculate the ratio of photon rate with
$Q^a\ne0$ to that in the perturbative limit, $Q^a=0$.
We will thus only keep track of $Q$-dependent factor 
$\sum_n (-1)^{n+1}\exp(-i(Q_b+Q_c-Q_d + (n-1)Q_e))/T)/n^2 $.
To proceed, we then need the form of the quark-gluon vertex
in the double line notation~\cite{Hidaka:2009hs}, 
appearing in the matrix element squared, Eq.~(\ref{eq:quark-gluon-vertex}).
We then multiply Eq.~\eqref{product_dist_fncs} by the product of
two quark-gluon vertices,
\begin{align}\label{poly_cmpt}
&\sum_{b,c,d,e}(T^{dc})_{be}(T^{cd})_{eb}
\;
\sum_{n=1}^\infty\frac{(-1)^{n+1}}{n^2}e^{-i(Q_b+Q_c-Q_d + (n-1)Q_e)/T} \no
=&
\sum_{b,c,d,e}
\frac{1}{2}\left(\dlt_{bd}\, \dlt_{ce}
-\frac{2}{\Nc}\, \dlt_{bd}\, \dlt_{ce}\, \dlt_{cd} \, \dlt_{be}
+\frac{1}{\Nc^2}\, \dlt_{cd}\, \dlt_{be} \right)
\sum_{n=1}^\infty\frac{(-1)^{n+1}}{n^2}
e^{-i(Q_b+Q_c-Q_d + (n-1)Q_e)/T} \no
=&\left(\frac{\Nc^2-1}{2\Nc}\right) 
\; \sum_{n=1}^\infty\frac{(-1)^{n+1}}{n^2} \; 
{\rm tr}\; {\bf L}^n \; .
\end{align}
When all $Q's$ are zero, this reduces to 
\begin{equation}
\left( \frac{\Nc^2-1}{2} \right)
\sum_{n=1}^\infty\frac{(-1)^{n+1}}{n^2}
= \left( \frac{\Nc^2-1}{2}\right) \; \frac{\pi^2}{12}
\; .
\label{compton_Q_B}
\end{equation}
For Compton scattering, the ratio of this contribution
when $Q^a \neq 0$, to that for $Q^a = 0$, is just the ratio of
Eqs.~\eqref{poly_cmpt} and \eqref{compton_Q_B},
\begin{equation}
f_{\text {Comp}}(Q) =
\frac{12}{\pi^2}\; \sum_{n=1}^\infty\frac{(-1)^{n+1}}{n^2}\; \ell_n \; ,
\label{compton}
\end{equation}
where $\ell_n$ is the $n$-{th} Polyakov loop
in Eq.~\eqref{define_polyakov_loop}.

The case of annihilation is similar.  In the presence of a background
color charge, the thermal distribution becomes
\begin{align}
&\int_{p+s/(4p)}^\infty dx \; 
e^{- (x+iQ_b-iQ_e)/T}\(1+\frac{1}{e^{(x-p+iQ_c-iQ_d)/T}-1}\) \no
=&\sum_{n=1}^\infty
\frac{1}{n} \; T \; e^{-p/T} \;
e^{-n s/(4\, p \, T)}e^{-i(Q_b-Q_e + (n-1)(Q_c-Q_d))/T} \; .
\end{align}
Again, the integration of $e^{-n s/(4\, p \, T)}$ 
over $s$ picks up an additional factor of $1/n$.
The color sum for scattering in the $t$-channel becomes
\begin{align}\label{poly_anni}
&\sum_{b,c,d,e}(T^{cd})_{be}(T^{dc})_{eb}
\sum_{n=1}^\infty\frac{1}{n^2}e^{-i(Q_b-Q_e + (n-1)(Q_c-Q_d))/T} \no
=&\frac{1}{2}\sum_{b,c,d,e}\left(\dlt_{bc}\, \dlt_{de}
-\frac{2}{\Nc}\, \dlt_{bc}\, \dlt_{de}\, \dlt_{cd} \, \dlt_{be}
+\frac{1}{\Nc^2}\, \dlt_{cd}\,\dlt_{be} \right)
\sum_{n=1}^\infty\frac{1}{n^2}
\; e^{-i(Q_b-Q_e + (n-1)(Q_c-Q_d))/T} \no
=&\frac{1}{2} \sum_{n=1}^\infty\frac{1}{n^2}\left( ({\rm tr} \;
{\bf L}^n)^2-1 \right) \; .
\end{align}
When all $Q's$ are zero, Eq.~\eqref{poly_anni} becomes
\begin{equation}
\left( \frac{\Nc^2-1}{2} \right)
\sum_{n=1}^\infty\frac{1}{n^2}=
\left( \frac{\Nc^2-1}{2}\right) \frac{\pi^2}{6} \; .
\end{equation}
Scattering in the $u$-channel gives a result identical to that
in the $t$-channel.  
Therefore, the suppression factor for annihilation is given by
\begin{align}
\label{pair_hard}
f_{\text{pair}}(Q)=
\frac{1}{\Nc^2-1}\; \frac{6}{\pi^2}\; 
\; \sum_{n=1}^\infty\frac{1}{n^2}\left(\Nc^2\; \ell_n^2-1\right) \; .
\end{align}
Remember that Compton scattering is $1/3$ of the total for $2 \rightarrow 2$
scatterings, Eqs.~\eqref{zeroQ_compton} and \eqref{zeroQ_hard_total}.
Summing over Compoton scattering and pair annihilation, 
to leading logarithmic order, we obtain the contribution from $2 \rightarrow 2$
scattering from hard momenta in the semi-QGP,
\begin{align}\label{eq:result-hard}
p \frac{d \varGamma}{d^3 p}
&=   \sum_f q^2_f \; \frac{1}{16}\; 
(\Nc^2-1) \; \alpha_\text{em} \; \alpha_s \; \frac{T^2}{\pi^2} \;
e^{-p/T} \; \ln\left(\frac{pT}{\mu^2}\right) \; 
f_\gamma(Q) \; ,\no
 f_\gamma(Q)&=\frac{1}{3}\left(f_{\text{Comp}}(q)+2f_{\text{pair}}(q)\right) \; .
\end{align} 
These expressions can be more simply expressed when $\Nc=3$ in terms 
of $Q^a = 2 \pi T (-q,0,q)$, Eq.~(\ref{eq:ln-Q}):
\begin{align}\label{ff_supress}
\nonumber
f_{\text{Comp}}(q)
&=   1 -\, 8\; q^2 \; ,\\
\nonumber
f_{\text{pair}}(q) & = 1 - 6 \, q + 9 \, q^2 \; ,\\
f_\gamma(Q)
%= \frac{1}{3}
%\left(f_{\text{Comp}}(q)+2f_{\text{pair}}(q)\right)
&=1 - 4 \, q + \frac{10}{3}\, q^2 \; .
\end{align}
The results for more than three colors are similar, simple 
quadratic polynomials in the $Q^a$'s.  That for $f_{\text{Comp}}(Q)$
involves the $Q^a$, while that for 
$f_{\text{pair}}(Q)$ is a function of the differences, $Q^a - Q^b$.

We also note that {\it exactly} the same functions of $q^a$ enter
into collisional energy loss for a heavy quark in the semi-QGP.
Because of the historical convention, the function 
for Compton scattering in photon production, $f_{\text{Comp}}(q)$,
is identical to that for Coulomb scattering of a heavy quark, Eq.~(33) of Ref.~\cite{Lin:2013efa}.
Similarly, the function for pair annhilation in photon production,
$f_{\text{pair}}(Q)$, is the same function as for Compton scattering
of a heavy quark, Eq.~(45) of Ref.~\cite{Lin:2013efa}.  
While these two functions are the same,
in detail they enter differently into collisional energy loss for
a heavy quark, times different logarithms of the energy.  

\subsection{Soft momentum exchange}

We now compute the contribution to photon production when the 
momentum exchanged is soft.  This case is simpler than when
the momentum exchanged is hard, and so we treat the case of a
nontrivial Polyakov loop at the outset.

We follow the analysis of Baier, Nakkagawa, Niegawa, and Redlich
\cite{Baier:1991em}.
We begin the computation in imaginary time, and then analytically
continue the external momentum.  
The photon self-energy in the imaginary time is
\begin{align}
\varPi^\mu_\mu(P)&= 2 \, e^2\; \sum_f q^2_f \; 
\; \sum_{a=1}^{\Nc} \; T \sum_{k^0}\int\frac{d^3k}{(2\pi)^3}\;
\Tr\left[\gamma^\mu \; S^*_a(K)\; \gamma_\mu \; S_a(K-P) \right] \; ,
\end{align}
The overall factor of two arises because $K$ or $K-P$ 
can be a soft momentum: we have chosen only $K$ to be soft.
Thus the momenta $K-P$ is hard, so we can use the bare 
quark propagator, $S_a(K-P)$.  For the quark with soft momenta
it is necessary to use a propagator, $S^*_a(K)$, which is
resummed with
Hard Thermal Loops (HTLs) in the presence of $Q^a \neq 0$ ~\cite{Hidaka:2009hs},
\begin{align}
S_a(K)&= \int^\infty_{-\infty}\frac{d\omega}{2\pi}
\frac{\rho(\omega,\vk)}{\omega-i\tilde{k}_0} \; ,\\
S^*_a(K)&= \int^\infty_{-\infty}\frac{d\omega}{2\pi}
\frac{\rho^*_a(\omega,\vk)}{\omega-i\tilde{k}_0} \; ,
\end{align}
with $\tilde{k}_0\equiv k_0+Q_a$ and the quark spectral functions are
\begin{align}
\label{eq:propagator-spectral-bare}
\rho(\omega,\vec{k})&=2\pi\varepsilon(\omega)\Slash{K} \delta(K^2) \; ,\\
\label{eq:propagator-spectral-HTL}
\rho^*_a(\omega,\vec{k})&= 
\frac{\gamma^0-\vgamma\cdot\hat{k}}{2}\rho^*_{a+}(\omega,\vec{k})
+ \frac{\gamma^0+\vgamma\cdot\hat{k}}{2}\rho^*_{a-}(\omega,\vec{k}) \; ,
\end{align}
where $\varepsilon(\omega)$ is the sign function.
We note that the bare quark spectral function $\rho(\omega,\vec{k})$ does not have its color index.
The HTL spectral functions are a sum of pole and cut terms,
\begin{align}
\label{eq:HTL-spectral}
\begin{split}
\rho^*_{a\pm}(\omega,\vk)&= 2\pi \left[Z_{\pm a}(k) \; 
\delta(\omega-\omega_{\pm a}(k))
+ Z_{\mp a}(k) \; \delta(\omega+\omega_{\mp a}(k)) \right]\\
&~~~+\theta(k^2-\omega^2)\rho^{\text{spacelike}}_{a\pm}(\omega,\vk) \; .
\end{split}
\end{align}
The quark quasi-particles have a 
thermal mass $\mf{}_a$, a dispersion relation 
$\omega_{\pm a}(k)$, and residue $Z_{\pm a}(k)$ ($k = |\vec{k}|$).
Explicitly,
\begin{align}
Z_{\pm a}(k)&=\frac{\omega^2_{\pm a}(k)-k^2}{2\, \mf^{2}{}_a} \; ,\\
\omega_{\pm a}(k)\mp k&= \frac{\mf^2{}_a}{k}
\left[\left(1\mp\frac{\omega_{\pm a}(k)}{k}\right)Q_0\left(\frac{\omega_{\pm a}(k)}{k}\right)\pm 1\right]\; ,
\end{align}
where 
\begin{equation}
Q_0(x)\equiv \frac{1}{2} \ln\left(\frac{x+1}{x-1}\right) \; .
\end{equation}
The explicit form of the cut term from Landau damping,
$\rho^{\text{spacelike}}_{a\pm}$, is irrelevant for our analysis.
The result for the quark quasi-particle mass $\mf{}_a$ is given later.

Introducing a spectral representation for the propagators,
\begin{align}
\varPi^\mu_\mu(P)=- \; 2\, e^2 \sum_f q^2_f\; 
\sum_a\int & \frac{d^3k}{(2\pi)^3}\;
\int\frac{d\omega_1}{2\pi}\; \int\frac{d\omega_2}{2\pi}\; \no
\times 
& \frac{\left(\widetilde{n}_{a}(\omega_2)-\widetilde{n}_{a}(\omega_1)\right)}{-ip^0+\omega_1-\omega_2}
\Tr\left[\gamma^\mu \rho^*_a(\omega_1,\vec{k})
\gamma_\mu \rho(\omega_2,\vk-\vp) \right] \; ,
\end{align}
Since for massless quarks their spectral density has only a vector component, 
\begin{align}
\Tr\left[\gamma^\mu \rho^*_a(\omega_1,\vk)\gamma_\mu \rho(\omega_2,\vk-\vp)
 \right]
&=- \, 2 \; \Tr\left[ \rho^*_a(\omega_1,\vk) \rho(\omega_2,\vk-\vp) \right]
\; .
\end{align}

Now we compute the discontinuity in the amplitude, as we analytically
continue the photon energy 
$p_0 \rightarrow - i E\pm \epsilon$, for infinitesimal $\epsilon$,
\begin{align}
\begin{split}
{\text{Disc}}\; \varPi^{\mu}_\mu(P)
=+ \; 2\, e^2 \; \sum_f q^2_f \; \sum_a\int\frac{d^3k}{(2\pi)^3}
\int\frac{d\omega}{2\pi}&
%\left(n_{Q_a}(\omega-E)-n_{Q_a}(\omega)\right)\\
\left(\widetilde{n}_{a}(\omega-E)-\widetilde{n}_{a}(\omega)\right)\\
&
\times
\Tr\left[ \rho^*_a(\omega,\vec{k}) 
\rho(\omega-E,\vec{k}-\vec{p}) \right]
\; ,
\end{split}
\label{discon}
\end{align}
where Disc$\varPi^\mu_\mu(P)\equiv 
[\varPi^\mu_\mu(E+i\epsilon,\vp)-\varPi^\mu_\mu(E-i\epsilon,\vp)]/(2i)$.
Since the photon is a singlet under color, there is no ambiguity
in how we do the analytic continuation for the photon energy.  
We have also used the fact that the spectral function is real.
When the $Q^a$'s vanish, this discontinuity is the same as the imaginary
part of the retarded self-energy.  
When the $Q^a \neq 0$, however, if we were to
compute the imaginary part, we would also obtain contributions from
the imaginary parts of the statistical distribution functions,
which are complex valued.  To us this is an unphysical
contribution which we neglect.  After all, the discontinuity is
directly related to the amplitude to produce physical particles, albeit
with an (imaginary) chemical potential for color.

By using the decomposition of the spectral functions, 
Eqs.~(\ref{eq:propagator-spectral-bare}) and 
(\ref{eq:propagator-spectral-HTL}),
\begin{align}
\Tr\left[ \rho^*_a(\omega,\vec{k}) 
\rho(\omega-E,\vec{k}-\vec{p}) \right]
=4\pi \epsilon(\omega-E)
\delta((P-K)^2)&\left(\rho^*_{+a}(\omega,\vec{k})
(\omega-E -k+\hat{k}\cdot \vec{p} \, )\right.\no
&\left.
+\rho^*_{-a}(\omega,\vec{k})(\omega-E+k-\hat{k}\cdot\vec{p}\, )  \right) \; ,
\end{align}
where  $p = |\vec{p}\, |$.

Since $k\ll T$, by using the assumption $p\gg T$, we find $p\gg k$.
Using this and $P^2=0$, 
\begin{align}
\epsilon(\omega-E)\delta((K-P)^2)&\simeq -\; \frac{1}{2 p k}
\delta\left(\cos\theta-\frac{\omega}{k}\right) \; .
\end{align}
Thus,
\begin{align}
\begin{split}
{\text{Disc}}\, \varPi^{\mu}_\mu(P)&\simeq
+\, 2\; \sum_f q^2_f \; e^2\; \sum_a\int\frac{d^3 k}{(2\pi)^2}
\int\frac{d\omega}{2\pi}
\left(\widetilde{n}_{a}(-E)-\widetilde{n}_{a}(0)\right)
%\frac{1}{p}
\frac{-1}{k}
\delta\left(\cos\theta-\frac{\omega}{k}\right) \\
&~~~\times\left(\rho^*_{+a}(\omega,\vec{k})
\left(-1+\frac{\omega}{k}\right)
+\rho^*_{-a}(\omega,\vec{k})
\left(-1-\frac{\omega}{k}\right)\right) \\
&= - \, 2\; \sum_f q^2_f\; e^2 \;
\frac{1}{2\pi}\sum_a \left(\widetilde{n}_{a}(-E)-\widetilde{n}_{a}(0)\right)
\int^{\mu}_0 dk \; k \; \int^{k}_{-k}\frac{d\omega}{2\pi}\\
&
\;\;\;\;\;\;\;\;\;\;\;\;\;\;\;\;\;\;\;\;\;\;\;\;\;\;\
\times \left(\rho^*_{+a}(\omega,\vec{k})\left(-1+\frac{\omega}{k}\right)
+\rho^*_{-a}(\omega,\vec{k})\left(-1-\frac{\omega}{k}\right)\right) \; ,
\end{split}
\end{align}
where we introduce an ultraviolet cutoff, $\mu$.
It is useful to use the sum rules~\cite{lebellac},
\begin{align}
\int^{\infty}_{-\infty}\frac{d\omega}{2\pi} 
\; \rho^*_{\pm a}(\omega,\vec{k})&= 1\;,\\
\int^{\infty}_{-\infty}\frac{d\omega}{2\pi} \; \omega
\; \rho^*_{\pm a}(\omega,\vec{k})&= \pm \; k \; .
\end{align}
Using the spectral functions in the time-like region, 
Eq.~(\ref{eq:HTL-spectral}), we obtain
\begin{align}
\begin{split}
{\text{Disc}}\, \varPi^{\mu}_\mu(P)&\simeq
+\; 2\; \sum_f q^2_f\; e^2 \; \frac{1}{2\pi} \sum_a 
\left(\widetilde{n}_a(-E)-\widetilde{n}_{a}(0)\right)\\
&\quad\times\int^{\mu}_0 dk \, k \; 2\;\left[Z_{+a}(k)\left(-1+\frac{\omega_{+a}(k)}{k}\right)
+Z_{-a}(k)\left(-1-\frac{\omega_{-a}(k)}{k}\right)\right] \; \\
  &\simeq + \; 2 \; \sum_f q^2_f \; e^2 \; \frac{1}{2\pi}\sum_a 
\left(\widetilde{n}_{a}(-E)-\widetilde{n}_{a}(0)\right)  \\
&\quad\times\left[\mu(\omega_{-a}(\mu)-\omega_{+a}(\mu))
+2\int^{\mu}_0 dk (\omega_{+a}(k)-\omega_{-a}(k))\right] \; ,
\end{split}
\end{align}
where we have used~\cite{lebellac}
\begin{align}
(\omega_\pm \mp k)(\omega^2_\pm-k^2)\frac{1}{\mf^2}&=
\omega_\pm- k \; \frac{d\omega_\pm}{dk} \; .
\end{align}
 The wave function constants and the mass shells are functions of the color index, $a$, but
we suppress this index for now to make it easier to read.
By using the asymptotic form for the mass shells at 
hard momenta, $k \gg gT$,
\begin{align}
\label{eq:HTL-dispersion-plus}
\omega_+&\simeq \; k + \; \frac{\mf^2}{k} \; ,\\
\label{eq:HTL-dispersion-minus}
\omega_-&\simeq \; k \; ,
\end{align}
we get
\begin{align}
\begin{split}
{\text{Disc}}\varPi^{\mu}_\mu(P) \, \simeq \, 
+ \; 2 \sum_f q^2_f \; e^2 & \frac{1}{2\pi}\sum_{a=1}^{\Nc} \mf^{2}{}_{a}
\left(\widetilde{n}_a(-E)-\widetilde{n}_a(0)\right)  \\
&\quad\times\left[-1
+2\int^{\mu}_0 dk \; \frac{\omega_{+a}(k)
-\omega_{-a}(k)}{\mf^{2}{}_a}\right]. 
\end{split}
\end{align}
We now make the dependence of the thermal quark mass 
on the color index $a$ manifest again.

Evaluating the integral by using Eqs.~(\ref{eq:HTL-dispersion-plus}) and (\ref{eq:HTL-dispersion-minus}) at the leading-log accuracy, 
\begin{align}
\begin{split}
{\text{Disc}}\varPi^{\mu}_\mu(P)&\simeq
+\; 2 \; \sum_f q^2_f \; e^2 \; \frac{1}{2\pi}
\sum_a \mf^{2}{}_a\left(\widetilde{n}_{a}(-E)-\widetilde{n}_{a}(0)\right)
\ln\left(\frac{\mu^2}{\cp^2 T^2}\right) \; . 
\end{split}
\end{align}
The lower limit of the integral comes from %$\mf^2 \sim g^2 T^2$.
$k \sim g T$, in which Eqs.~(\ref{eq:HTL-dispersion-plus}) and (\ref{eq:HTL-dispersion-minus}) becomes unreliable.

To leading logarithmic order, then, the $Q^a$'s only enter
through the statistical distribution functions of the quarks,
and the quark thermal mass.
By using Eqs.~(\ref{eq:W-ImPiR}) and (\ref{eq:photon-W}), the contribution to the production rate for photons from soft quarks is found to be
\begin{align}
p \frac{d \varGamma}{d^3p}
&=  f_\gamma^\text{soft}(Q) 
\; \left. p \frac{d \varGamma}{d^3p}\right|_{\text{pQGP}} \; .
\end{align}
The result in the perturbative QGP~\cite{Baier:1991em} is
%
%{\bf\color{green}{SL: in the final result, it is better to replace $E$ by $p$}}
\begin{align}
\nonumber 
\left. p \frac{d \varGamma}{d^3 p}\right|_{\text{pQGP}}
&= - \; \sum_f q^2_f\; \frac{1}{8} \; \alpha_\text{em} \alpha_s\; 
\frac{T^2}{\pi^2}
(\Nc^2-1)
\left(\frac{\widetilde{n}(-p)-1/2}{1 - e^{p/T}}  \right)
\ln\left(\frac{\mu^2}{\cp^2 T^2}\right)\; .
\end{align}

In the semi-QGP, this is modified by a $Q$-dependent factor, 
\begin{align}\label{fgm}
f_\gamma^{\text{soft}}(Q)= \frac{1}{\Nc}
\frac{\sum_a \mf^{2}{}_a\left(\widetilde{n}_{a}(-p)-\widetilde{n}_{a}(0)\right)}
{m^2_\text{qk}\left(\widetilde{n}(-p)-\widetilde{n}(0)\right)} \; ,
\end{align}
where $\mf$ is the thermal mass 
when $Q^a=0$, whose expression will be written later.

To evaluate the photon production rate in the semi-QGP, 
we need the explicit form of the thermal quark mass when $Q^a \neq 0$.
From Ref.~\cite{Hidaka:2008dr},
\begin{equation}
\mf^{2}{}_{a} = \frac{g^2 }{24} \;
\left( \sum_{b=1}^{\Nc} \left(\Acal(Q^a - Q^b) - 
\widetilde\Acal(Q^b ) \right)
- \frac{1}{\Nc} \left( \Acal(0) - \widetilde{\Acal}(Q^a) \right) \right) \; .
\label{thermal_quark_mass}
\end{equation}
The function $\Acal(Q)$ is given by
\begin{equation}
\Acal(Q) = \frac{3}{\pi^2} \; \int^\infty_0
dE \; E \; \left( \frac{1}{e^{(E+iQ)/T} - 1 } 
+ \frac{1}{e^{(E-iQ)/T} - 1 } \right) \; ,
\label{first_def_acal}
\end{equation}
and $\widetilde\Acal(Q)\equiv \Acal(Q+\pi T)$.
Note that $\Acal(Q)$ is an even function of $Q$.

Our definition
of $\Acal(Q)$ differs by $T^2$ from that in Ref.~\cite{Hidaka:2008dr},
which we do to emphasize the physics in 
the following section, Eqs.~(\ref{sum_gluon_dist_AcalB}) and
(\ref{sum_qk_dist_AcalB}).
Also for the purposes of this discussion to follow, we note that
in Eq.~(\ref{thermal_quark_mass}) the terms involving 
$\sum_b \Acal(Q^a - Q^b)$ and $\Acal(0)$ are from the gluon distribution
functions, while $\widetilde\Acal(Q^b)$ and $\widetilde\Acal(Q^a)$
are from the quark distribution functions.

In the perturbative QGP, the thermal quark mass squared is
\begin{equation}
\mf^2 = \frac{\cp^2}{24} \; \left( \Nc - \frac{1}{\Nc} \right) 
T^2 \; \left( 1 - \left( - \frac{1}{2} \right) \right)
= \left( \frac{\Nc^2 -1}{2\Nc} \right) 
\frac{\cp^2 T^2}{8} \; .
\label{pert_quark_mass}
\end{equation}
In the first expression the $1$ is from the gluon distribution functions,
while the $+1/2$ is from the quark distribution functions.  

It is direct to evaluate $\Acal(Q^a)$
in terms of the dimensionless variable $q^a = Q^a/(2 \pi T)$, 
Eq.~(\ref{define_q_Q}),
\begin{equation}
\Acal(Q) = \left( 1 - 6 \, |q|_{{\rm mod} \; 1} (1 - |q|_{{\rm mod} \; 1}) 
\right)
T^2 \; \; .
\label{defineAcal}
\end{equation}
While nominally a quadratic polynomial in $q$, some care must be
taken in using this expression.
Only the absolute value of $q$ enters because by construction
Eq.~(\ref{first_def_acal}) is even in $Q$.
Secondly, $q$ is defined modulo one, since
only $\exp(\pm 2 \pi i q^a)$ enters into 
the Bose-Einstein distribution functions in 
Eq.~(\ref{first_def_acal}), so the $q^a$ are manifestly periodic variables.

Equation~\eqref{fgm} can be simplified for large photon energy $p \gg T$.
In this case, $\widetilde{n}_a(-p) \sim \widetilde{n}(-p)\sim 1$, 
independent of $q$.  
We further make use of the fact that all $Q$'s pair up 
as in Eq.~\eqref{eq:Q-distribution} and the 
corresponding thermal quark masses are identical 
for the components in the pair, $\mf^2{}_a=\mf^2{}_{N+1-a}$. 
Consequently, we have
\begin{align}\label{fgm_sum}
\sum_{a=1}^N\mf^{2}{}_a\left(\widetilde{n}_{a}(-p)-\widetilde{n}_{a}(0)\right)\simeq\sum_{a=1}^{N/2}\mf^{2}{}_a\left(2-\widetilde{n}_{a}(0)-\widetilde{n}_{N+1-a}(0)\right)=\sum_{a=1}^{N}\mf^{2}{}_a\left(1-\widetilde{n}(0)\right) \; .
\end{align}
This allows us to express $f_\gamma^{\text{soft}}(Q)$ as the ratio of avergae thermal quark mass sqaured when $Q^a\ne0$ to the perturbative thermal quark mass:
\begin{align}\label{fgm_avg}
f_\gamma^{\text{soft}}(Q)\simeq\frac{1}{\Nc}
\frac{\sum_a \mf^{2}{}_a}
{m^2_\text{qk}} \; .
\end{align}
We note that Eq.~\eqref{fgm_avg} is derived assuming an even $N$. The conlusion holds for odd $N$ also.

For three colors, taking the eigenvalues as in 
Eq.~(\ref{three_colors_eigenvalues}),
the components of thermal mass read
\begin{align}
\nonumber
m_\text{qk}^{2}{}_1=m_\text{qk}^{2}{}_3
&= \frac{g^2T^2}{6}
\left(1 - \frac{9}{2} \, q + 5 \, q^2 \right) \; , \\ 
\nonumber
m_\text{qk}^{2}{}_2 &= 
\frac{g^2T^2}{6} \left(1 - 3 \, q \right) \; .
\end{align}
The suppression factor is then
\begin{align}
f_\gamma^\text{soft}(Q)&= \frac{1}{3}
\left[ \left(1 - 3 q\right)
+\left(1 -\frac{9}{2}\, q +  5\, q^2 \right)
\frac{\widetilde{n}_{1}(-p)-\widetilde{n}_{1}(0)+\widetilde{n}_{3}(-p)-\widetilde{n}_{3}(0)}{\left(\widetilde{n}(-p)-\widetilde{n}(0)\right)}
\right] \; .
\end{align}
For large energy, we obtain a simple polynomial in $q$,
\begin{align}
\begin{split}
f^\text{soft}_\gamma(Q)&\simeq \frac{1}{3}
\left[ \left(1 - 3 q \right)
+2\left(1 -\frac{9}{2}\, q + 5 \, q^2\right)\right]
= 1 - 4 \, q + \frac{10}{3} \, q^2 \; ,
\end{split}
\end{align}
which agrees with the suppression factor for the hard contribution, $f_\gamma(Q)$.
Altogether, the photon production rate from soft momentum exchange is 
\begin{align}
\label{eq:result-soft}
p \frac{d \varGamma}{d^3p}
&=   \sum_f q^2_f\;
\frac{1}{2} \; \alpha_\text{em} \alpha_s \; \frac{T^2}{\pi^2} \; e^{-p/T} \;
\ln\left(\frac{\mu^2}{\cp^2 T^2}\right)
\; f_\gamma(Q) \; .
\end{align}

\begin{figure}
\includegraphics[width=0.75\textwidth]{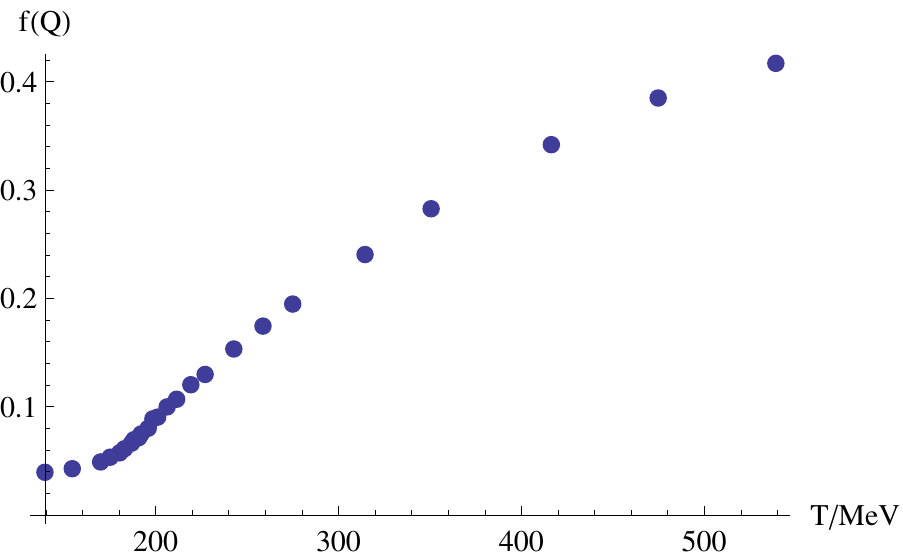}
\caption{The suppression factor $f_\gamma(Q)$ 
versus temperature, with the loop from Ref.~\cite{Lin:2013efa}.}
\label{fig_fQ}
\end{figure}

Comparing the hard contribution in Eq.~\eqref{eq:result-hard} 
to the soft contribution in Eq.~\eqref{eq:result-soft}, 
we see that the dependence upon the momentum cutoff $\mu$ cancels.
This is a nontrivial check of our computation.
The sum of the two contributions is
\begin{align}\label{2to2_rate}
p \frac{d \varGamma}{d^3p}
&=  f_\gamma(Q) \; \left. p \frac{d \varGamma}{d^3p}\right|_{\text{pQGP}} \; ,
\end{align}
where
\begin{align}
\left. p \frac{d \varGamma}{d^3p} \right|_{\text{pQGP}}
&=   {\color{black}{\sum_f q^2_f}}\; \frac{1}{2}\;  
\alpha_\text{em} \alpha_s \; \frac{T^2}{\pi^2} \;
e^{-p/T}
\ln\left(\frac{p}{\cp^2T}\right) \; .
\end{align}
We can extract 
$Q$ from lattice results of Polyakov loop and obtain 
$f(Q)$ as a function of the temperature. 
The result is shown in Fig.~(\ref{fig_fQ})
.

\subsection{Why so few photons are produced in the semi-QGP}

For dilepton production we found a moderate enhancement near $T_c$.
In contrast, Fig.~(\ref{fig_fQ}) shows 
that photon production is {\it strongly} suppressed in the semi-QGP,
versus the perturbative QGP.  
To understand the suppression of photons, as in Sec.~\ref{more_dilepton_sec}~
it helps to generalize the computation to an arbitrary number of colors.
In the calculation of the contribution from the Compton scattering, 
the following product of the distribution function appears, 
as was discussed in Sec.~\ref{hard_photon_prod}:
\begin{align}
\frac{1}{\Nc^2}\sum_{b,c} e^{-(E_1-iQ_b)/T}e^{-(E_2-iQ_{c}+iQ_{b})/T}
(1-\widetilde{n}_{c}(E')) \; .
\end{align}
Here the Boltzmann approximation was applied to the initial state, and 
we took the large-$\Nc$ limit, in which we ignore the second term of 
$(T^{dc})_{be}$ (Eq.~(\ref{eq:quark-gluon-vertex})) appearing in the 
matrix element squared.
The factor $1/\Nc^2$ was multiplied for normalization.
The quantity above becomes
\begin{align}
\frac{1}{\Nc}\sum_{c} e^{- E_1/T}e^{-(E_2-iQ_{c})/T}(1-\widetilde{n}_{c}(E'))
\end{align}
after partial cancellation of the phase of the distribution functions 
in the initial state.
Here we note that this cancellation is not complete unlike the dilepton case:
The phase $i Q_c/T$ still remains in the present case while the 
phase completely cancels for dilepton production
in the Boltzmann approximation~\cite{previous}.
By performing the sum as in the dilepton case and 
using Eq.~(\ref{eq:ln-confined}), this expression can be rewritten as
\begin{align}
e^{- E/T} \; \widetilde{n}(\Nc E')
\end{align}
in the confined phase.
We see that this expression vanishes in the 
$\Nc\rightarrow \infty$ limit, unlike the dilepton case.
The origin of this behavior can be tracked to the fact that 
the cancellation of the phase of the distribution functions 
for the initial state is only partial, and not complete.
This is because that the initial state 
for photon production is not a color singlet, as it is for
dilepton production.

For the contribution from pair annihilation,
the product of the distribution functions is, 
in the confined phase and the large-$\Nc$ limit, 
again $e^{- E/T} n(\Nc E')$.
We note that previously~\cite{previous}, 
we gave a similar but simpler analysis, 
using the Boltzmann approximation to both the final as well as the initial
state.

Next, let us discuss more quantitative point: 
the origin of the $1/\Nc^2$ dependence of the suppression factor in the confined phase.
For hard photons, with $E \gg T$, we have 
shown that the ratio of photon production
in the semi-QGP, to that in the perturbative QGP, is just the ratio
of the thermal quark masses squared, of course summed over color:
\begin{align}
f_\gamma(Q)= \frac{1}{m_\text{qk}^2} \; \frac{1}{\Nc \, } 
\sum_{a=1}^{\Nc} \mf^2{}_a  \; .
\label{photon_supp}
\end{align}
This result is not surprising, as
the photon production rate is usually written 
\cite{Baier:1991em, *Kapusta:1991qp, Aurenche:2000gf, Arnold:2002ja, *Arnold:2001ba,*Arnold:2001ms, *Arnold:2002ja} as proportional to the thermal
quark mass squared.
In the perturbative QGP this is somewhat trivial, however, as photon
production is naturally proportional to 
$\sim e^2 g^2 T^2$.   This relation is less trivial in the semi-QGP, since
then the thermal quark mass is a function of the $Q^a$'s.
Of course Eq.~(\ref{photon_supp}) holds only to the order
at which we compute, which is leading logarithmic
order.

To illustrate how large photon suppression can be, we take the
most extreme case,  the confined
phase.  It is most useful to use the form of 
$\Acal(Q)$ in Eq.~(\ref{first_def_acal}), as an integral over the
energy, $E$.  The gluon distribution enters as
\begin{equation}
\frac{1}{\Nc^2} \; \sum_{a,b = 1}^{\Nc} \Acal(Q^a - Q^b)
= \frac{6}{\pi^2} \int^\infty_0
\; dE \; E \; \frac{1}{\Nc^2} \; 
\sum_{a,b = 1} \frac{1}{e^{(E - i (Q^a-Q^b))/T} - 1}\; .
\label{sum_gluon_dist_Acal}
\end{equation}
In the perturbative QGP, $Q^a=0$, this integral is ${\cal A}(0)=T^2$.
In the confined phase, we use Eq.~(\ref{sum_gluon_dist}) to obtain
\begin{equation}
\frac{1}{\Nc^2} \; \sum_{a, b = 1}^{\Nc} 
\Acal(Q^a_{{\rm conf}} - Q^b_{{\rm conf}})
\; = \; \frac{6}{\pi^2} \int^\infty_0
\; dE \; E \; \frac{1}{e^{\Nc E/T} - 1}
\; = \; \frac{T^2}{\Nc^2} \; .
\label{sum_gluon_dist_AcalB}
\end{equation}
Notice that the integral over $E$ is {\it exactly} the same as when
$Q^a = 0$.  The {\it only} difference
is that because only loops which are multiples of $\Nc$ enter,
the energy enters not as $E/T$, but as $\Nc E/T$.
Hence in the confined phase we can replace 
$T$ by $T/\Nc$: as the integral is $\sim T^2$, 
this term is suppressed by $1/\Nc^2$ 
relative to that in the perturbative QGP.

From Eq.~\eqref{sum_gluon_dist_AcalB}, we see that the terms involving the gluon distribution
function in the thermal quark mass squared, Eq.~(\ref{thermal_quark_mass}), 
cancel identically.  This leaves only the
terms from the quark distribution functions, which are functions of the
color index $a$.  However, photon production
only depends only upon the sum over colors, Eq.~(\ref{photon_supp}), 
and so we compute
\begin{align}
\frac{1}{\Nc} \; \sum_{b = 1}^{\Nc} \widetilde\Acal(Q^b_{{\rm conf}})
& = - \frac{6}{\pi^2} \int^\infty_0
\; dE \; E \; \frac{1}{\Nc} \; 
\sum_{b = 1} \frac{1}{e^{(E - i Q^b_{{\rm conf}})/T} + 1} \no
\; & = - \; \frac{6}{\pi^2} \int^\infty_0
\; dE \; E \; \frac{1}{e^{\Nc E/T} + 1}
\; = \; - \; \frac{T^2}{2 \Nc^2} \; ,
\label{sum_qk_dist_AcalB}
\end{align}
by using Eq.~(\ref{sum_qk_dist}).  Again, this result is precisely
$1/\Nc^2$ times the result for $Q^a = 0$.
We thus find that in the confined phase, 
the square of the thermal quark masses, summed over color, is
\begin{equation}
\frac{1}{\Nc} \sum_{a=1}^{\Nc} \mf^2 (Q_{{\rm conf}})
= \left( \frac{\Nc^2 -1}{2 \Nc} \right) 
\; \frac{\cp^2 \, T^2}{24} \frac{1}{\Nc^2} \; . 
\end{equation}
Comparing to the thermal quark mass 
in the perturbative QGP, Eq.~(\ref{pert_quark_mass}), we obtain
\begin{equation}
f_\gamma(Q_{{\rm conf}}) = \frac{1}{3 \Nc^2} \; .
\label{ratio_pert_semi_photon}
\end{equation}
The coefficient of $1/3$ arises as follows.
As discussed following Eq.~(\ref{pert_quark_mass}), 
for the thermal quark mass squared in the perturbative QGP,
the gluon terms contribute two thirds (the $1$), and the quarks,
one third (the $1/2$).  
In the confined vacuum the gluon distributions cancel identically, while
the quark terms are precisely 
$1/\Nc^2$ times that for $Q^a = 0$, or $1/(3 \Nc^2)$ in all.

This shows that photon production is {\it strongly} suppressed in the confined
phase, by $\sim 1/\Nc^2$.  
Because the coefficient is small, $= 1/3$,
even for three colors the suppression
is significant, $= 1/27$.  This is why the suppression in 
Fig.~(\ref{fig_fQ}) is so dramatic.

The above analysis applies to the soft contribution to photon production.
It can also be computed from 
the hard contribution to photon production, 
since the suppression factor is common.
As demonstrated in Sec.~\ref{hard_photon_prod}, there are two
contributions.  That from Compton scattering is given in Eq.~(\ref{compton}), where by definition,
$f_{\text {Comp}}(0) = 1$ in the perturbative QGP.  
To compute its value in the confined phase of the pure
gauge theory, we remember that
the only nonzero loops are those which wrap around a multiple of $\Nc$
times, Eq.~(\ref{eq:ln-confined}).   Hence
\begin{align}
f_{\text {Comp}}(Q_{{\rm conf}}) =
\frac{12}{\pi^2}\; \sum_{n=1}^\infty\frac{(-1)^{n+1}}{n^2}\; 
\ell_n(Q_{{\rm conf}}) \; = \; 
\frac{12}{\pi^2}\; \sum_{j=1}^\infty\frac{(-1)^{j+1}}{(j \Nc)^2}
= \frac{1}{\Nc^2} \; .
\label{compton_conf}
\end{align}
The contribution of pair annihilation is given by 
$f_{\text{pair}}(Q)$ in Eq.~(\ref{pair_hard}), where
$f_{\text{pair}}(0) = 1$.  In the confined phase,
\begin{align}
\label{pair_hard_conf}
f_{\text{pair}}(Q_{{\rm conf}})=
\frac{1}{\Nc^2-1}\; \frac{6}{\pi^2 }\; \left(
\Nc^2 \;\sum_{j=1}^\infty\frac{1}{(j \Nc)^2} 
- \sum_{n=1}^\infty\frac{1}{n^2}\right) \; = 0 \; ,
\end{align}
and the contribution from pair annihilation vanishes identically.

In the confined phase, then, 
the hard part of photon production only receives
a contribution from Compton scattering.
From Eq.~(\ref{ff_supress}), relative to the perturbative QGP
photon production in the semi-QGP
is one third the sum of Compton scattering, plus
equal contributions from pair annihilation in the $t$ and $u$ channels.
Since pair annihilation vanishes in the confined phase, Eq.~(\ref{pair_hard_conf}), 
and the contribution from the Compton scattering 
is just $1/\Nc^2$ times that
of the perturbative QGP, Eq.~(\ref{compton_conf}), in all we obtain
a relative suppression factor of $1/(3 \Nc^2)$, 
Eq.~(\ref{ratio_pert_semi_photon}).

\section{Collinear rate}
\label{sec:photon-collinear}
\subsection{Review of AMY's calculation of photon production}

\subsubsection{Photon self-energy in RA basis}
\begin{figure}
\includegraphics[width=0.5\textwidth]{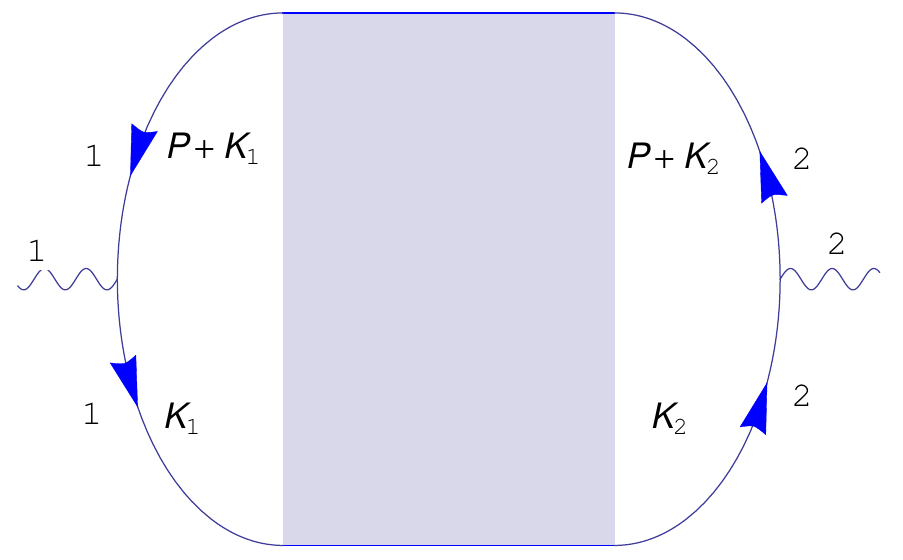}
\caption{The diagram of four point function 
with labels indicating fields insertions 
on different branches of the Keldysh contour.}\label{fig_G1122}
\end{figure}

First we recapitulate the detailed analysis by 
Arnold, Moore, and Yaffe (AMY)
~\cite{Arnold:2002ja, *Arnold:2001ba,*Arnold:2001ms, *Arnold:2002ja} 
for the collinear contribution to the photon production, in the case 
that $Q_a=0$.
We start with the expression for differential photon emission rate, 
Eq.~(\ref{eq:photon-W}).
In the $1/2$ basis in the real time formalism, $W_{\mu \nu}$ is given by
\begin{align}\label{Wmn}
W_{\mu\nu}&= e^2\sum_fq_f^2 \int\frac{d^4\lpnew_1}{(2\pi)^4}
\int \frac{d^4\lpnew_2}{(2\pi)^4}
\; (\lknew+2\lpnew_1)_\mu(\lknew+2\lpnew_2)_\nu  \; \notag\\
&\quad\times G_{1122}(-\lpnew_1,\lknew+\lpnew_1,-\lknew-\lpnew_2,\lpnew_2) \; .
\end{align}
As will be justified in the next subsection, $(P+2K_1)_\mu$ and 
$(P+2K_2)_\nu$ come from quark-photon vertices. 
$G_{1122}(-\lpnew_1,\lknew+\lpnew_1,-\lknew-\lpnew_2,\lpnew_2)$ is the 
Fourier transform of the four-point function $G_{1122}(x_1,x_2,y_1,y_2)$.  
The labels $1,2$ distinguish different field insertions on the 
Keldysh contour. 
Fig.~(\ref{fig_G1122}) summarizes the field labeling and momenta flow,
with convention that the momenta flow from right to left in propagators.
It is easier to calculate the four-point function in 
the RA basis, which is defined for quarks and gluons as
\begin{align}\label{ra_def}
&\psi_R=\frac{\psi_1+\psi_2}{2} \; , \; \; \psi_A=\psi_1-\psi_2 \; ; \\
&A^\mu_R=\frac{A^\mu_1+A^\mu_2}{2}\; , \;\; A^\mu_A=A^\mu_1-A^\mu_2 \; .
\end{align}
In this basis, 
$G_{RA}$ and $G_{AR}$ correspond to the
retarded and advanced propagators, respectively.  The
propagator $G_{AA}$ vanishes, while vertices with an odd number of R
indices vanish.  The latter is true for quark-gluon vertices only,
but only these are relevant to the calculation of the collinear rate.
To perform the calculation in the RA basis, 
we need to decompose $G_{1122}$ in terms of four-point functions. 
While there are in total 16 four-point functions, 
only 7 of them are independent \cite{Wang:1998wg}. 
The decomposition into an independent set has 
been done for neutral scalar in Ref.~\cite{Wang:1998wg}. 
It is easily generalized to the case of fermions with $\mu=0$ as
\begin{align}\label{12_ra}
G_{1122}&=\af_1G_{AARR}+\af_2G_{AAAR}+\af_3G_{AARA}+\af_4G_{ARAA}\no
&\quad+\af_5G_{RAAA}+\af_6G_{ARRA}+\af_7G_{ARAR} \no
&\quad+\bt_1G_{AARR}^*+\bt_2G_{AAAR}^*+\bt_3G_{AARA}^*+\bt_4G_{ARAA}^*\no
&\quad+\bt_5G_{RAAA}^*+\bt_6G_{ARRA}^*+\bt_7G_{ARAR}^* \; .
\end{align}
Detailed analysis by AMY~\cite{Arnold:2002ja, *Arnold:2001ba,*Arnold:2001ms, *Arnold:2002ja}
shows that the collinear rate receives contributions only from $G_{AARR}$.
Thus the only coefficients which we need are $\af_1$ and $\bt_1$,
\begin{align}\label{af_bt}
\af_1=\nf(p_1)\nf(p_2),\; 
\bt_1=-(1-\nf(p_3))(1-\nf(p_4))
\frac{-1+\nf(p_1)+\nf(p_2)}{-1+\nf(p_3)+\nf(p_4)} \; .
\end{align}
In our case 
$(p_1,p_2,p_3,p_4)=
(-\lpnew_1,\lknew+\lpnew_1,-\lknew-\lpnew_2,\lpnew_2)$. %, and $\nf(E)$ is the Fermi-Dirac distribution function.

\begin{figure}
\includegraphics[width=0.5\textwidth]{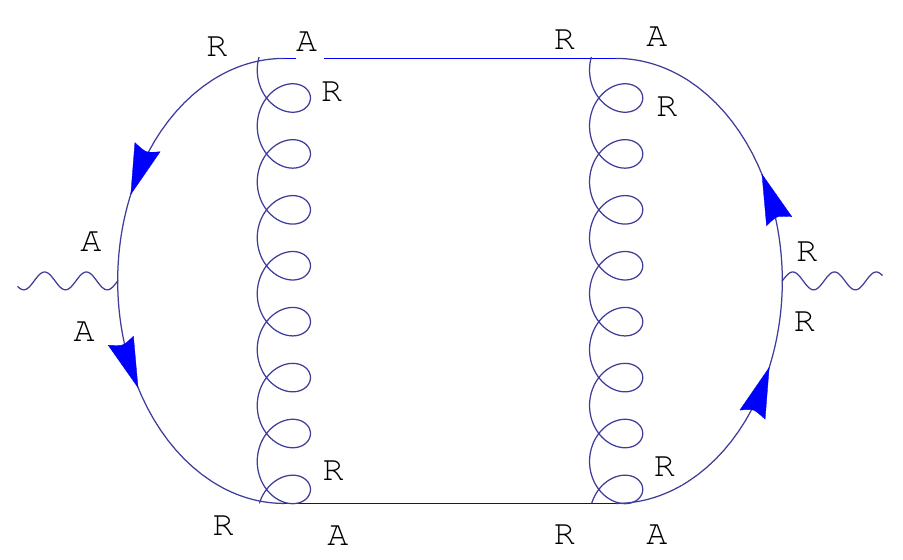}
\caption{The four point function $G_{AARR}$ with labels of all internal lines uniquely fixed in $RA$ basis.}\label{fig_Gaarr}
\end{figure} 
For the four-point function $G_{AARR}$, 
the RA labeling is uniquely fixed as in Fig.~(\ref{fig_Gaarr}). 
The contribution to the collinear regime 
arises from the kinematic regime where hard quark in the loop 
is nearly collinear with the photon: $\spnew^0\simeq \pl\gtrsim T$,
$\pt\sim gT$, with $\para$ and $\perp$ defined with 
respect to photon momentum $p$. 
The gluon exchanged between the quark lines are soft: 
$q^0\sim gT$, $q\sim gT$. From the collinear scattering of quarks and gluons,
the energy of the quarks remain unmodified at order $T$.
With this kinematic simplification, Eq.~\eqref{af_bt} reduces to
\begin{align}
\af_1\simeq\bt_1\simeq \nf(\pl+\sknew)(1-\nf(\pl)) \; ,
\end{align}
and therefore,
\begin{align}\label{1122aarr}
G_{1122}=2 \; \af_1 \; {\rm Re}\; G_{AARR} \; .
\end{align}

\subsubsection{Reduction of spinor structure}

To proceed, we need to see how collinear enhancement works. 
To do so, consider the convolution of two quark propagators, 
which enters as a unit upon inserting an additional gluon scattering 
into $G_{AARR}$:
\begin{align}\label{int_ss}
\int\frac{d\spnew^0}{2\pi}S_{AR}(\lknew+\lpnew)S_{RA}(\lpnew) \; .
\end{align}
$S_{AR}$ and $S_{RA}$ are advanced and retarded dressed quark propagators:
\begin{align}
-S=\frac{1}{{\slashed \spnew}-{\slashed \Sig}}
=\frac{1}{2}\frac{\gm^0-{\vec\gm}\cdot{\hat \spnew}}{A_0-A_s}
+\frac{1}{2}\frac{\gm^0+{\vec\gm}\cdot{\hat \spnew}}{A_0+A_s} \; ,
\end{align}
where 
\begin{equation}
A_0=\spnew^0-\Sig^0 \; , \; 
A_s=|{\vec \spnew}-{\vec \Sig}| \; .
\end{equation}
Here $\slashed \Sig=\Sigma^\mu\gamma_\mu$ is the retarded 
or advanced quark self-energy, and $\Sigma^\mu$ does not have a 
spinor structure.
Note that due to rotational symmetry, 
${\vec \spnew}\para{\vec \Sig}$, so $A_s=|\spnew-\Sig|$. 
The advanced and retarded propagators differ only in the sign of the 
damping rate, which corresponds to the imaginary part of self-energy $\Sig$. 
Both $S_{AR}$ and $S_{RA}$ have two poles with positive and negative energies. 
The collinear enhancement occurs when two poles coming from the two 
propagators pinch the real axis of $k_0$ plane. 
Thus, it suffices to consider the pole contribution:
\begin{align}\label{prop_pole}
-S(\spnew)\simeq \left\{\begin{array}{ll}
\frac{1}{2}\frac{\gm^0-{\vec\gm}\cdot{\hat \spnew}}{A_0-A_s}
\simeq \frac{{\slashed \spnew}}{2\spnew^0(A_0-A_s)}& \text{for}\;\spnew^0>0\\
\frac{1}{2}\frac{\gm^0+{\vec\gm}\cdot{\hat \spnew}}{A_0+A_s}
\simeq \frac{{\slashed \spnew}}{2\spnew^0(A_0+A_s)}& \text{for}\; \spnew^0<0
\end{array}
\right..
\end{align}
It is useful to write ${\slashed \spnew}$ in terms of spinor sums:
\begin{align}\label{pslash}
{\slashed \spnew}&=\sum_su_s(\spnew){\bar u}_s(\spnew) \; ,
\;\text{for}\; \spnew^0>0 \; , \no
{\slashed \spnew}&=\sum_tv_t(\spnew){\bar v}_t(\spnew) \; ,
\;\text{for}\; \spnew^0<0 \; ,
\end{align}
where $u$ and $v$ refer to the spinor basis
\begin{align}
u&=\(\sqrt{\spnew\cdot\sig}\xi_s,\sqrt{\spnew\cdot{\bar \sig}}\xi_s\)^T,\no
v&=\(\sqrt{\spnew\cdot\sig}\eta_t,-\sqrt{\spnew\cdot{\bar \sig}}\eta_t\)^T \; ,
\end{align}
with
\begin{align}
\xi_s=(\dlt_{s1},\dlt_{s2})^T,\;\eta_t=(\dlt_{t1},\dlt_{t2})^T
,\;\;s,t=1,2 \; .
\end{align}
Here, $\sigma_{\mu}=(1,\sigma^i)$, $\bar{\sigma}_{\mu}=(1,-\sigma^i)$
with Pauli matrices $\sigma^i$.
Note that $\spnew^0,\spnew\gtrsim T$ and  $\Sig\sim gT$, 
so we can take
\begin{align}\label{A_approx}
A_0-A_s\simeq \spnew^0{\pm}\frac{i}{2}\Gm_\spnew-E_\spnew \; ,\; \;\;
A_0+A_s\simeq \spnew^0{\pm}\frac{i}{2}\Gm_\spnew+E_\spnew \; , \;\;\;
E_\spnew=\sqrt{\spnew^2+{\color{black}{\minf^2}}
%\frac{m^2}{2\spnew}
} \; .
\end{align}
The asymptotic thermal mass $\minf$ and damping rate 
$\Gm_k/2$ are of order $\minf\sim gT$ and $\Gm_\spnew\sim g^2T$,
and the explicit expressions of these quantities will be given later.
For the retarded (advanced) propagator, we take 
the positive (negative) sign, respectively. 
It is not difficult to find that the pinching of poles 
occurs when $\spnew^0\simeq \pl$, with $\pl$ defined with respect to photon momentum $p$. With the approximation in Eqs.~\eqref{A_approx} and
\eqref{int_ss} evaluates to
\begin{equation}\label{int_ss2}
\int\frac{d\spnew^0}{2\pi}S_{AR}(\lknew+\lpnew)S_{RA}(\lpnew) 
\simeq \left. 
\frac{({\slashed \lknew}+{\slashed \lpnew})\,{\slashed \lpnew}}
{4\pl(\pl+\sknew)
\(\Gm+i\dlt E\)}\right|_{\spnew^0=\pl} \; ,
\end{equation}
where
\begin{equation}
\Gm\equiv\frac{1}{2}(\Gm_\spnew+\Gm_{\spnew+\sknew}) \; ,
\dlt E\equiv E_\spnew \; {\rm sgn}(\pl)+\sknew-
E_{\spnew+\sknew} \; {\rm sgn}(\pl+\sknew) \; .
\end{equation}
Note that ${\slashed \lknew}+{\slashed \lpnew}$ and 
${\slashed \lpnew}$ in the numerator of Eq.~\eqref{int_ss2} 
carry independent spinor indices, which are to be contracted 
with quark-gluon and quark-photon vertices. 
Contracting each quark-gluon vertex with two spinors from the propagators
joining it, with Eq.~\eqref{pslash} there is one of two situations,
depending on the sign of $\spnew^0$.  Since
\begin{align}\label{qg_vertex}
{\bar u}_t(\lpnew)\gm^\mu u_s(\lpnew)=2 \; \lpnew^\mu\; \dlt_{ts}
\;\;\; , \;\;\;
{\bar v}_t(\lpnew)\gm^\mu v_s(\lpnew)=2 \; \lpnew^\mu \; \dlt_{ts} \; ,
\end{align}
each gives the same result.  
We have neglected the momentum of the exchanged soft gluon $Q$, 
since it is negligible compared with $K$.
Cross terms between $u$ and $v$ are not allowed because multiple scatterings 
with soft gluons do not change the sign of $\spnew^0$. 
We have not included the coupling constant $g$ and color factors, 
which will be discussed separately in the next subsection. 
According to Eq.~\eqref{qg_vertex}, each quark-gluon vertex 
gives rises to $2\lpnew^\mu$, while maintaining the quark's chirality.

Now consider the quark-photon vertex, contracting 
the left/right quark-photon vertex with two spinors 
from the propagators joining them. 
As an example, consider $\spnew^0>0$:
\begin{align}
{\bar u}_t(\lpnew)\gm^\mu u_s(\lpnew+\lknew)
\; , \;\;\; {\bar u}_s(\lpnew+Q+\lknew)\gm^\nu u_t(\lpnew+Q) \; .
\end{align}
Summing over spinor indices and (transverse) photon polarizations, 
after some algebra~\cite{Arnold:2002ja, *Arnold:2001ms, *Arnold:2002ja} we obtain
\begin{align}\label{qp_vertex}
\sum_{s,t=1,2}\; \sum_{i=\perp} \; &
{\bar u}_t(\lpnew)\gm^i u_s(\lpnew+\lknew)
\; {\bar u}_s(\lpnew+Q+\lknew)\gm^i u_t(\lpnew+Q) \no
&= \; 4\; \spnew^0\; 
(\spnew^0+\sknew)\; 
\pt\cdot(\pt+q_\perp)
\left(\frac{(\spnew^0)^2+(\spnew^0+\sknew)^2}{(\spnew^0)^2(\spnew^0+\sknew)^2}
\right)
\;  \no
&\simeq  
4\pt\cdot(\pt+q_\perp)
\left(\frac{\pl^2+(\pl+\sknew)^2}{\pl(\pl+\sknew)}\right)
 \; .
\end{align}
The other cases are similar, with the same result as
Eq.~\eqref{qp_vertex}. Note that by definition $\sknew_\perp=0$, 
so we can write Eq.~\eqref{qp_vertex} as
\begin{align}\label{scalar_vertex}
\sum_{s,t=1,2}\; \sum_{i=\perp}\;
& {\bar u}_t(\lpnew)\gm^i u_s(\lpnew+\lknew)
{\bar u}_s(\lpnew+Q+\lknew)\gm^i u_t(\lpnew+Q) \no
&= \left(\frac{\pl^2+(\pl+\sknew)^2}{\pl(\pl+\sknew)}\right)
\sum_{i=\perp}\eps^i_\mu\eps^i_\nu
\; (2\lpnew+\lknew)^\mu(2\lpnew+2Q+\lknew)^\nu \; .
\end{align}
Apart from an overall factor $(\pl^2+(\pl+\sknew)^2)/(\pl(\pl+\sknew))$, 
Eq.~\eqref{scalar_vertex} allows us to interpret 
$(2\lpnew+\lknew)^\mu$ and $(2\lpnew+2Q+\lknew)^\nu$ 
as quark-photon verices on the left and right of self-energy diagrams.
We have thus shown in Eqs.~\eqref{qg_vertex} and \eqref{scalar_vertex} that quark-gluon and quark-photon vertices can be simplified as $2(K_1+K_2)^\mu$ with $K_1$ and $K_2$ being the incoming and outgoing momenta of quarks.
We note that $\sum_{i=\perp}\eps^i_\mu \eps^i_\nu = -g_{\mu\nu}$
as in Eq.~(\ref{eq:photon-W}).

\subsubsection{Color structure in the double line basis}
The color structure of the gluon propagator is given by $P^{ab}_{cd}$, 
as in Eq.~(\ref{prop_ra}).
Thus the color sum which appears when a gluon propagator 
is sandwiched between two quark-gluon vertices is
\begin{align}\label{unit_color}
\(T^{ab}\)_{ef}
\;
P^{ab}_{cd}
\;
\(T^{dc}\)_{gh}=\frac{1}{2} \; P^{hg}_{ef} \; .
\end{align}
This can be simplified further by noting that 
in the photon self-energy, Eq.~\eqref{unit_color} 
is sandwiched with the quark-photon vertex in the vertices
which are all the way to the left or all the way to the right.
Starting from the left hand side gives
\begin{align}\label{dlt_P}
\frac{1}{2} \; \dlt^e_h  \; P^{hg}_{ef}
=\frac{1}{2}\left(\Nc-\frac{1}{\Nc}\right) \; \dlt^f_g=C_F \; \dlt^f_g \; ,
\end{align}
where $C_F = (\Nc^2-1)/(2\Nc)$ is the quadratic Casimir for the fundamental
representation.  If this is iterated further, each quark-gluon vertex
preserves the Kronecker delta in color, and generates an additional
factor of $C_F$.  After the last quark-gluon scattering, the delta
function is color is contracted with the right most quark-photon vertex,
giving an overall factor of $\Nc$.

\subsubsection{Resummation of infinite self-energy diagrams}

\begin{figure}
\includegraphics[width=0.5\textwidth]{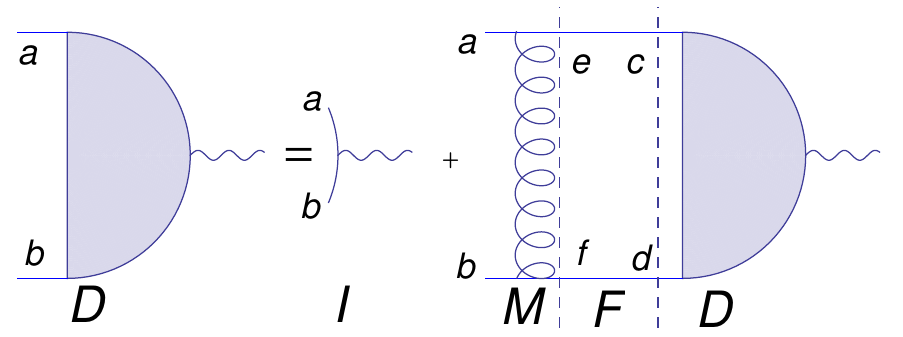}
\caption{The diagrammatic equation in 
terms of graphical elements $D$, $I$, $M$ and $F$.}
\label{fig_diagram_eq}
\end{figure} 

We next resum diagrams with arbitrary quark-gluon scatterings. 
This is done by solving the 
integral equation illustrated in Fig.~(\ref{fig_diagram_eq}).
The graphical elements are the same as those defined by
Arnold, Moore, and Yaffe \cite{Arnold:2002ja, *Arnold:2001ms, *Arnold:2002ja},
except that we use the double
line notation for future convenience.
The 
integral equation shown in Fig.~(\ref{fig_diagram_eq}) becomes
\begin{align}\label{diag_eq_color}
D^{\mu}_{ab}(\lpnew,\lknew)=
I^{\mu}_{ab}(\lpnew,\lknew)+
\int\frac{d^4q}{(2\pi)^4} \; M(\lpnew,Q,\lknew)_{ab,ef} \; 
F(\lpnew+Q,\lknew)_{ef,cd} \; D^{\mu}_{cd}(\lpnew+Q,\lknew) \; .
\end{align}
The color structure can be taken as
\begin{align}\label{ge_color}
&I^{\mu}_{ab}(\lpnew,\lknew)= I^\mu(\lpnew,\lknew) \; \dlt_{ab} \; , \no
&M(\lpnew,Q,\lknew)_{ab,ef}=\frac{1}{2C_F}
\; {\cal P}^{ae}_{bf} \; M(\lpnew,Q,\lknew) \; , \no
&F(\lpnew+Q,\lknew)_{ef,cd}= \; \dlt_{ce} \; \dlt_{df} 
\; F(\lpnew+Q,\lknew) \; , \no
& D^{\mu}_{ab}(\lpnew,\lknew)=D^\mu(\lpnew,\lknew) \; \dlt_{ab} \; ,
\end{align}
so that Eq.~\eqref{diag_eq_color} simplifies 
\begin{align}\label{diag_eq}
D^{\mu}(\lpnew,\lknew)=I^{\mu}(\lpnew,\lknew)+
\int\frac{d^4q}{(2\pi)^4} \; M(\lpnew,Q,\lknew)
\; F(\lpnew+Q,\lknew) \; D^{\mu}(\lpnew+Q,\lknew) \; .
\end{align}
Note the color factor from Eq.~\eqref{dlt_P} cancels the factor of
$1/(2C_F)$ in Eq.~\eqref{ge_color}.

As discussed previously, we regard the quark-gluon vertices 
and quark-photon vertices as $(\lpnew+Q)^\mu$, with 
$\lpnew$ and $Q$ being the incoming and outgoing momenta. As a result, 
\begin{align}
I^\mu(\lpnew,\lknew)=(2\lpnew+\lknew)^\mu \; ,
\end{align}
and the function $F(\lpnew,\lknew)$ equals
\begin{align}
F(\lpnew,\lknew)&=(-i)^2
\left. G_{AR}(\lknew+\lpnew)
\; G_{RA}(\lpnew)\right|_{\text{pinch}} \no
&\simeq \left( \frac{-1}{4\pl(\pl+\sknew)}\right)
\left( \frac{1}{\Gm+i\dlt E}\right)
4\pi\; 
\dlt\left(2\spnew^0+\sknew-E_\spnew 
{\rm sgn}(\pl)-E_{\spnew+\sknew}{\rm sgn}(\pl+\sknew) \right) \no
&\simeq \left(\frac{-1}{4\pl(\pl+\sknew)}\right)
\left(\frac{1}{\Gm+i\dlt E}\right) 2\pi \; \dlt(\spnew^0-\pl) \; .
\end{align}
The pinching condition is enforced by the delta function. The rung of the
ladder equals
\begin{align}\label{M}
M(\lpnew,Q,\lknew)&=
ig^2 \; C_F \; (2\lpnew+Q+2\lknew)^\mu \; (2\lpnew+Q)^\nu 
\; G^{RR}_{\mu\nu}(Q) \no
%&=4ig^2 \; C_F \; (\lpnew+\lknew)^\mu \; \lpnew^\nu \; G^{RR}_{\mu\nu}(Q) \no
&\simeq 4ig^2 \; C_F \;
\pl \; (\pl+\sknew) \; {\widehat P}^\mu \; {\widehat P}^\nu \;
G^{RR}_{\mu\nu}(Q) \; ,
\end{align}
with ${\widehat \lknew}^\mu=(1,{\hat \sknew})$. 
The Ward identity and the fact that $\lpnew^\mu$ is almost collinear with 
$\lknew^\mu$ was used to simplify Eq.~\eqref{M}. 
To further simplify Eq.~\eqref{diag_eq}, we define
\begin{align}\label{f_def}
f^\mu(\bp,\sknew)\equiv-4\pl(\pl+\sknew)\int
\frac{d\spnew^0}{2\pi}F(\lpnew,\lknew)D^\mu(\lpnew,\lknew) \; ,
\end{align}
%
%We have used $\bp$ as argument because of the delta function $\dlt(k^0-k_\para)$ in $F(\lpnew,\lknew)$.
%From Eq.~\eqref{diag_eq}, we derive an integral equation for $f^\mu({\bp},\sknew)$,
which leads us to
\begin{align}\label{f_eq}
(\Gm+i\dlt E)f^\mu({\bp},\sknew)=
(2\lpnew+\lknew)^\mu+\int\frac{d^3q}{(2\pi)^3}
\; C({\bq},\sknew)\; f^\mu(\bp+\bq,\sknew) \; ,
\end{align}
where
\begin{align}\label{coll}
C(\bq,\sknew)=g^2C_F\int
\frac{dq^0}{2\pi}\; 2\pi\; \dlt(q^0-q_\para)
(-iG^{RR}_{\mu\nu}(Q){\widehat \lknew}^\mu{\widehat \lknew}^\nu) \; .
\end{align}
The delta function in Eq.~\eqref{coll} again results from $\dlt(k^0+q^0-k_\para-q_\para)$ in $F(\lpnew+Q,\lknew)$.
We can further simplify Eq.~\eqref{f_eq} 
using the explicit expression of the damping rate $\Gm$,
\begin{align}\label{damping}
\Gm_\spnew=g^2C_F\int
\frac{d^3q\; dq^0}{(2\pi)^4}\; 2\pi\; \dlt(q^0-q_\para)
(-iG^{RR}_{\mu\nu}(Q){\widehat \lknew}^\mu{\widehat \lknew}^\nu) \; .
\end{align}
As this is independent of $\spnew$, 
$\Gm=(\Gm_\spnew+\Gm_{\spnew+\sknew})/2=\Gm_\spnew$. 
This allows us to write Eq.~\eqref{f_eq} as
\begin{align}\label{f_inteq}
i\; \dlt E \; f^\mu(\bp,\sknew)=
(2\lpnew+\lknew)^\mu+\int
\frac{d^3q}{(2\pi)^3}\;
C({\bq},\sknew)\; [f^\mu(\bp+\bq,\sknew)-f^\mu(\bp,\sknew)] \; .
\end{align}
As is clear from Eq.~(\ref{scalar_vertex}),
only the transverse components of $f^\mu(\spnew,\sknew)$ are 
needed, so we can project Eq.~\eqref{f_inteq} onto the transverse plane,
\begin{align}\label{f_inteq_tr}
i\; \dlt E \; {\bf f}_\perp(\bp,\sknew)
=2{\bf \spnew}_\perp+\int\frac{d^3q}{(2\pi)^3}
\; C({\bq},\sknew)\; [{\bf f}_\perp(\bp+\bq,\sknew)-{\bf f}_\perp(\bp,\sknew)] \; .
\end{align}
The last element is to determine the propagator,
$G^{RR}_{\mu\nu}(Q){\widehat \lknew}^\mu{\widehat \lknew}^\nu$. 
Since $q\sim gT$, we use the HTL-resummed propagator, 
\begin{align}\label{grr}
-iG^{RR}_{\mu\nu}(Q){\widehat \lknew}^\mu{\widehat \lknew}^\nu
=\frac{2T}{q_0}\left(1-\frac{q^2_\parallel}{q^2}\right)
{\text{Im}}\(\frac{1}{Q^2-\Pi^R_T(Q)}
-\frac{1}{Q^2-\Pi^R_L(Q)}\) \; ,
\end{align}
where we have taken Feynman gauge, and used $q_0\simeq q_\parallel$.
$\Pi_L$ and $\Pi_T$ are the retarded longitudinal 
and transverse self-energies of the gluon:
\begin{align}
\Pi^R_L(Q)&=\frac{Q^2}{q^2}M^2\left[1-\frac{q_0}{2q}
\ln\left(\frac{q_0+q}{q_0-q}\right)\right] \; ,\\
\Pi^R_T(Q)&= \frac{M^2}{2}\left[\left(\frac{q_0}{q}\right)^2
-\frac{Q^2}{q^2}\frac{q_0}{2q}\ln\left(\frac{q_0+q}{q_0-q}\right)
\right] \; ,
\end{align}
where the gluon Debye mass is given by 
\begin{align}
\label{eq:Debye-Q=0}
M^2= g^2T^2 \left(\frac{\Nc}{3}+\frac{\Nf}{6}\right).
\end{align}
The Wightman correlator for two electromagnetic currents can be
expressed as 
\begin{align}\label{W_full}
W^{\mu\nu}&=(-)2\Nc e^2\sum_f q_f^2\af_1 \; 
{\rm Re}\int\frac{d^4\lpnew}{(2\pi)^4}
\; I^\mu(\lpnew,\lknew)\; F(\lpnew,\lknew)\; D^\nu(\lpnew,\lknew)
\left( \frac{\pl^2+(\pl+\sknew)^2}{\pl(\pl+\sknew)} \right) \no
&=\Nc  e^2\sum_f q_f^2 \; \int\frac{d^3 \spnew}{(2\pi)^3} \; A(\pl,\sknew) \; 
{\rm Re} [I^\mu(\lpnew,\lknew)f^\nu(\lpnew,\lknew)] \; ,
\end{align}
where
\begin{align}
A(\pl,\sknew)=\nf(\pl+\sknew)(1-\nf(\pl))
\left( \frac{\pl^2+(\pl+\sknew)^2}{2\pl^2(\pl+\sknew)^2} \right) \; .
\end{align}
Note that an overall factor of $(\pl^2+(\pl+\sknew)^2)/(\pl(\pl+\sknew))$ 
in Eq.~\eqref{scalar_vertex} is inserted into Eq.~\eqref{W_full} 
along with $-1$ from the fermion loop. 
From Eq.~(\ref{eq:photon-W}), $W^{\mu\nu}$ is contracted with 
$-g_{\mu\nu}$ to give the collinear rate:
\begin{align}\label{LPM}
\sknew\frac{d\vGm_\gm}{d^3\sknew}=\frac{\af_\text{em}\Nc \sum_fq_f^2}{4\pi^2}
\int\frac{d^3\spnew}{(2\pi)^3}\; A(\pl,\sknew)
\; {\rm Re} [2{\bf \pt}\cdot {\bf f}_\perp(\spnew,\sknew)] \; .
\end{align}
Note there is an additional factor of $\Nc$ for each color of $W^{\mu\nu}$.

\subsection{Photon self-energy with nontrivial Polyakov loop}

\subsubsection{Quark and gluon thermal masses with background color charge}

Now we compute the modification of the results in the previous section
in the presence of a nontrivial Polyakov loop. 
In this case, quantities like the thermal mass, the damping rate,
and so on are all dependent on the background color charge.  This
changes the color structure of the self-energy diagrams. 

The quantities relevant for the problem at hand are the quark asymptotic thermal mass and the resummed gluon propagator. 
The asymptotic quark thermal mass is $\sqrt{2}$ 
times the quark thermal mass in Eq.~(\ref{thermal_quark_mass}):
\begin{align}
\label{eq:quark-asymptotic-mass}
\minf^2_a=2\mf^2{}_a \; .
\end{align}
%
%
%\begin{align}\label{asym_mass}
%m_a^2(Q\ne0)=\frac{g^2}{24}\big[\sum_{b=1}^\Nc\(\Acal(Q^{ab})-\widetilde{\Acal}(Q^b)\)-\frac{1}{\Nc}\(\Acal(0)-\widetilde{\Acal}(Q^a)\)\big] .
%\end{align}
%
%At $Q=0$, it reduces to the familiar thermal mass
%
%\begin{align}
%&m^2(Q=0)=\frac{C_Fg^2T^2}{8}.
%\end{align}
%

\begin{figure}
\includegraphics[width=0.5\textwidth]{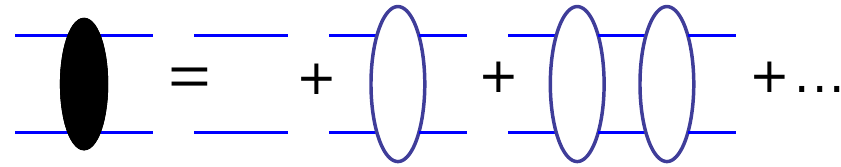}
\caption{The resummed gluon propagator as an infinite series of 
propagators with arbitrary self-energy insertions. 
Each unfilled circle represents a self-energy insertion $\Pi$.
The double line notation is used.}\label{fig_resummed}
\end{figure}

Next we consider the HTL-resummed gluon propagator. 
The resummed gluon propagator consists of bare gluon propagators 
with arbitrary number of self-energy insertions, Fig.~(\ref{fig_resummed}). 
The bare gluon propagator in the RA basis of the real time formalism, 
in the presence of background color charge, is proportional to ${\cal P}^{ab}_{cd}$:
\begin{align}\label{prop_ra}
&G_{RA,ab,cd}^{\mu\nu}=
\frac{g^{\mu\nu}}{(\spnew^0+i\eps)^2-\spnew^2}{\cal P}^{ab}_{cd}\; ,\;\; 
G_{AR,ab,cd}^{\mu\nu}=\frac{g^{\mu\nu}}{(\spnew^0-i\eps)^2-\spnew^2}
{\cal P}^{ab}_{cd}\; , \no
&G_{RR,ab,cd}^{\mu\nu}=-i\pi\eps(\spnew^0)
(1+2\nb_{ab}(\spnew^0))
\dlt(\lpnew^2)g^{\mu\nu}{\cal P}^{ab}_{cd}\; ,\;\; G_{AA}^{\mu\nu}=0 \; .
\end{align}

Here we need to recall that in the analysis in the case of
$Q_a=0$, 
Bose-Einstein enhancement was essentially important for the 
collinear contribution to be as large as the $2 \rightarrow2$ contribution: 
for soft gluons with $\spnew^0\sim gT$,
$n(\spnew^0)\sim 1/g$. 
This is no longer true in the presence of hard background 
charge $Q_a\sim T$.  
The only exception is for diagonal gluons, $Q_a=Q_b$, 
where Bose-Einstein enhancement is still operative. 
Therefore, we only need the expression of the diagonal components of the gluon propagator. 
Thus we contract the bare gluon propagator 
in Eq.~\eqref{prop_ra} with $\dlt_{ab}$. 
As a result, the soft diagonal gluon carries only one index,
\begin{align}
\dlt_{ab}{\cal P}^{ab}_{cd}=\dlt_{ab}\dlt_{cd}{\cal P}_{ac}\; ,
\end{align}
where we defined a color projection operator for one-index gluons,
\begin{align}
&{\cal P}_{ac}\equiv \dlt_{ac}-\frac{1}{\Nc}\; , 
\end{align}
which satisfies ${\cal P}_{ab}{\cal P}_{bc}={\cal P}_{ac}$.
In terms of one-index projection operator, %schematically the
the resummed gluon propagator is given by the sum of the following terms:
\begin{align}\label{series}
\frac{1}{\lqnew^2}{\cal P}_{ab} 
+ 
\frac{1}{\lqnew^2}{\cal P}_{ac}\Pi_{cd}
\frac{1}{\lqnew^2}{\cal P}_{db}
+ 
\frac{1}{\lqnew^2}
{\cal P}_{ac}\Pi_{cd}\frac{1}{\lqnew^2}
{\cal P}_{de}\Pi_{ef}\frac{1}{\lqnew^2}{\cal P}_{fb}
+ \ldots \; ,
\end{align}
where the Lorentz indices are suppressed for the time being.
%Here $\Pi_{cd}$ is given by $\Pi^{\mu\nu}_{cd}(Q)\equiv \Pi^{\mu\nu}_{cc,dd}(Q)$, with $\Pi^{\mu\nu}_{cd,ef}(Q)$ is the gluon self-energy.
Here $\Pi_{cd}$ is given by $\Pi_{cd}(Q)\equiv \Pi_{cc,dd}(Q)$, with $\Pi_{cd,ef}(Q)$ is the gluon self-energy.
Each color projection operator is accompanied by a momentum 
dependent part of the bare propagator $1/{\lqnew^2}$. 
The color structure of one-index gluon self-energy is 
\begin{align}\label{Pi_color}
\Pi_{ab}=\dlt_{ab}F_a-\frac{1}{\Nc}\; G_{ab}\; .
\end{align}

By restoring the Lorentz indices, the gluon self-energy in the HTL approximation~\cite{Hidaka:2009hs} is
given by 
\begin{align}
\Pi^{\mu\nu}_{ab,cd}(Q)= M_{ab,cd}^2\dlt\Pi^{\mu\nu}\; .
\end{align}
Here
\begin{equation}
\dlt\Pi^{\mu\nu}= 
\left( -\delta^{0\mu}\delta^{0\nu}
+q^0\; \int \frac{d\Omega_q}{4 \pi} \; \frac{\widehat{Q}^\mu \widehat{Q}^\nu}
{Q \cdot \widehat{Q} \pm i\epsilon} \right)
\end{equation}
with $\widehat{Q}\equiv (1,\widehat{\bf{q}})$.
The sign in the denominator is plus (minus) 
if the self-energy is retarded (advanced). 
The form of $\dlt\Pi^{\mu\nu}$ is identical to that in perturbative QGP. The Polyakov loop dependence is entirely in the gluon Debye mass:
%The gluon Debye mass is
%
\begin{align}\label{gluon_Pi}
M_{ab,cd}^2=&\frac{g^2}{6}\left[\dlt_{ac}\dlt_{bd}
\left(\sum_{e=1}^\Nc\(\Acal(Q^{ae})+\Acal(Q^{eb})\)
-\Nf\(\widetilde{\Acal}(Q^a)+\widetilde{\Acal}(Q^b)\)\right) \right.\no
&-\; \left. 2\dlt_{ab}\dlt_{cd}\left(\Acal(Q^{ac})
-\frac{\Nf}{\Nc}\(\widetilde{\Acal}(Q^a)+\widetilde{\Acal}(Q^c)-\frac{1}{\Nc}
\sum_{e=1}^\Nc\widetilde{\Acal}(Q^e)\)\right)\right]\; ,
\end{align}
which for $Q_a=0$ reduces to Eq.~(\ref{eq:Debye-Q=0}), namely
\begin{align}
M_{ab,cd}^2=
g^2T^2\(\frac{1}{3}\Nc +\frac{1}{6}\Nf\){\cal P}^{ab}_{cd} \; .
\end{align}
Equation~(\ref{gluon_Pi}) leads us to
\begin{align}
&F^{\mu\nu}_a = \frac{g^2}{3}\(\sum_{e=1}^\Nc\Acal(Q^{ae}{)}-\Nf\widetilde{\Acal}(Q^a)\) \delta\Pi^{\mu\nu}(Q)\; ,\no
&G^{\mu\nu}_{ab} = \frac{g^2}{3}\(\Nc\Acal(Q^{ab})-\Nf\(\widetilde{\Acal}(Q^a)+\widetilde{\Acal}(Q^b)\)-\sum_{e=1}^\Nc\widetilde{\Acal}(Q^e)\)\delta\Pi^{\mu\nu}(Q)\; ,
\end{align}
where we have restored the Lorentz indices.

Formally the two terms in Eq.~\eqref{Pi_color}
are of the same order if we regard $\dlt_{ab}\sim 1/\Nc$. 
However, we show in Appendix B that a naive large $\Nc$ limit is justified.
This allows us to disregard the term proportional to $G$, 
so that the gluon Debye mass becomes
\begin{align}\label{mg2_L}
\mD_{a}^2%=2m_a^2(Q\ne0)
=\frac{g^2}{3}
\left[\sum_{e=1}^\Nc\Acal(Q^{ae})-\Nf\widetilde{\Acal}(Q^a)\right],
\end{align}
where $\mD^2_a$ is defined as $\mD^2_{aa,bb}= \mD^2_a \delta_{ab}$.
When $Q^a =2\pi Tq\ne0$, 
the explicit form of $\mD^2_{a}$ for $\Nc=\Nf=3$ are
\begin{align}\label{mD_explicit}
&\mD^2_{1}=\mD^2_{3}=g^2T^2\(\frac{3}{2}-6q+4q^2\), \no
&\mD^2_{2}=g^2T^2\(\frac{3}{2}-4q+4q^2\).
\end{align}

%
%Since the only place the thermal gluon mass enters in 
%Eq.~\eqref{grr} is through the Debye mass, which related to \eqref{mg2} by $m_D^2=2m_a^2(Q=0)$, we conclude the effect of Polyakov loop is replace it by:
%\begin{align}\label{debye}
%m_{D,a}^2=\frac{g^2T^2}{\pi^2}\sum_{n=1}^\infty\frac{1}{n^2}2\cos(n\bt Q_a)(NTrL^n+(-1)^{n+1}\Nf).
%\end{align}
%Interestingly \eqref{debye} coincides with standard thermal gluon mass. We note however the color projection operator has changed from ${\cal P}_{ab}$ to $\dlt_{ab}$. The approximation becomes exact only in the large $N$ limit.
% I don't see the point of the discussion above, which repeats the exact
%same expression for the Debye mass.

%This paragraph was incorporated into the next.
%From the above discussion, quarks running around the loop 
%carry the same color index. The same holds when the Polyakov loop
%is nontrivial, except that the color index also enters into 
%various quantities such as thermal mass, damping rate etc. In the next subsection, we will formulate an integral equation with an explicit color index.

\subsubsection{AMY's integral equation with one color index}

\begin{figure}
\includegraphics[width=0.5\textwidth]{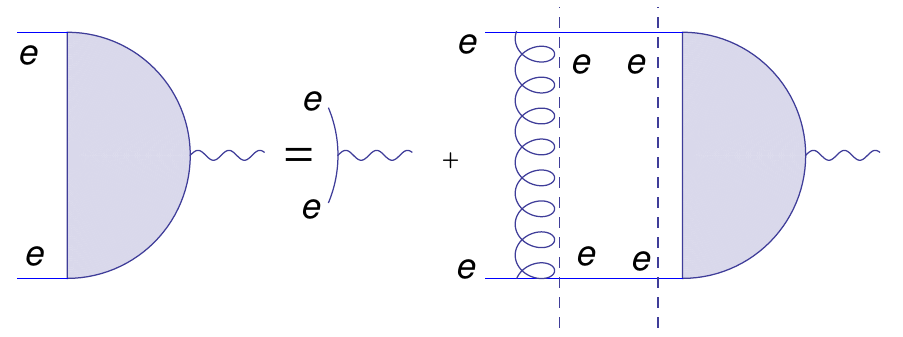}
\caption{The diagrammatic equation for a color index $e$, in
the presence of nontrivial Polyakov loop.}\label{fig_diagram_e}
\end{figure}

We next generalize the integral equation of AMY for a nontrivial Polyakov loop
in the limit of a large number of colors.
From the discussion of the previous subsection, 
all elements of the graph carry one color index,  
as in Fig.~(\ref{fig_diagram_e}). 

\begin{figure}
\includegraphics[width=0.4\textwidth]{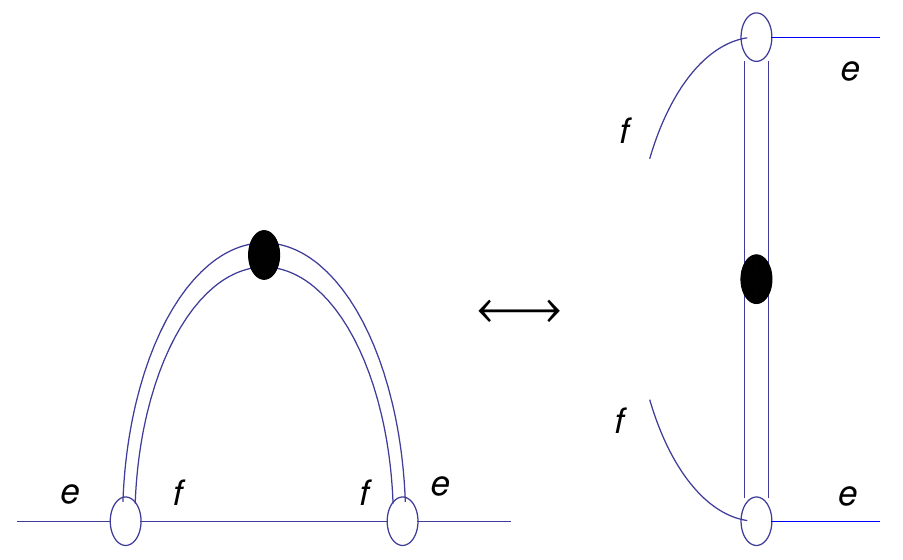}
\caption{The correspondence between the diagrams for
quark damping and for soft gluon exchange.
The left-hand side is the quark self-energy, 
whose imaginary part gives the damping rate of the quark.
The right-hand side is the diagram from gluon exchange, 
which is the cut diagram on the left-hand side.
The double line notation is used.
}
\label{fig_damping}
\end{figure}

The integral equation analogous to Eq.~\eqref{diag_eq} is
\begin{align}
D_e^{\mu}(\lpnew,\lknew)=I_e^{\mu}(\lpnew,\lknew)
+\int\frac{d^4q}{(2\pi)^4}
\; M_e(\lpnew,Q,\lknew) \; F_e(\lpnew+Q,\lknew)\; D_e^{\mu}(\lpnew+Q,\lknew)
\; .
\end{align}
Most quantities only need trivial modifications:
\begin{align}
&I^\mu_e(\lpnew,\lknew)=(2\lpnew+\lknew)^\mu \; , \no
&F_e(\lpnew,\lknew)=\frac{-1}{4\pl(\pl+\sknew)}\; 
\frac{1}{\Gm_e+i\dlt E_e}\; 2\pi\; \dlt(\spnew^0-\pl) \; , \no
&M_e(\lpnew,Q,\lknew)=
4ig^2\frac{1}{2}\pl(\pl+\sknew){\widehat \lknew}^\mu{\widehat \lknew}^\nu 
G_{\mu\nu,e}^{RR}(Q) \; .
\end{align}
A distinct difference is the color factor $C_F$ in $M_e$ changes to 
$1/2$.  This follows from enforcing color neutrality
on the soft gluon and dropping terms $1/\Nc$ 
in the gluon self-energy. 
Apart from this, the color index $e$ enters $\Gm$ and $\dlt E$ 
through quark asymptotic thermal mass 
Eq.~\eqref{eq:quark-asymptotic-mass} and gluon Debye mass Eq.~\eqref{mg2_L}. 
Explicitly,
\begin{align}\label{Egrr_e}
\dlt E_e&=E_\spnew {\rm sgn}(\pl)+\sknew-E_{\spnew+\sknew}
\; {\rm sgn}(\pl+\sknew)\simeq \frac{\sknew}{2\pl(\pl+\sknew)}(\pt^2+m_e^2)\; , \no
-iG^{RR}_{\mu\nu e}(Q){\widehat \lknew}^\mu{\widehat \lknew}^\nu
&=\frac{2T}{q_0}\left(1-\frac{q^2_\parallel}{q^2}\right)
{\text{Im}}\(\frac{1}{Q^2-\Pi^R_{T,e}(Q)}
-\frac{1}{Q^2-\Pi^R_{L,e}(Q)}\) \; ,
\end{align}
with
\begin{align}\label{Pi_e}
\Pi^R_{L,e}(Q)&=\frac{Q^2}{q^2}\mD^2_e\left[1-\frac{q_0}{2q}
\ln\left(\frac{q_0+q}{q_0-q}\right)\right]\; ,\\
\Pi^R_{T,e}(Q)&= \frac{\mD^2_e}{2}\left[\left(\frac{q_0}{q}\right)^2
-\frac{Q^2}{q^2}\frac{q_0}{2q}
\ln\left(\frac{q_0+q}{q_0-q}\right)
\right]\; .
\end{align}
Following the case with $Q=0$, we define
\begin{align}\label{f_def_e}
f^\mu_e(\bp,\sknew)=-4\pl(\pl+\sknew)
\int\frac{d\spnew^0}{2\pi}\; F_e(\lpnew,\lknew)\; D_e^\mu(\lpnew,\lknew) \; .
\end{align}
Similarly, 
\begin{align}\label{f_eq_e}
(\Gm_e+i\dlt E_e)f_e^\mu({\bp},\sknew)=(2\lpnew+\lknew)^\mu
+\int\frac{d^3q}{(2\pi)^3}\; C_e({\bq},\sknew)\; f_e^\mu(\bp+\bq,\sknew) \; ,
\end{align}
where
\begin{align}\label{coll_e}
C_e(\bq,\sknew)=\frac{g^2}{2}
\int\frac{dq^0}{2\pi}\; 2\pi\; \dlt(q^0-q_\para)
(-iG^{RR}_{\mu\nu,e}(Q){\widehat \lknew}^\mu{\widehat \lknew}^\nu) \; .
\end{align}
The term proportional to $\Gm_e$ can be written in terms of $C_e$,
\begin{align}\label{Gm_e}
\Gm_e=\frac{g^2}{2}\int\frac{d^3q\; dq_0}{(2\pi)^4}
\; 2\pi\; \dlt(q_0-q_\para)
(-iG^{RR}_{\mu\nu,e}(Q){\widehat \lknew}^\mu{\widehat \lknew}^\nu) \; .
\end{align}
Physically, this is because quark damping is due to 
scattering off of soft and diagonal gluons.
We note that, by using $q\sim gT$ and $G^{RR}_{\mu\nu,e}\sim T/q^3$, 
$\Gamma_e$ is of order $g^2T$.
This is suppressed by $1/\Nc$ compared with $\Gamma\sim g^2\Nc T$, 
which is the damping rate when $Q=0$.
The diagrams of quark damping are the same as gluon rung $M$, 
as illustrated in Fig.~(\ref{fig_damping}). Note that the Bose-Einstein enhancement fixes the color indices as $f=e$.

Therefore we have
\begin{align}\label{f_inteq_e}
i\dlt E_e \; f_e^\mu(\bp,\sknew)=
(2\lpnew+\lknew)^\mu+\int\frac{d^3q}{(2\pi)^3}\;
C_e({\bq},\sknew)\; [f_e^\mu(\bp+\bq,\sknew)-f_e^\mu(\bp,\sknew)] \; .
\end{align}
We again need only an equation for projected ${\bf f_e}$,
\begin{align}\label{f_inteq_etr}
i\dlt E_e \; {\bf f_{e\perp}}(\bp,\sknew)=2{\bf \spnew_\perp}+
\int\frac{d^3q}{(2\pi)^3}
\; C_e({\bq},\sknew)\; [{\bf f_{e\perp}}(\bp+\bq,\sknew)
-{\bf f_{e\perp}}(\bp,\sknew)]\; .
\end{align}
The last changes are for Eqs.~\eqref{12_ra} and \eqref{1122aarr}. 
Following the derivation of Ref.~\cite{Wang:1998wg}, 
Eq.~\eqref{12_ra} becomes
\begin{align}
G_{1122}&=\af_1G_{AARR}+\af_2G_{AAAR}+\af_3G_{AARA}+\af_4G_{ARAA}\no
&\quad+\af_5G_{RAAA}+\af_6G_{ARRA}+\af_7G_{ARAR}\no
&\quad+\bt_1G_{AARR}^\Dlt+\bt_2G_{AAAR}^\Dlt
+\bt_3G_{AARA}^\Dlt+\bt_4G_{ARAA}^\Dlt\no
&\quad+\bt_5G_{RAAA}^\Dlt
+\bt_6G_{ARRA}^\Dlt+\bt_7G_{ARAR}^\Dlt \; ,
\end{align}
where $\Dlt$ is defined as complex conjugation together with 
charge conjugation, {\it i.e.} 
flipping the sign of background color charge. 
The relevant coefficients are 
\begin{align}
\af_1\simeq \bt_1\simeq \nf_e(\pl+\sknew)(1-\nf_e(\pl))\; .
\end{align}
As a result, Eq.~\eqref{1122aarr} becomes
\begin{align}
G_{1122}=\af_1G_{AARR}(Q_a)+\af_1G_{AARR}^*(-Q_a)\; .
\end{align}
Note that the background charge enters the integral equation 
Eq.~\eqref{f_inteq_etr} only through Eqs.~\eqref{eq:quark-asymptotic-mass} and 
\eqref{mg2_L}, which are independent of the sign of $Q_a$, 
as ${\cal A}$ is an even function.
We still have
\begin{align}
G_{1122}=2\af_1 \; {\rm Re} \, G_{AARR}(Q_a) \; .
\end{align}
Finally, the collinear rate is given by
\begin{align}\label{LPM_e}
\sknew\frac{d\vGm_\gm}{d^3\sknew}=\frac{\af_\text{em}\sum_fq_f^2}{4\pi^2}
\int\frac{d k_\parallel}{2\pi}\sum_e \; A_e(\pl,\sknew)\;
\int\frac{d^2 k^2_\perp}{(2\pi)^2}{\rm Re} [2{\bf \pt}\cdot {\bf f_e}_\perp(\spnew,\sknew)]\; ,
\end{align}
where
\begin{equation}
\label{eq:A-def}
A_e(\pl,\sknew)=\nf_e(\pl+\sknew)(1-\nf_e(\pl))
\left(\frac{\pl^2+(\pl+\sknew)^2}{2\pl^2(\pl+\sknew)^2}\right) \; .
\end{equation}
Note that the factor $\Nc$ in Eq.~\eqref{LPM} is replaced by a 
sum over color index $e$ in Eq.~\eqref{LPM_e}. 

To summarize, the collinear rate in the presence of nontrivial Polyakov loop 
is given by Eq.~\eqref{LPM_e}, with 
${\bf f_e}_\perp(\spnew,\sknew)$ the solution of
Eq.~\eqref{f_inteq_etr}.  All quantities which depend upon
the background charge are defined in Eqs.~\eqref{eq:quark-asymptotic-mass}, 
\eqref{mg2_L},
\eqref{Egrr_e}, \eqref{Pi_e}, and \eqref{coll_e}.
We note that the Polyakov loop effect enters separately in both 
the longitudinal and the transverse parts, as can be seen from 
Eqs.~(\ref{Pi_e}), (\ref{LPM_e}) and (\ref{eq:A-def}).
In the longitudinal part, 
the Polyakov loop dependence is reflected in the distribution 
function factor $\nf_e(\pl+\sknew)(1-\nf_e(\pl))$.
In the transverse part, 
the Polyakov loop effect appears in the asymptotic quark thermal 
mass $m_e$ and the gluon Debye mass $\mD_e$.

\subsection{Photon rate in the collinear regime at large N }

To obtain the photon rate in collinear regime, we need to solve Eq.
\eqref{f_inteq_etr}. In the limit of large $\Nc$, 
the collision term in Eq.~\eqref{f_inteq_etr} is suppressed by $1/\Nc$. 
This can be understood as follows.
Since $\delta E_e\sim m^2_e/k_\parallel\sim g^2T^2N/k_\parallel$, 
the left-hand side of Eq.~\eqref{f_inteq_etr} 
is of order $g^2T^2\Nc/k_\parallel {\bf f_{e\perp}}$.
On the other hand, by using Eq.~(\ref{Gm_e}) 
the terms containing $C_e$ in the right-hand side are 
of order $\Gamma_e {\bf f_{e\perp}}\sim g^2T {\bf f_{e\perp}}$.
Thus, at sufficiently large $\Nc$, 
terms in the latter are small compared to the former.

This allows us to solve Eq.~\eqref{f_inteq_etr} 
perturbatively. 
The solutions to zeroth and first order in terms of $C_e$ 
are easily obtained.  
In the argument of all quantities, we suppress
$\sknew$ but indicate $\bp$:
\begin{align}
&{\bf f_{e\perp}}^{(0)}=\frac{2\bp_\perp}{i\dlt E_e(\bp)} \; , \no
&{\bf f_{e\perp}}^{(1)}=\frac{1}{i\dlt E_e(\bp)}\int\frac{d^3q}{(2\pi)^3}
\; C_e(q)\; 
\bigg[\frac{2(\bp_\perp+\bq_\perp)}{i\dlt E_e(\bp+\bq)}
-\frac{2\bp_\perp}{i\dlt E_e(\bp)}\bigg]
\; .
\end{align}
Only the solution to first order 
contributes to the photon rate. 
The relevant combination is
\begin{align}\label{pt_int_f}
\int\frac{d^2\pt}{(2\pi)^2}
\; {\rm Re}[2\bp_\perp\cdot {\bf f_{e\perp}}(\bp)]
=4\int\frac{d^2\pt}{(2\pi)^2}\int\frac{d^3q}{(2\pi)^3}
\; C_e(q)\;
\bigg[\frac{\pt^2}{\dlt E_e(\bp)^2}
-\frac{\bp_\perp\cdot(\bp_\perp+\bq_\perp)}{\dlt E_e(\bp)\dlt E_e(\bp+\bq)}\bigg] \; .
\end{align}
We note that, because of this truncation, the LPM effect is suppressed.
Using the sum rules of Refs.~\cite{Aurenche:2000gf,Aurenche:2002pd},
\begin{align}
\int\frac{dq^0dq^\parallel}{2\pi}
\; \dlt(q^0-q^\parallel)
\; (-iG^{RR}_{\mu\nu,e}(Q){\widehat \lknew}^\mu{\widehat \lknew}^\nu)
=T\(\frac{1}{q_\perp^2}-\frac{1}{q_\perp^2+\mD_{e}^2}\) \; ,
\end{align}
Eq.~\eqref{pt_int_f} simplifies to
\begin{align}\label{dpdq}
\int\frac{d^2\pt}{(2\pi)^2}
\; {\rm Re}[2\bp_\perp\cdot {\bf f_{e\perp}}(\bp)]&=
2g^2\(\frac{2\pl(\pl+\sknew)}{\sknew}\)^2
\int\frac{d^2\pt}{(2\pi)^2}\int\frac{d^2q_\perp}{(2\pi)^2}
\; \frac{T\mD_{e}^2}{q_\perp^2(q_\perp^2+\mD_{e}^2)} \no
&\quad\times\frac{1}{\pt^2+m_e^2}\(\frac{\pt^2}{\pt^2+m_e^2}
-\frac{\bp_\perp\cdot(\bp_\perp+\bq_\perp)}{|\bp+\bq|^2+m_e^2}\) \; \nonumber \\
&= -\frac{2g^2T}{(2\pi)^2}\(\frac{\pl(\pl+\sknew)}{\sknew}\)^2
\int d\pt^2\int dq^2_\perp
\; \frac{\mD_{e}^2}{q_\perp^2(q_\perp^2+\mD_{e}^2)} \no
&\quad\times\frac{1}{\pt^2+m_e^2}\(\frac{m_e^2}{\pt^2+m_e^2}
-\frac{q^2_\perp+2m^2_e}{2\sqrt{(k^2_\perp+q^2_\perp+m^2_e)^2-(2k_\perp q_\perp)^2}}\) \; ,
\end{align}
where in the second line we have performed the two angular integrations, 
and used the formula~\cite{Aurenche:2000gf},
\begin{align} 
\int^\infty_0 d\pt^2 \(\frac{1}{\pt^2+m_e^2}
-\frac{1}{\sqrt{(k^2_\perp+q^2_\perp+m^2_e)^2-(2k_\perp q_\perp)^2}}\) 
=0\; .
\end{align}
Here $\pt$ and $q_\perp$ are of order $gT$. 
Nevertheless, since
the integrand is convergent in both the infrared and the ultraviolet,
we can extend the range of the integrations of $\pt$ and $q_\perp$ 
to $[0,\infty]$. 
The result can be expressed in terms of a dimensionless function of the mass ratio $\mD_{e}/m_e$:
\begin{align}\label{fg_defining}
\int\frac{d^2\pt}{(2\pi)^2}\;
{\rm Re}[2\bp_\perp\cdot {\bf f_{e\perp}}(\bp)]&\equiv
2g^2\(\frac{2\pl(\pl+\sknew)}{\sknew}\)^2\; T \; 
\fg\left(\frac{\mD_{e}}{m_e}\right) \; ,
\end{align}
where 
\begin{align}
\begin{split}
\fg\left(\frac{\mD_{e}}{m_e}\right)
&\equiv -\frac{1}{(4\pi)^2} \int^\infty_0 dk^2_\perp
\Biggl[\frac{m^2_e}{(k^2_\perp+m^2_e)^2}\ln\left|\frac{m^2_e\mD^2_e}{(k^2_\perp+m^2_e)^2}\right| \\
&~~~-\frac{\mD^2_e-2m^2_e}{2\sqrt{A}(k^2_\perp+m^2_e)}
\ln \frac{\mD^2_e(m^2_e-k^2_\perp-\mD^2_e-\sqrt{A})}{\mD^2_e(m^2_e-k^2_\perp-\mD^2_e)+A-(k^2_\perp+m^2_e)\sqrt{A}}\Biggr] \; ,
\end{split}
\end{align}
with $A\equiv (\mD^2_e)^2-2\mD^2_e(m^2_e-k^2_\perp)+(k^2_\perp+m^2_e)^2$.
The function $\fg(\mD_{e}/m_e)$ can be determined numerically. 

Consequently, the collinear rate can be expressed as
\begin{align}\label{dpl}
\sknew\frac{d\vGm_\gm}{d^3\sknew}=\frac{\af_\text{em}\sum_f q_f^2}{4\pi^2}\sum_e
\int \frac{d\pl}{2\pi}\; {\color{black}{\nf_{e}(\pl+\sknew)(1-\nf_{e}(\pl))}}
\; 4g^2T \; \fg\left(\frac{\mD_{e}}{m_e}\right)
\left(\frac{\pl^2+(\pl+\sknew)^2}{\sknew^2} \right) \; .
\end{align}
The final $\pl$-integral can be done as follows:
\begin{align}
&\int\frac{d\pl}{2\pi}\left(\frac{\pl^2+(\pl+\sknew)^2}{\sknew^2}\right)
\; {\color{black}{\nf_e(\pl+\sknew)\; (1-\nf_{e}(\pl))}} \no
=&\int_{-\infty+iQ_e}^{+\infty+iQ_e}
\frac{dl}{2\pi}\left(\frac{2l^2
+\sknew^2/2-2Q_e^2-4i l\, Q_e}{\sknew^2}\right)
\nf\left(l{\color{black}{+}}\frac{\sknew}{2}\right)
\nf\left(-l{\color{black}{+}}\frac{\sknew}{2}\right) \; ,
\end{align}
where $l=\pl+\sknew/2+iQ_e$. The integrand is 
exponentially suppressed as $\text{Re}\,l\to\pm\infty$, 
which allows us to shift the integration contour to the real axis. 
The following integration formulas are useful:
\begin{align}\label{l_integrals}
&\int\frac{dl}{2\pi}\nf\(l{\color{black}{+}}\frac{\sknew}{2}\)
\nf\(-l{\color{black}{+}}\frac{\sknew}{2}\)
=\frac{1}{2\pi} \; \frac{\sknew}{e^{\sknew/T}-1}\; , \no
&\int\frac{dl}{2\pi}l^2\nf\(l{\color{black}{+}}
\frac{\sknew}{2}\)\nf\(-l{\color{black}{+}}
\frac{\sknew}{2}\)=\frac{1}{2 \pi} \; 
\frac{\sknew(4\pi^2T^2+\sknew^2)}{12(e^{\sknew/T}-1)} \; .
\end{align}
They can be obtained by integrating $
l\;\nf\(l+\frac{\sknew}{2}\)\nf\(-l+\frac{\sknew}{2}\)$ 
and $l^3\nf\(l+\frac{\sknew}{2}\)\nf\(-l+\frac{\sknew}{2}\)$ 
along the rectangular contour bounded by $-\infty$, $\infty$, 
$\infty+2\pi Ti$ and $-\infty+2\pi Ti$.
Using Eq.~\eqref{l_integrals}, we obtain the 
collinear rate from Eq.~\eqref{dpl}
\begin{align}\label{l_result}
%&\int\frac{d\pl}{2\pi}\left(\frac{\pl^2+(\pl+\sknew)^2}{\sknew^2}\right)
%\; \nf_e(\pl+\sknew)\; (1-\nf_{e}(\pl)), \no
\sknew\frac{d\vGm_\gm}{d^3\sknew}
=\frac{\af_\text{em}\af_s\sum_f q_f^2}{\pi^2} \; 
2\,  T \; \sum_e\fg\left(\frac{\mD_{e}}{m_e}\right)\;
\frac{2\pi^2T^2+2\sknew^2-6Q_e^2}{3\sknew(e^{\sknew/T}-1)} \; .
\end{align}
For a hard photon, where $\sknew\gg T, Q_e$, the collinear rate is simplified to
\begin{align}\label{rate_N}
\sknew\frac{d\vGm_\gm}{d^3\sknew}&\simeq
\frac{\af_\text{em}\af_s \sum_f q_f^2}{\pi^2} \; 
\frac{4\, T\sknew}{3} \; \sum_e\fg\left(\frac{\mD_{e}}{m_e}\right)\;
e^{- \sknew/T}\;  ,
\end{align}
whose parametric behavior is a Boltzmann factor times a term linear
in $\sknew$.  
This $p$-dependence is consistent with AMY's analysis 
without the LPM mechanism, and the analysis at 
two-loop order~\cite{Aurenche:1998nw}.  

\begin{figure}
\includegraphics[width=0.75\textwidth]{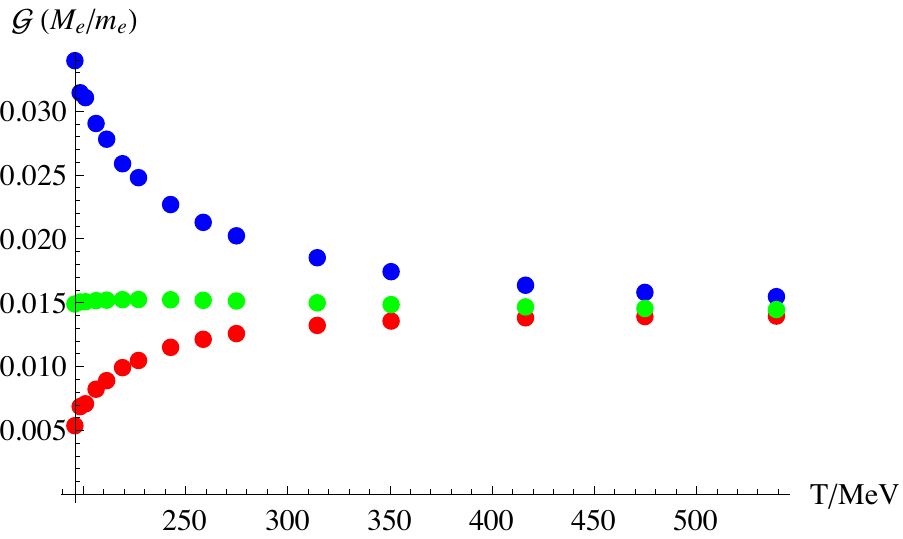}
\caption{Temperature dependence of function $\fg({M_{e}}/{m_e})$. 
In $\Nc=3$, the background color charge is parametrized as 
$Q_e=(-Q,0,+Q)$. Points of different colors in the figure 
correspond to $e=1$ (red), $e=2$ (blue) 
and $\fg$ averaged over three colors (green).}\label{fig_mr}
\end{figure}

Note that the Polyakov loop only enters through the sum
$\sum_e \fg(\mD_{e}/m_e)$. 
In Fig.~(\ref{fig_mr}) we show the temperature dependence of this 
function when $\Nc=\Nf=3$. While each individual term 
$\fg(\mD_{e}/m_e)$ changes with temperature, especially near $T_c$,
the sum is remarkably flat, with
$\sum_e \fg(\mD_{e}/m_e) \simeq 3\times 0.015$ 
over a wide range of temperature. 

\begin{figure}
\includegraphics[width=0.30\textwidth]{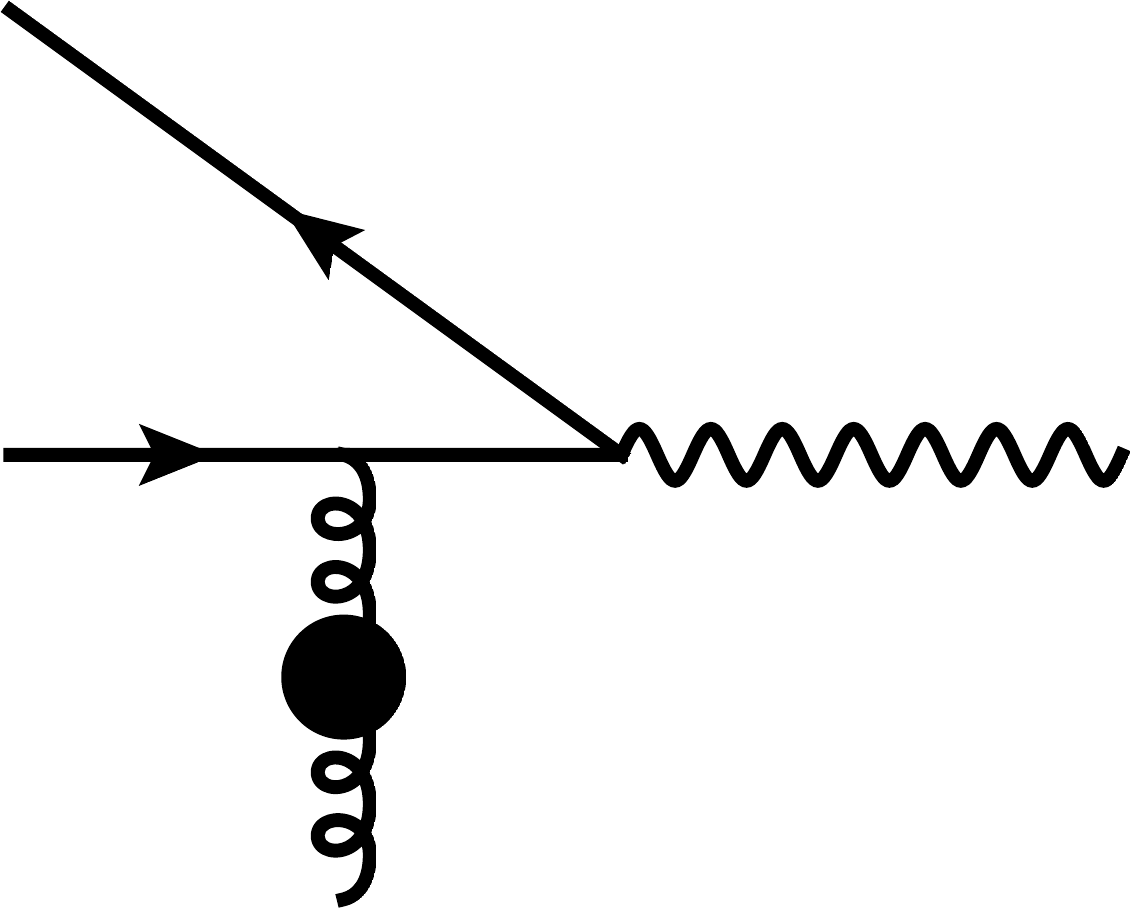}
\caption{The collinear pair annihilation process.
The curly line with a blob is the gluon HTL-resummed propagator.}
\label{fig:pair-LPM}
\end{figure}

From Eq.~(\ref{rate_N}) the collinear rate is not suppressed
in the confined phase.   At first this is a surprising result, and
it is worth discussing in some detail.
It happens because the soft gluon which is radiated is diagonal
in color space, so the quarks in the initial and final state have
the same color indices.
The distribution factor which appears in Eq.~(\ref{dpl}) is 
$\nf_e(\pl+\sknew)\; (1-\nf_{e}(\pl))
= \nb(p)(\nf_e(\pl)-\nf_e(\pl+\sknew))$.
For large $p > 0$, this factor is nonzero only when 
$k_\parallel+p$ is positive, and $k_\parallel$ is negative
~\cite{Arnold:2001ms, Aurenche:1998nw}.
This corresponds to pair annihilation, as illustrated in
Fig.~(\ref{fig:pair-LPM}); the other processes correspond to bremsstrahlung,
and do not contribute in this limit.

Since $k_\parallel+p$ is positive, $k_\parallel$ is not only negative,
but large.  Consequently, as $p \gg T$, we can use a Boltzmann
approximation for the statistical distribution functions:
\begin{align}
\label{eq:LPM-distributions}
\begin{split}
\frac{1}{\Nc}\sum_e\nf_e(\pl+\sknew)\; (1-\nf_{e}(\pl))
&= \frac{1}{\Nc}\sum_e\nf_e(\pl+\sknew)\; \nf_{\overline{e}}(-\pl) \\
& \simeq \frac{1}{\Nc}\sum_e e^{-(\pl+\sknew-iQ_e)/T}\; e^{-(-\pl+iQ_e)/T}\\
&= e^{-\sknew/T} \; ,
\end{split}
\end{align}
Thus the collinear contribution is not suppressed in the confined phase
because the phases cancel between the quark and anti-quark.
This is exactly the same cancellation as found for dilepton production,
and rather unlike the color flow for the
contribution to photon production from $2 \rightarrow 2$ scattering.

This completes our derivation of photon rate in the semi-QGP,
with a nontrivial Polyakov loop at large $\Nc$.
The result is a sum of leading logarithmic term from the rate for
$2\lra2$, Eq.~\eqref{2to2_rate}, and 
the collinear rate in the large $\Nc$ limit, Eq.~\eqref{l_result}. 
We emphasize that rates for $2 \lra 2$ and collinear emission
depend upon the Polyakov loop in completely different ways.
When the Polyakov loop is small, the rate for $2 \lra 2$ is
suppressed while that for collinear emission is not.
We note that our results are valid only for
small values of $g^2$ and large $\Nc$. 
At moderate values of the coupling constant, 
corrections due to the constant under the logarithm
become important. At moderate values of $\Nc$, 
the LPM effect becomes relevant, and will produce cancellations 
between diagrams with different number of loops, suppressing 
the photon rate in the collinear regime.

\subsection{Why the LPM effect is suppressed in the semi-QGP}

A nontrivial Polyakov loop is understood as from quantum fluctuations
in $A_0$ of order $T/g$. This background gauge field 
affects quarks and gluons in different ways.  
As shown in Ref.~\cite{Lin:2013efa}, it
reduces the density of hard quarks.  
It also acts as a Higgs effect for gluons,
giving mass of order $T$ to off-diagonal gluons, while
leaving diagonal gluons massless. 
The only gluons which scatter off of quarks
in the large $\Nc$ limit are diagonal, and so are 
reduced by $1/\Nc$.

In the perturbative QGP, the 
LPM effect is relevant because the photon formation time, $t_F$, 
is comparable to the mean free path, $\lam$,  of a 
quark undergoing multiple scattering with gluons in the medium. 
The formation time is the time scale when a collinear photon can 
be well separated from the quark, which is 
\begin{align}
t_F\sim\frac{1}{\dlt E}\sim \frac{T}{k_\perp^2+m^2}\sim \frac{1}{g^2\Nc T} \; .
\end{align}
On the other hand, the mean free path has the same order of magnitude as the 
damping rate of a quark in the thermal bath, 
with $\lam\sim 1/\Gm \sim 1/(g^2\Nc T)$, which is comparable to $t_F$.

A nontrivial Polyakov loop modifies the two scales differently. 
The thermal mass $m$ results from interactions of a quark with 
hard thermal gluons. The Polyakov loop 
suppresses the quark and the gluon density, and thus also $m$, 
by a loop dependent factor. 
The damping rate is due to the scattering off of soft gluons,
but as these are suppressed for $Q^a \sim T$,
only the scattering off of soft, diagonal gluons matters.
Consequently, $\lam \sim 1/(g^2 T)$ times a loop dependent factor,
so at large $\Nc$, $\lam \gg t_F$. This implies that quarks rarely scatter 
more than once during the emission of a photon, 
and thus the LPM effect can be ignored.

\section{Summary and Concluding Remarks}

We calculated the production rates of the dilepton and the real photon
in a matrix model of the semi-QGP.
The main results of this paper are Eqs.~(\ref{define_fll}), 
(\ref{2to2_rate}), and (\ref{rate_N}).
The dilepton production rate was found to be slightly 
enhanced in the confined phase due to a cancellation in the 
phases of the statistical distribution functions for the quark and anti-quark
~\cite{previous}.
By contrast, the photon production rate due to the $2 \rightarrow2$ 
scattering is strongly suppressed for small values of the Polyakov loop,
as the phases in the distribution functions do not cancel.
We showed that 
the collinear contribution to the photon production is suppressed
at large $\Nc$ in the semi-QGP, since when the Polyakov loop is
small, the $Q^a$'s are large, and off-diagonal gluons do not experience
Bose-Einstein enhancement.
We computed the collinear contribution at large $\Nc$, and 
found that because of a cancellation
of phases, like dilepton production
it is not suppressed even in the confined phase.

These results will modify the theoretical predictions for
thermal production in heavy ion collisions.
Certainly the production rates for dileptons and photons are altered.
The elliptic flow for these particles are similarly modified,
as the total elliptic flow is an
average over all the phases, from the initial state, to the QGP,
to hadrons.
These effects were previously discussed in Ref.~\cite{previous}.
However, in that work 
the modifications of photon production from 
$2 \rightarrow2$ scattering and from collinear emission
were {\it{not}} considered separately.  Clearly a more realistic treatment
is called for.  

In the current analysis, the effect of the confinement is taken into 
account as a nontrivial value of the Polyakov loop.
It is also interesting to consider the effect of the chiral symmetry 
restoration as well as confinement~\cite{Islam:2014sea, Satow-Weise}. 

For the future, besides doing a more complete analysis of photon
production, the most urgent problem is to compute radiative energy loss
for light quarks.  This is closely related to collinear photon emission,
and so we expect that near $T_c$, it will be dominated by diagonal
gluons for processes in which the color phases cancel.

\begin{acknowledgments}
R.D.P. would like to
thank C. Islam, S. Majumder, N. Haque, and M. Mustafa for discussions
about their work on dilepton production
in the PNJL model \cite{Islam:2014sea} in Mumbai, at
the workshop 
\href{http://theory.tifr.res.in/~qcd2015}{``QCD at high density''},
and in Kolkata, at the
\href{http://indico.vecc.gov.in/indico/internalPage.py?pageId=37&confId=29}{``7th International Conference on Physics and Astrophysics of the Quark-Gluon Plasma''}. 
S.L. would like to thank B. Wu and L. Yaffe for useful discussions.
Y.H. is supported by JSPS KAKENHI (Grants No. 24740184), and 
by the RIKEN iTHES Project.
S.L. is supported by the RIKEN Foreign Postdoctoral Researchers Program.
R.D.P. is supported by DOE Contract DE-SC0012704 and by the 
RIKEN/BNL Research Center.
D.S. is supported by JSPS Strategic Young
Researcher Overseas Visits Program
for Accelerating Brain Circulation (No. R2411).

\end{acknowledgments}

\vskip 1cm

\appendix

\section{Corrections to Boltzmann approximation to thermal distribution functions}

We will argue that correction to \eqref{boltzmann} is suppressed by 
additional exponential. We illustrate this in case of Compton scattering.
The exact thermal distribution factors can be expressed as
\begin{align}\label{exact}
&\frac{1}{e^{\bt E_1}+1}\frac{1}{e^{\bt E_2}-1}\(1-\frac{1}{e^{\bt E'}+1}\) \no
=&\sum_{m,n=0}^\infty(-1)^me^{-(m+n+{2})\bt x/2-(m-n)\bt y/2}\(1-\frac{1}{e^{\bt(x-E)}+1}\) \;.
\end{align}
Now the $y$-integral becomes
\begin{align}
\int\frac{dye^{-(m-n)\bt y/2}}{\sqrt{ay^2+by+c}}=\frac{\pi}{\sqrt{-a}}e^{-(m-n)\bt y_0/2}I_0(\frac{m-n}{2}\bt\Dlt y) \; ,
\end{align}
where
\begin{align}
y_0&=\frac{t-u}{t{+}u}(x-2E) \;, \no
\Dlt y&=-\frac{2\sqrt{tu(t+u+4E(x-E))}}{t+u},
\end{align}
{and $I_0(z)$ is the modified Bessel function of the first kind.}
Note that the leading logarithmic contribution comes from $t\sim\mu^2$, $s\sim p\,T$, which implies $\Dlt y\sim \mu$. Therefore we may set $I_0((m-n)\bt\Dlt y/2)=1$. This leads to the following $x$-integral
\begin{align}
&\int_{p+\frac{s}{4p}}^\infty dxe^{-(\bt+\dlt) x}\(1-\frac{1}{e^{\bt(x-p)}+1}\) \no
=&\frac{e^{-\bt p-\dlt x}}{\dlt}F(1,-\frac{\dlt}{\bt},1-\frac{\dlt}{\bt},-e^{\bt(x-p)})\vert_{x=p+s/(4p)} \no
%=&\frac{e^{-\bt p-\dlt(p+s/(4p))}}{\dlt}\frac{\Gm(1-\frac{\dlt}{\bt})\Gm(-1-\frac{\dlt}{\bt})}{\Gm(-\frac{\dlt}{\bt})^2}e^{-\bt s/(4p)}F(1,1+\frac{\dlt}{\bt},2+\frac{\dlt}{\bt},-e^{-\bt s/(4p)}),\no
=& {\frac{e^{-(\bt+\dlt )(p+s/(4p))}}{\delta+\beta}F(1,1+\frac{\dlt}{\bt},2+\frac{\dlt}{\bt},-e^{-\bt s/(4p)})\; ,}
\end{align}
with $\dlt={\beta(m+n)/2+\beta(m-n)(t-u)/(2(t+u))}>0$ unless $m=n=0$. 
{Here, $F(a,b,c,z)$ is the hypergeometric function.}
For non-vanishing $\dlt$, there is an additional exponential suppression factor $e^{-\dlt(p+s/(4p))}$. Therefore, we conclude any terms with non-vanishing $m$ or $n$ is negligible in {Eq.}~\eqref{exact}, leaving only the term with $m=n=0$, which corresponds to the Boltzmann approximation.

\section{Thermal gluon mass in the presence of Polyakov loop}

We regard $\Nf\sim \Nc$ as a large number. 
%According to {Eq.}~\eqref{gluon_Pi}, $F(Q_a)$ and $G(Q_a,Q_b)$ are of the same order in $\Nc$:
%\begin{align}
%&F(Q_a)\simeq \frac{g^2}{3}\(\sum_{e=1}^\Nc\Acal(Q^{ae}{)}-\Nf\widetilde{\Acal}(Q^a)\) \no
%&G(Q_a,Q_b)\simeq \frac{g^2}{3}\(\Nc\Acal(Q^{ac})-\Nf\(\widetilde{\Acal}(Q^a)+\widetilde{\Acal}(Q^c)\)-\sum_{e=1}^\Nc\widetilde{\Acal}(Q^e)\).
%\end{align}
Naively, the $F$ and $G$ terms give the same order contribution in $\Nc$ because $\dlt_{ab}\sim1/\Nc$, however, as we show below, the $G$ term is suppressed by $1/\Nc$ compared to the $F$ term. By plugging {Eq.}~\eqref{Pi_color} into {Eq.}~\eqref{series}, we obtain the first few terms explicitly:
\begin{align}
&\frac{1}{Q^2}\(\dlt_{ab}-\frac{1}{\Nc}\) \; , \no
&\frac{1}{(Q^2)^2}\(\dlt_{ab}F_a-\frac{1}{\Nc}L_1(F,G)\) \; , \no
&\frac{1}{(Q^2)^3}\(\dlt_{ab}F^2_a-\frac{1}{\Nc}L_2(F,G)\) \; .
\end{align}
Here $L_1$($L_2$) are complicated functions linear(quadratic) in $F$ or $G$. By induction, we can obtain the form of propagator with $n$ self-energy insertions
\begin{align}
&\frac{1}{(Q^2)^{n+1}}\(\dlt_{ab}F^n_a-\frac{1}{\Nc}L_n(F,G)\).
\end{align}
It is easy to see the $\dlt_{ab}$ term can be summed as a geometric series, while the $1/\Nc$ term is not summable in simple manner. In any case, the resummed gluon propagator has the following color structure
\begin{align}\label{resum_AB}
\dlt_{ab}A_a(Q)-\frac{1}{\Nc}B_{ab}(Q) \; ,
\end{align}
with $A_a(Q)$ and $B_{ab}(Q)$ of the same order in $\Nc$. Note $A_a(Q)=(Q^2-F_a)^{-1}\dlt_{ab}$ is entirely from $F$, while $B_{ab}(Q)$ has contribution from both $F$ and $G$.

Now we insert the resummed propagator into the graphical element $M$. Focusing again on the color structure, we obtain after summing over gluon color indices:
\begin{align}\label{M_L}
&(T^{aa})_{ef}(T^{bb})_{hg}\big[\dlt_{ab}A_a-\frac{1}{\Nc}B_{ab}\big] \no
&=\dlt_{ef}\dlt_{hg}\frac{1}{2}\bigg[\dlt_{fg}A_f-\frac{1}{\Nc}(A_f+A_g)
+\frac{1}{\Nc^2}\sum_cA_c-\frac{1}{\Nc}B_{fg} \no
&\quad+\frac{1}{\Nc^2}\sum_c(B_{fc}+B_{cg})-\frac{1}{\Nc^3}\sum_{cd}B_{cd}\bigg] \; .
\end{align}
In the above, we have suppressed the $P$ dependence of $A$ and $B$ for notational simplicity. Formally all terms are of the same order if we regard $\dlt_{fg}\sim1/\Nc$ and sum as $\sim \Nc$. However, we have learned from the case without background color charge that the structure of $M$ is ultimately contracted with $\dlt_{eh}$ on the left, which brings {Eq.}~\eqref{M_L} into the following form
\begin{align}
&\dlt_{fg}\bigg[\dlt_{fg}A_f-\frac{1}{\Nc}(A_f+A_g)+\frac{1}{\Nc^2}\sum_cA_c-\frac{1}{\Nc}B_{fg} 
+\frac{1}{\Nc^2}\sum_c(B_{fc}+B_{cg})-\frac{1}{\Nc^3}\sum_{cd}B_{cd}\bigg] \; .
\end{align}
We see the $\dlt_{fg}$ in the first term of the bracket becomes redundant. We can replace it by $1$. Consequenctly, all other terms are suppressed by $1/\Nc$. We will keep only the first term, which is fortunately easy to calculate. This approximation amounts to dropping the $1/\Nc$ term in {Eq.}~\eqref{gluon_Pi}, leading to the gluon Debye mass, Eq.~(\ref{mg2_L}).
%\begin{align}
%M_a^2(Q\ne0)=\frac{g^2}{3}\(\sum_{e=1}^\Nc\Acal(Q^{ae}{)}-\Nf\widetilde{\Acal}(Q^a)\)\,.
%\end{align}

\bibliography{v8_LongPhoton}

\end{document}